\documentclass[aip,jcp,reprint,citeautoscript,floatfix]{revtex4-1}
\usepackage{natbib}
\usepackage{graphicx}
\usepackage[caption=false,lofdepth,lotdepth]{subfig}
\usepackage{import}
\usepackage{lmodern}
\usepackage[colorinlistoftodos,color=black!30]{todonotes}
\usepackage{amsmath, amsthm, amssymb, mathtools}
\usepackage[T1]{fontenc}
\usepackage{microtype}
\usepackage[utf8]{inputenc}
\usepackage{url}
\usepackage{enumerate}
\usepackage{physics} 
\usepackage{bm} 
\usepackage[hidelinks]{hyperref} 
\usepackage{booktabs,tabularx}
\usepackage{cleveref} 
\usepackage{accents}
\usepackage{layouts}
\usepackage{booktabs}
\usepackage{float}

\crefname{equation}{Eq.}{Eqs.}
\crefname{section}{Sec.}{Secs.}
\crefname{table}{Tab.}{Tabs.}
\crefname{figure}{Fig.}{Figs.}
\crefname{subfigure}{Fig.}{Figs.}

\usepackage[nolist,nohyperlinks]{acronym}

\usepackage{pgfplots}
\usetikzlibrary{pgfplots.groupplots}

\pgfplotsset{compat=1.5}

\usepackage{siunitx} 
\usepackage{multirow} 

\begin{document}

\title{
Enhanced path sampling using subtrajectory Monte Carlo moves
}
\author{Daniel T. Zhang}
\affiliation{Norwegian University of Science and Technology, Department of Chemistry,  NO-7491 Trondheim, Norway}
\author{Enrico Riccardi}
\affiliation{Department of Informatics, UiO, Gaustadall\'{e}en 23B, 0373 Oslo, Norway}
\author{Titus S. van Erp}
\affiliation{Norwegian University of Science and Technology, Department of Chemistry,  NO-7491 Trondheim, Norway}

\date{20 September 2022}
\begin{abstract}
Path sampling allows the study of rare events like chemical reactions, nucleation and protein folding via a Monte Carlo (MC) exploration in path space. Instead of configuration points, this method samples short molecular dynamics (MD) trajectories with specific start- and end-conditions.  As in configuration MC, its efficiency highly depends on the types of MC moves.  Since the last two decades, the central MC move for path sampling has been the so-called shooting move in which a perturbed phase point of the old path is propagated backward and forward in time to generate a new path. Recently, we proposed the subtrajectory moves, stone-skipping (SS) and web-throwing (WT), that are demonstrably more efficient. However, the one-step crossing requirement makes them somewhat more difficult to implement in combination with external MD programs or when the order parameter determination is expensive. In this article, we present strategies to address the issue. The most generic solution is a new member of subtrajectory moves, wire fencing (WF), that is less thrifty than the SS, but more versatile. This makes it easier to link path sampling codes with external MD packages and provides a practical solution for cases where the calculation of the order parameter is expensive or not a simple function of geometry. We demonstrate the WF move in a double well Langevin model, a thin film breaking transition based on classical force fields, and a smaller ruthenium redox reaction at the ab initio level in which the order parameter explicitly depends on the electron density.
\end{abstract}

\maketitle
\section{Introduction}

\label{sec:introduction}
Rare event simulation techniques aim to sample
events that 
require an exceedingly long CPU/wall time to be simulated with standard 
molecular dynamics (MD).  
In classical full atom simulations of protein folding, for example, the longest reported~\cite{howfast} MD runs
generated
by the special-purpose molecular dynamics Anton 1 supercomputer are around 1 ms,
allowing the study of fast-folding proteins. 
The most recently released Anton 3 supercomputer
is even able to generate 100 $\mu$s/day for a 
million atom system~\cite{Anton3}.
Despite this  a remarkable speed,
it is still not fast enough to study the folding of all proteins. 
For instance, 
the tryptophan synthase $\beta_2$ subunit has an
experimentally measured~\cite{slowfold}
folding rate of $k_f=0.001$ s$^{-1}$. Hence,
the protein 
needs on average 1000 seconds to fold. The Anton 3 computer
would thus need  27,379 wall time years to generate one single transition.
For ab initio MD (AIMD) the situation is even worse
as the quantum mechanical force evaluation is orders of magnitude slower than 
computing
 the gradient of 
a classical force field potential. 
In addition,
no special purpose AIMD computers exist today. 

Yet, 
rare event simulations allow
the calculation of rate constants and the study  reaction mechanisms  
orders of magnitude faster than MD, oftentimes 
without sacrificing any molecular-level resolution.~\cite{BaronBook}
(Replica exchange) transition interface sampling (RE)TIS~\cite{TIS, RETIS}
is such a method that exploits the idea of
transition path sampling (TPS)~\cite{TPS98}
to focus the CPU time on the actual barrier crossing event via 
a Monte Carlo (MC) sampling of MD paths.

RETIS and TIS 
employ a
series of path sampling simulations, each sampling a different path ensemble. 
The path ensembles differ with respect to a minimal progress 
requirement, i.e.
 the number of interfaces (defined by fixed values of the reaction coordinate/order parameter) that have  to be crossed.\cite{Raffa} 
Combining the results of all path ensembles 
allows the computation of rate constants and other properties 
with an exponentially reduced CPU time compared to a single MD simulation.

For instance, a classical simulation study on 
methane hydrate formation~\cite{methanehydrate}
using TIS and RETIS reports on a crystallisation rate of $10^{-17}$ nuclei per second per simulation volume.
In other words,  
in a system as small as those used in atomistic simulations,
the process for forming a single critical
nucleus takes physically 3 years.
Naturally, 
the hypothetical wall time for reaching this with MD 
is astronomical for any supercomputer.
Likewise, RETIS simulations~\cite{PNASwater}
 reproduced rate constant of water dissociation at the AIMD level 
in reasonable agreement with experiments suggesting 
it happens once per 11 hours for each water molecule.~\cite{eigen1958, natzle1985}
 As it required 
 30 minutes to produce  1 ps MD time in the
 32 water molecules system, 
 a naive straightforward AIMD 
approach would need 0.7 billion centuries wall time to 
generate a single dissociation.

Despite 
being
orders of magnitude times faster than plain 
MD,
simulations like the above are still 
computationally expensive and  can require months to years 
to obtain 
satisfactory
statistical accuracy.
A further increase in efficiency is therefore desirable. 
There are essentially three approaches to achieve this: 
i) reducing the cost of the MC moves, ii)
reducing the number of required trajectories, and iii) 
parallelization of the algorithms. 
Partial path sampling (PPTIS)~\cite{PPTIS} and 
milestoning~\cite{Milestoning} 
can be viewed as realisations of i) by 
sampling more restrictive path ensembles with a
reduced
average path length. 
 Unfortunately, this introduces additional approximations.
 Strategies ii) and iii), on the other hand, allow for a speed-up while still 
 producing exact results, 
identical 
 to those from 
 hypothetical unattainably long MD simulations.
In fact, RETIS  successfully 
employs strategy ii) by complementing 
the shooting moves with
replica exchange moves between path ensembles. 
RETIS is thus more CPU efficient compared to 
TIS. 
However,
regarding strategy iii), TIS has the advantage 
that 
path ensembles can be run in parallel completely independently,
while
replica exchange moves  require
the progress of the sampling in the
path ensembles to be synchronized such that
processing units do not have to wait for each other.
As a result, RETIS might not always outperform TIS 
based on wall time, which is the reason why previously mentioned hydrate formation study was partly based on TIS.~\cite{methanehydrate}
The recently introduced $\infty$RETIS algorithm~\cite{infRETIS} is expected to solve this 
issue for future studies based on a fundamentally new replica exchange 
technique for cost-unbalanced replicas.

In fact, $\infty$RETIS  implicitly applies 
the cost-free replica exchange moves 
an infinite number of times 
after 
each shooting move.
Still, replica exchange moves 
alone are not ergodic and should, therefore,
only be used in combination
with another MC move like shooting.~\cite{shoot} 
To further push strategy ii),
the principle MC move should be changed
in order to reduce both the rejection rate and the resemblance between accepted paths.
This is exactly what subtrajectory moves aim to establish.
These MC moves resemble PPTIS~\cite{PPTIS} and milestoning~\cite{Milestoning} in the sense
that they evolve via series of shorter paths 
(subtrajectories/subpaths), but differently to those methods, these subpaths are just intermediates between sampled paths that are extended to their full lengths. Sampled paths, therefore, have no configuration 
point in common with the previous path and the statistical inefficiency is typically reduced 
by a factor equal to the number of intermediate subtrajectories. 
So while the creation of a full new path becomes more expensive,
this is more than offset by the fact that far fewer trajectories are needed to achieve a certain statistical accuracy.
In addition, the approach can be combined with a high-acceptance protocol, which minimizes
the number of rejections. As a result, most path ensembles obtain a nearly 
100\% acceptance.\cite{riccardi_fast_2017}

The two moves presented in Ref.~\onlinecite{riccardi_fast_2017}, stone skipping (SS) and 
web throwing (WT), however, have  one element, the one-step crossing condition,
which can hinder the practical implementation with external MD programs or
when the calculation of the order parameter is computationally expensive. 
In SS and WT, the subtrajectories are launched from a configuration point of a previous (sub) path that is just before or after the path ensemble's interface. At this configuration 
point, velocities are generated such that the interface is crossed again
within a single time step.
The velocity randomization and one-step crossing test
is reiterated several times until the
condition is fulfilled. 
The procedure is based on the idea that generation of new random velocities 
followed by a one-step crossing test is relatively cheap compared to generating MD steps,
especially if the test can be performed without   new force calculations.
This might not always be the case.
Present path sampling codes~\cite{OPS1,OPS2,PyRETIS1,PyRETIS2} 
use external MD codes for performing the MD steps.
PyRETIS version 2 has for instance couplings to 
Gromacs,\cite{GROMACS} 
Lammps,\cite{lammps1995} openmm\cite{openmm}, and CP2K\cite{CP2K}. In order to reduce 
the number of stop/restart calls to these programs, a ``time step'' in the 
RETIS program is often
several (10-1000) MD steps by the external MD engine.
This complicates the one-step crossing condition as it actually involves not one, but several 
steps which is costly and not easy to predict without actually performing these steps. 
Another issue arises when the calculation of the order parameter is expensive such as 
the those used in
nucleation studies.~\cite{nucleation1, CPUnucleation2}

In this article, we discuss several approaches to tackle this issue. The most generically applicable solution is a new member of the subtrajectory family called
wire fencing (WF). The approach is slightly more wasteful with respect to
the number of
MD steps compared to SS, but very versatile and does not require any code modifications of the external engines.
We illustrate the WF move on three model systems, a simple 1D double well potential,
a Gromacs thin film breakage application, and a CP2K study on ruthenium redox reactions.

\section{subtrajectory moves}
\begin{figure}[ht!]
    \includegraphics[width=0.31\textwidth]{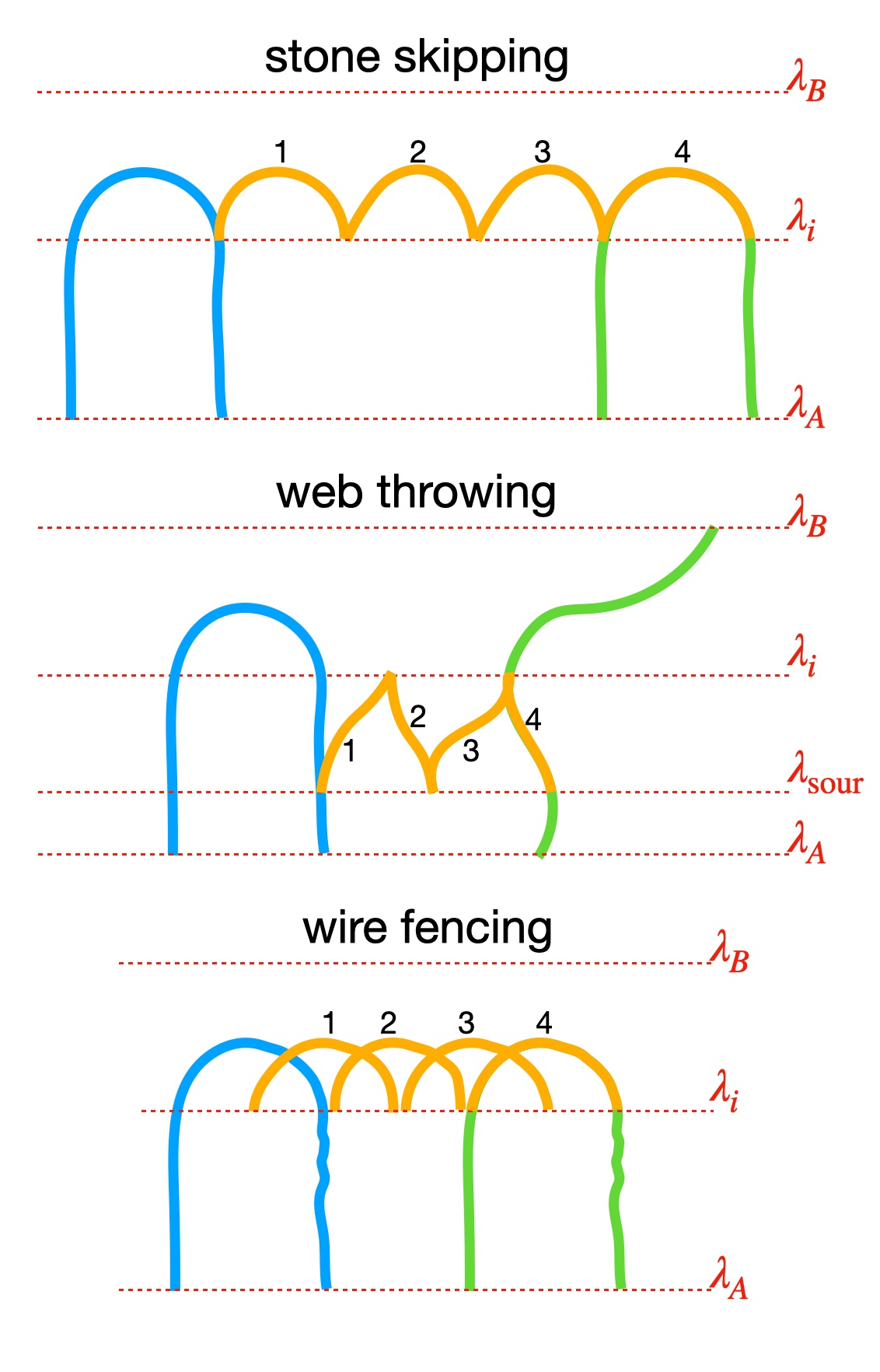}
    \centering
    \caption{Cartoon representation of the three subtrajectory moves: stone skipping, web throwing, and wire fencing. The old path is shown in blue. Four subtrajectories are shown in orange. The final new path consists of the fourth subtrajectory and its extensions colored in green. }
    \label{fig:3moves}
\end{figure}

The schematic main idea of the three subtrajectory moves is shown in Fig.~\ref{fig:3moves}. 
These are the stone skipping (SS), web throwing (WT), and the new wire fencing (WF) move.
The commonality is that an arbitrary number of partial trajectories (subtrajectories/subpaths) are generated before the completion of a new full trajectory. 
The subtrajectories obey different start- and end-conditions and are, due to this, considerably shorter than full trajectories. The subtrajectories are not part of the sampling, but just intermediate steps between one full trajectory to another.
The  $[i^+]$ path ensemble that is being sampled in Fig.~\ref{fig:3moves} consists of paths starting at $\lambda_A$, crossing $\lambda_i$ at least once, and ending at either $\lambda_A$ or $\lambda_B$.
In Fig.~\ref{fig:3moves}, the old full trajectory is colored blue. In the example, the new trajectory is generated via 4 subtrajectories.
The first subtrajectory is obtained from a shooting
move from the old trajectory. Then, a next subtrajectory is generated from the previous one until the number of predetermined subtrajectories (4 in this case, colored in orange) is reached.
The final subtrajectory is extended backward and forward in time until reaching a stable state. The new full
trajectory
comprises the last subtrajectory and the extensions 
colored in green.
The difference between  the three moves lies in the way the shooting of subpaths is executed. 

The SS move resembles a flat stone that collides with the water's surface after a skilful throw. The move starts by selecting randomly any of the crossing points
of the old path with $\lambda_i$, generates new velocities that also establish a crossing, and then proceeds until $\lambda_B$ is crossed or 
$\lambda_i$ is crossed again. The process is then repeated by selecting  the subpath's last crossing with
$\lambda_i$
for shooting off the next subpath.
Finally, the last subpath is extended and possibly accepted or rejected.\cite{riccardi_fast_2017}

The WT move has been named after a gesture of the famous Marvel character  swinging between skyscrapers. Here, an additional interface 
needs to be defined, \emph{the surface of unlikely return} (SOUR), at the state $A$ side of the $\lambda_i$ interface. If this interface, $\lambda_{\rm sour}$, is crossed towards the direction of state $A$, it is assumed to be highly unlikely that the MD trajectory will end up in state $B$ rather than $A$ (defined by the last interface, $\lambda_B$, and the first interface, $\lambda_A$, respectively). 
The first subpath is then shot from a random crossing point with either  
$\lambda_{i}$ or $\lambda_{\rm sour}$ at a path segment of the old path that
connects these two interfaces.
After the velocities of the system's atoms are re-set, like in the SS move, the subpath is continued till  $\lambda_{\rm sour}$ or $\lambda_{i}$ is crossed, but is only kept if the subpath connects $\lambda_{\rm sour}$ and $\lambda_{i}$ again like the segment of the old path. If not, the subpath is rejected and a new crossing point is taken randomly from the same segment.  
If both $\lambda_{\rm sour}$ and $\lambda_{i}$ are crossed, the subpath replaces the segment.  The process is repeated until the selected number of subpaths, accepted or rejected, has been completed. 
The final accepted subpath is extended in both 
time-directions to
make a full new path.
Note that a rejection of a subpath does not imply a rejection of the MC move itself, but just redirects the process of achieving a new path
 from an old path. The time-direction is chosen such that from $\lambda_{\rm sour}$ the trajectory is propagated backward in time and from $\lambda_{i}$ forward in time. Due to the placement of 
$\lambda_{\rm sour}$, it is nearly guaranteed that the backward extension reaches state $A$.
As $\lambda_{i}$ is also crossed, it is ensured that the path is valid for the $[i^+]$ ensemble, thought it might still be rejected due to a final acceptance/rejection step, as required by detailed balance~\cite{Metroplois}.

The WF move, further discussed in Sec.~\ref{sec:WF_move}, differs with the other moves by its location of the shooting points.  In the WF move, these might be any point with a corresponding value
of the reaction coordinate that is larger than $\lambda_i$ and lower than
$\lambda_B$ (or $\lambda_{\rm cap}$ if a so-called \emph{cap interface} is set, see Sec.~\ref{sec:WF_move}).
From this point, no specific requirements are needed for the velocities so that they are most conveniently generated
from a Maxwell-Boltzmann distribution for the temperature of interest.
From the new phase point, MD steps are generated forward and backward in time until $\lambda_B$ (or $\lambda_{\rm cap}$) or $\lambda_i$ is crossed.
The subpath is accepted unless it reaches $\lambda_B$ (or $\lambda_{\rm cap}$) in
both time directions. In that case it would be rejected and the next shot is taken again from latest accepted subpath or the previous segment of the old path
if no accepted subpaths yet exist.
After finishing the number of desired subpaths, the last accepted one
is extended to the stable states,
like in SS and WT.
While the WF move is slightly more wasteful with respect to the MD moves compared to SS, the velocity generation is much simpler which can 
have both practical and fundamental advantages compared to SS and WT.
These are further discussed in Sec.~\ref{sec:one-step}.
The name of the WF move
is derived from the visual resemblance between the set of full paths and subpaths and the top of a wire fence.

The  subtrajectory moves go against strategy i) as these MC moves require more MD steps than just the number of MD steps for generating a new path. These moves are nevertheless more efficient 
because 
they 
 utilize strategy ii): the statistical inefficiency of the sampling is reduced and, therefore, fewer trajectories are needed to achieve a desired statistical error. 
Like with the standard shooting move, 
a final acceptance/rejection step should ensure
that the correct statistical distribution of paths is sampled.
However, due to the complexity of the subtrajectory move,
the design and mathematical validation 
of the acceptance rule
is substantially more complex and is derived from 
the so-called
superdetailed balance~\cite{FrenkelBook} principle.

\section{Superdetailed balance}
\label{sec:supdetbal}
The term superdetailed balance
was first introduced within the context of 
configurational bias MC (CBMC),~\cite{Siepmann92,Vlugt99,FrenkelBook}
which is an effective method to study the adsorption of polymers in nanoporous materials such as zeolites.
In this algorithm, polymers are removed and then regrown atom by atom such that any overlap between the polymer and the zeolite's walls and  other polymers
is avoided. In this growth process, several attempted branch formations are tested and potentially rejected. Therefore, a specific final accepted configuration could, in principle,
be obtained from the old configuration via an infinite number
of ways (construction paths). As a result, the Metropolis-Hastings~\cite{Hastings} rule for deriving acceptance probabilities 
becomes impractical as it requires the knowledge on the 
generation probabilities of
all these branches, accepted and rejected, that need to be summed up.
This issue is overcome in CBMC using the superdetailed balance principle, which can be formulated in terms of
a construction path $\chi$ and its inverse
$\overline{\chi}$.\cite{riccardi_fast_2017}
That is, we not only require detailed balance
between any possible old state and  new state, 
but we require this
for any specific 
route 
that connects these two states:
\begin{align}
P_{\rm acc}=
{\rm min}\left[ 1,
\frac{P({\rm path}^{(n)})
P_{\rm gen}({\rm path}^{(n)} \rightarrow 
{\rm path}^{(o)} \textrm{ via } \overline{\chi})
}{
P({\rm path}^{(o)})
P_{\rm gen}({\rm path}^{(o)} \rightarrow {\rm path}^{(n)} \textrm{ via } \chi)
}
\right]
\label{eq:supdet}
\end{align}
where $
P_{\rm gen}({\rm path}^{(o)} \rightarrow 
{\rm path}^{(n)} \textrm{ via } {\chi})
$ is the generation probability to
to generate the new state (path in our case) from
the old state via construction path $\chi$
and
$P_{\rm gen}({\rm path}^{(n)} \rightarrow 
{\rm path}^{(o)} \textrm{ via } \overline{\chi})
$
is 
the generation probability to
to generate the old state  from
the new state via the reverse construction path $\overline{\chi}$.

In subtrajectory moves the ``construction'' path 
does not only describe
the MD extensions
of the final path, but also the sequence of subtrajectories 
including the failed ones.
For SS and WT, 
the unsuccessful velocity generations, that do not obey the one-step crossing condition, should also be 
considered as part of the construction path
$\chi$.
In other words, $\chi$ consists of several steps and 
the generation probability ``via $\chi$'' is given by the product of generation probabilities of each step.

For each construction path $\chi$ there should exist an unique
reverse construction path $\overline{\chi}$. Roughly said,
when $\chi$ represents a sequence of algorithmic steps, $\overline{\chi}$ will typically consist of 
the reverse steps in reverse order. However, some groups of consecutive steps
might actually happen in the same order.
In fact, there is no unique way to define ``a reverse'', 
but for a given definition 
there will be a one-to-one relation between any possible $\chi$ and its reverse $\overline{\chi}$ and with that,
valid acceptance/rejection rules can be derived based on the superdetailed balance, Eq.~\ref{eq:supdet}.

Yet, the definition of the reverse should be chosen such that
the acceptance probability is computable and not negligibly small in the majority of cases. Therefore, the mathematical definition for the inverse is taken such that the
probabilities of most of the algorithmic steps in the expressions for 
$
P_{\rm gen}({\rm path}^{(o)} \rightarrow 
{\rm path}^{(n)} \textrm{ via } {\chi})
$ and
$P_{\rm gen}({\rm path}^{(n)} \rightarrow 
{\rm path}^{(o)} \textrm{ via } \overline{\chi})
$
will cancel. 

For instance, if we represent 
the construction path
as vector containing the different steps in chronological order, $\chi$ 
could look like
\begin{align}
    \chi=[s^0,t^1, t^2, s^3, s^4, t^5, s^6]
    \label{eq:chi}
\end{align}
which shows that there were 6 subtrajectories generated
of which there were 3 failed trials $t^1, t^2$ and $t^5$.
The initial step involves cutting out the very first 
subtrajectory $s^0$ from the old path, while the final step implies not only the generation of the last 
subtrajectory $s^6$ but also its extension to a full trajectory. 
The reverse construction path in this case is conveniently 
defined as 
\begin{align}
    \overline{\chi}=[s^6, s^4, t^5, s^3,s^0,t^1, t^2]
    \label{eq:invchi}
\end{align}
So the order of the steps is not completely reversed, 
but the reverse order takes place on groups of consecutive steps, 
 a group being a successful subtrajectory  with all its failed trials
that follow.
The reason for this inverse  is that
Eq.~\ref{eq:chi} shows   that
trial trajectory $t^5$ can be generated starting from
$s^4$, but this is not necessarily the case from $s^6$.
Reversely, as $s^6$ was generated from $s^4$, they 
share a common configuration point, 
which makes it possible to 
 to generate $s^4$ from $s^6$. There is, however, no reason 
whatsoever that $s^6$ and $t^5$ share a common configuration point.
Hence, if we would consider the reverse to be
$\overline{\chi}=[s^6, t^5, s^4, \ldots]$,
$P_{\rm gen}({\rm path}^{(n)} \rightarrow 
{\rm path}^{(o)} \textrm{ via } \overline{\chi})
$ would most likely be zero as $\overline{\chi}$ itself cannot be generated. In contrast,
the inverse based on the grouped reordering, Eq.~\ref{eq:invchi},
contains generation probabilities 
like the probability to generate
$t^5$ given $s^4$ which appear both in $\chi$ 
and $\overline{\chi}$. Therefore, all the
generation probabilities of failed trajectories
cancel in Eq.~\ref{eq:supdet}. 
Likewise, all failed velocity generations in SS and WT that
do not obey the one-step crossing condition cancel out for the same reason
as is shown in Ref.~\onlinecite{riccardi_fast_2017}.

Excluding all the failed steps that will cancel
in Eq.~\ref{eq:supdet}, we can write for
$
P_{\rm gen}({\rm path}^{(o)} \rightarrow 
{\rm path}^{(n)} \textrm{ via } {\chi})
$:
\begin{align}
    P_{\rm gen}({\rm path}^{(o)} \rightarrow 
{\rm path}^{(n)} \textrm{ via } {\chi})
\propto 
P_{\rm sel}(s^0|{\rm path}^{(o)}) \times
\nonumber \\
P_{\rm sel}(r^{0,3}|s^0) P_{\rm gen}(v^{0,3})
P_{\rm MD}(s^3|x^{0,3}) \times \nonumber \\
P_{\rm sel}(r^{3,4}|s^3) P_{\rm gen}(v^{3,4})
P_{\rm MD}(s^4|x^{3,4}) \times \nonumber \\
P_{\rm sel}(r^{4,6}|s^4) P_{\rm gen}(v^{4,6}) 
P_{\rm MD}(s^6|x^{4,6}) \times \nonumber \\
P_{\rm sel}({\rm td})
P_{\rm MD}(
{\rm path}^{(n)}
|s^6) 
\label{eq:P1propto}
\end{align}

Here, $P_{\rm sel}(s^0|{\rm path}^{(o)})$
is the probability for selecting $s^0$ from the old path 
${\rm path}^{(o)}$ and $P_{\rm sel}(r^{0,3}|s^0)$
is the selection probability 
of choosing point $r^{0,3}$ from the subpath $s^0$ as the shooting point.
Since $r^{0,3}$ is a shooting point to go from $s^0$ to $s^3$, it is a configuration point that 
 $s^0$ and $s^3$ have in common.
 $P_{\rm gen}(v^{0,3})$ is the probability 
 for generating the velocities $v^{0,3}$
 that are the velocities of $s^3$ at the 
 corresponding configuration point $r^{0,3}$.
 $P_{\rm MD}(s^3|x^{0,3})$ is the chance that starting
 from phase point $x^{0,3}=(r^{0,3},v^{0,3})$, the MD integrator
 produces subpath $s^3$ by integrating the equations of motion forward and backward in time. The MD integrator
 can be based on actual Newtonian MD, Langevin, Brownian, etc.
 Likewise, 
 $P_{\rm MD}(
{\rm path}^{(n)}
|s^6)$ is the chance that the new path ${\rm path}^{(n)}$
is produced by extending the final subpath $s^6$.
Finally, $P_{\rm sel}({\rm td})$ is the 
selection probability for 
direction of time
along the new path. 
Note that the direction of time along the 
subpaths is irrelevant in WT and WF. In SS,  
subpaths do have a sort of direction as
 the next shooting always takes place at the last $\lambda_i$ crossing.\cite{riccardi_fast_2017}

For the reverse construction path, 
Eq.~\ref{eq:invchi}, 
we can write 
\begin{align}
    P_{\rm gen}({\rm path}^{(n)} \rightarrow 
{\rm path}^{(o)} \textrm{ via } \overline{\chi})
\propto 
P_{\rm sel}(s^6|{\rm path}^{(n)}) \times
\nonumber \\
P_{\rm sel}(r^{6,4}|s^6) P_{\rm gen}(v^{6,4})
P_{\rm MD}(s^4|x^{6,4}) \times \nonumber \\
P_{\rm sel}(r^{4,3}|s^4) P_{\rm gen}(v^{4,3})
P_{\rm MD}(s^3|x^{4,3}) \times \nonumber \\
P_{\rm sel}(r^{3,0}|s^3) P_{\rm gen}(v^{3,0}) 
P_{\rm MD}(s^0|x^{3,0}) \times \nonumber \\
P_{\rm sel}({\rm td})
P_{\rm MD}(
{\rm path}^{(o)}
|s^0) 
\label{eq:P2propto}
\end{align}

Now, it becomes apparent that most terms will cancel out
in Eq.~\ref{eq:supdet} when we take the ratio between
 Eq.~\ref{eq:P2propto} and
Eq.~\ref{eq:P1propto}.
First of all, the time-direction is chosen with a 50\% probability such that $P_{\rm sel}({\rm td})=0.5$.
Then,
we can use the fact that a path probability 
can be written in terms of a phase point probability
times the MD generation probability
\begin{align}
    P({\rm path})=\rho(x) P_{\rm MD}({\rm path}|x)
\end{align}
where $\rho(x)$
is the phase space equilibrium density 
for any phase point $x$ that is part of the path.~\cite{riccardi_fast_2017}
For a phase point $x=(r,v)$ this can be split into
\begin{align}
    \rho(x)=\rho_r(r) \rho_v(v)
\end{align}
 where 
$\rho_r$ and $\rho_v$ are, respectively, 
the configuration (Boltzmann) 
distribution and the velocity Maxwell-Boltzmann 
distribution (possibly subjected to bond- and angle constraints if applicable).
Further, as generating new velocities in Eqs.~\ref{eq:P1propto} and \ref{eq:P2propto} is based 
on the velocity distribution, $P_{\rm gen}(v)=\rho_v(v)$,
we can substitute all $P_{\rm MD}$ terms in
Eqs.~\ref{eq:P1propto} and \ref{eq:P2propto}, e. g.:
\begin{align}
    P_{\rm gen}(v^{4,6})
P_{\rm MD}(s^6|x^{4,6}) &= 
\frac{ P_{\rm gen}(v^{4,6}) P(s^6) }{\rho(x^{4,6})} 
\nonumber \\
&=
\frac{ \rho_{v}(v^{4,6})P(s^6)}{\rho(x^{4,6})}=
\frac{P(s^6) }{\rho_r(r^{4,6})} \nonumber \\
P(s^6) P_{\rm MD}(
{\rm path}^{(n)}
|s^6)&=P({\rm path}^{(n)})
\label{eq:PgenPmd}
\end{align}
Applying these operations 
to Eqs.~\ref{eq:P1propto} and 
\ref{eq:P2propto}, we get
\begin{align}
    P_{\rm gen}({\rm path}^{(o)} \rightarrow 
{\rm path}^{(n)} \textrm{ via } {\chi})
\propto 
P_{\rm sel}(s^0|{\rm path}^{(o)}) 
P_{\rm sel}({\rm td}) \times
\nonumber \\
P_{\rm sel}(r^{0,3}|s^0) 
P_{\rm sel}(r^{3,4}|s^3) 
P_{\rm sel}(r^{4,6}|s^4) \times
\nonumber \\
P(s^3) P(s^4) P({\rm path}^{(n)})/[
\rho_r(r^{0,3}) 
\rho_r(r^{3,4})  \rho_r(r^{4,6})
] \nonumber \\
 P_{\rm gen}({\rm path}^{(n)} \rightarrow 
{\rm path}^{(o)} \textrm{ via } \overline{\chi})
\propto 
P_{\rm sel}(s^6|{\rm path}^{(n)}) 
P_{\rm sel}({\rm td})
\times
\nonumber \\
P_{\rm sel}(r^{6,4}|s^6) 
P_{\rm sel}(r^{4,3}|s^4) 
P_{\rm sel}(r^{3,0}|s^3)
\times \nonumber \\
P(s^4) P(s^3) P({\rm path}^{(o)})
/[\rho_r(r^{6,4}) \rho_r(r^{4,3})  \rho_r(r^{3,0}) ]
\label{eq:P12propto}
\end{align}

In the ratio of these two equations more terms will cancel out as $r^{\alpha,\beta}=r^{\beta, \alpha}$.
Further, $P_{\rm sel}({\rm td})=0.5$ as stated before.
In all  subtrajectory moves,
$P_{\rm sel}(r|s^\alpha)$  is either a fixed number (SS and WT) or it 
depends on $s^\alpha$, but not on $r$ (WF).
In SS, the shooting point is selected from the last crossing
with $\lambda_i$ 
and therefore $P_{\rm sel}(r|s^\alpha)=2$ (the phase point
just before or after $\lambda_i$).
In WT, it is randomly chosen from a crossing with either
$\lambda_i$ or $\lambda_{\rm sour}$ and therefore 
$P_{\rm sel}(r|s^\alpha)=4$. With stochastic dynamics
one can also opt to choose only the inner points~\cite{riccardi_fast_2017} such that $P_{\rm sel}(r|s^\alpha)=2$. In WF any point of the subpath 
that lies between $\lambda_i$ and $\lambda_B$ (or $\lambda_{\rm cap}$) can be chosen. In all these cases the 
$P_{\rm sel}(r|s^\alpha)$ terms with identical $s^\alpha$ 
cancel out in the ratio.
That means that the only terms 
that remain depend on
the first and last subpath ($s^0$ and $s^6$),
or on the full paths (${\rm path}^{(o)}$ and ${\rm path}^{(n)}$):
\begin{align}
\frac{
P_{\rm gen}({\rm path}^{(n)} \rightarrow 
{\rm path}^{(o)} \textrm{ via } \overline{\chi})
}{
 P_{\rm gen}({\rm path}^{(o)} \rightarrow 
{\rm path}^{(n)} \textrm{ via } {\chi})
}&=
\label{eq:ratio}
\\
\frac{
P_{\rm sel}(s^6|{\rm path}^{(n)}) 
P_{\rm sel}(r^{6,4}|s^6) 
P({\rm path}^{(o)})
}{
P_{\rm sel}(s^0|{\rm path}^{(o)}) 
P_{\rm sel}(r^{0,3}|s^0) 
P({\rm path}^{(n)})
}&=
\nonumber \\
\frac{
P_{\rm sel}(r^{6,4}| 
{\rm path}^{(n)}
) 
P({\rm path}^{(o)})
}{
P_{\rm sel}(r^{0,3}|{\rm path}^{(o)}) 
P({\rm path}^{(n)})
} &=
\frac{
P({\rm path}^{(o)})
/M^{(n)}
}{
P({\rm path}^{(n)})
/M^{(o)}
}\nonumber
\end{align}
where in the third expression 
we contracted the selection probabilities
involving the two-steps (first selecting $s^0$ or $s^6$, 
then selecting $r^{0,3}$ or $r^{6,4}$) to the chance 
of selecting the very first successful crossing point
from the existing full path.
Finally, the latter was replaced by $1/M^{(n)}$ and
$1/M^{(o)}$ where $M^{(n)}$ and  
$M^{(o)}$
are the number of different equal probable 
possibilities
to select a shooting point for generating a subtrajectory from the
new and old full path, respectively.

If we substitute Eq.~\ref{eq:ratio} into Eq.~\ref{eq:supdet},
we obtain a rather simple expression for the acceptance:
\begin{align}
    P_{\rm acc}=
    {\rm min}\left[ 1,
\frac{M^{(o)}
}
{
M^{(n)}
}
\right]
\label{eq:supdet2}
\end{align}
In SS, $M^{(o)}$ and
$M^{(n)}$ are simply proportional to 
the number of crossing points
of the old and new path
with $\lambda_i$,
while for WT these are proportional to the 
number segments that can be cut out
of these trajectories that connect 
$\lambda_{\rm sour}$ and $\lambda_{i}$.~\cite{riccardi_fast_2017}
In WF, these relate to the number of points
between $\lambda_i$ and $\lambda_B$. 
If a so-called \emph{cap}-interface is defined, 
$M^{(o)}$ and
$M^{(n)}$ relate to the number of points 
between $\lambda_i$ and $\lambda_{\rm cap}$ excluding any 
points lying on a segment 
$\lambda_{\rm cap} \rightarrow \lambda_{\rm cap}$ without 
crossing $\lambda_i$.

Eq.~\ref{eq:supdet2} can also be combined with an early
rejection scheme as was introduced in Ref.~\onlinecite{TIS}.
In the standard approach one would complete the MC move, compute the acceptance probability, Eq.~\ref{eq:supdet2},
take a uniform random number $\alpha$ between 0 and 1, and then accept if $\alpha < P_{\rm acc}$ and reject otherwise.
In the early rejection scheme, the random number
$\alpha$
is taken
first
and the move is rejected as soon as $M^{(n)} >
M^{(o)}/\alpha$.
In normal shooting, this provides a considerable 
speed up since long paths have a high chance to get rejected. 
Using the early rejection scheme
a lot of unnecessary MD steps can be avoided as
these path can be stopped whenever they exceed 
the predetermined maximum length. Yet,
for the subtrajectory moves the \emph{high-acceptance}
scheme is preferable as we discuss in Sec~\ref{sec:highacc}.
In the next section we show why the subtrajectory moves
allow us to sample fewer trajectories than with standard shooting via a reduction of the statistical inefficiency.

\section{Statistical inefficiency}
\label{sec:statineff}
The principal property that is computed in the $[i^+]$ ensemble is the local 
crossing probability $P_A(\lambda_{i+1}|\lambda_{i})$. This is 
the history dependent conditional probability that the system, 
given it crosses $\lambda_A$ and 
then crosses $\lambda_i$, crosses $\lambda_{i+1}$ before $\lambda_A$.
In the \emph{post hoc} analysis, this local crossing probability is
simply the fraction of sampled path in the $[i^+]$ ensemble that happen to cross
$\lambda_{i+1}$ in addition to $\lambda_{i}$. 
Once these are accurately enough determined, the global crossing probability
$P_A(\lambda_{B}|\lambda_{A})$ is obtained from~\cite{TIS,Raffa}:
\begin{eqnarray}
P_A(\lambda_{B}|\lambda_{A})=\prod_{i=0}^{n-1}
P_A(\lambda_{i+1}|\lambda_{i})
\label{eq:Pcross}
\end{eqnarray}
where $\lambda_0=\lambda_A$ and $\lambda_n=\lambda_B$.
The above expression is exact since the local crossing
probabilities include the full history dependence ($\lambda_A \rightarrow \lambda_i$) in their condition.~\cite{TitusRev}
An alternative approximate expression 
for the global crossing probability
is used in partial path TIS~\cite{PPTIS}
in which the amount of spatial  memory is reduced though not set to zero, as in milestoning~\cite{Milestoning}.
The global crossing probability gives the rate of the transition 
when multiplied with $f_A$, the conditional flux through $\lambda_A$.

In TIS, the flux is calculated by straightforward MD where the system is prepared in state $A$ and then the number of crossings with $\lambda_A$ per time 
unit
is computed.
If a spontaneous transition to state $B$ takes place, which is unlikely for a rare event,  the simulation is paused, reinitiated in state $A$ and then continued. RETIS computes the flux term differently as it does not use a single continuous MD simulation.
Instead, 
the $[0^-]$ path ensemble is introduced to explore the $A$ state, and the
flux is derived from the average path lengths in $[0^-]$ and $[0^+]$.\cite{RETIS}
In addition to rate constants, the overall crossing probability can also be used to compute permeability coefficients~\cite{permeability} 
and activation energies.\cite{actE, Elab}

Considering
the $j$-th path in the simulation for 
path ensemble  
$[i^+]$,
the main output of sample $j$ (the generated path)
that is relevant for the computation of the crossing probability is
simply the observation of whether it crosses $\lambda_{i+1}$ or not.
We can describe this by a characteristic function $h_j$ which equals 1 if 
$\lambda_{i+1}$ is crossed and 0 otherwise. 
The simulation estimate of the local crossing probability, 
$p(m)$,  after $m$ MC moves
is then expressed as
\begin{align}
    p(m) = \frac{1}{m} \sum_{j=0}^{m-1} h_j \approx
    P_A(\lambda_{i+1}|\lambda_{i})
    \label{eq:pm}
\end{align}
where the index counter starts from zero for mathematical convenience.

For finite $m$, the value of $p(m)$ will not be exact and 
the absolute error, $\epsilon_a$, is defined as the 
the standard deviation of the mean $\sigma_{p(m)}$.
This is essentially the
standard deviation in possible $p(m)$
results 
if the simulation experiment would be carrier out multiple times.
Mathematically we can write this as
\begin{align}
    \epsilon_a=\sigma_{p(m)}=
    \sqrt{
    \left \langle (p(m)-p)^2 \right \rangle
    }
    \label{eq:ea}
\end{align}
where $p=p(\infty)=P_A(\lambda_{i+1}|\lambda_{i})$ and the 
brackets $\left \langle \cdot \right \rangle$
refer to the perfect ensemble sampling average.
This can be viewed as the hypothetical average 
that is obtained 
after repeating the simulation an infinite number of times starting
with initial conditions that are randomly drawn form a perfect statistical equilibrium 
distribution.
In other words, we have $\left \langle p(1) \right \rangle=\left \langle p(m) \right \rangle=p$. Further, since detailed balance MC moves conserve the equilibrium distribution,~\cite{FrenkelBook} the absolute value of the index $j$ is irrelevant and
$\left \langle h_0 \right \rangle=\left \langle h_1 \right \rangle
=\left \langle h_j \right \rangle=p$ and   
$\left \langle h_j h_k \right \rangle=\left \langle h_0 h_{k-j} \right \rangle$
for any $j,k$.
Using this, one can show that~\cite{TISeff}:
\begin{align}
   \sigma_{p(m)}^2 &= \frac{
   \sigma_{p(1)}^2
   }{m} {\mathcal N}, \quad {\mathcal N}=\left[ 
   1 +2 n_c \right]
   \label{eq:spm2}
   \end{align}
where ${\mathcal N}$ is called the statistical inefficiency and $n_c$ is the correlation number 
which is the integral of the correlation function
$C(j)$:
   \begin{align}
   n_c=\sum_{j=1}^\infty C(j), \quad
   C(j) =
   \frac{\left \langle (h_0-p)(h_j-p) \right \rangle}{
   \left \langle (h_0-p)^2 \right \rangle
   }
   \label{eq:nc}
\end{align}
As the output $h_j$ of a single sample is either 1 with a probability $p$ or 0 with a probability $(1-p)$, the sample standard deviation
$\sigma_{p(1)}$
can be simplified
\begin{align}
   \sigma_{p(1)}^2&=\left \langle 
   (p(1)-p)^2 \right \rangle
   =\left \langle 
   (h_0-p)^2
   \right \rangle \nonumber \\
   &=p (1-p)^2+(1-p)(0-p)^2=p(1-p) 
   \label{eq:s2}
\end{align}
Via Eqs.~\ref{eq:spm2}, ~\ref{eq:nc}, and ~\ref{eq:s2}, we
can write for the relative error:
\begin{align}
  \epsilon_r=\frac{\epsilon_a}{p}=\sqrt{\frac{1-p}{p}
  \frac{\mathcal N}{m}
  }
   \label{eq:er}
\end{align}
Eq.~\ref{eq:er} shows that for a fixed number of MC moves $m$,
the larger the local crossing probability $p=P_A(\lambda_{i+1}|\lambda_{i})$, 
the lower the relative error. Hence, 
the result in simulation $[i^+]$ converges faster
when the difference between $\lambda_i$ and $\lambda_{i+1}$ is small, but
this will obviously increase the number of path ensembles needed.
Analytical results on model systems suggest that the optimum placement 
of interfaces in TIS is achieved when $p\approx 0.2$ for all ensembles.\cite{TISeff}
In RETIS the optimum is expected to be slightly 
higher
as this would lead 
to more successful swaps.
Likewise, the optimum is also slightly higher if
the weighted histogram analysis method (WHAM)~\cite{WHAM}
is used instead of single-point matching
to determine 
the total crossing probability. 
In this approach, the crossing statistics of path ensemble
$[i^+]$ 
is not limited 
to the fraction of paths crossing
$\lambda_{i+1}$, but also 
 the fractions for crossing
$\lambda_{i+2}$, $\lambda_{i+3}$, etc. are used to get a slightly more accurate estimate 
of Eq.~\ref{eq:Pcross}.\cite{predictive, Rogal}

If the sampling between successive MC moves is completely uncorrelated, we have that
$\left \langle (h_0-p)(h_j-p) \right \rangle=\left \langle (h_0-p) \right \rangle \cdot \left \langle (h_j-p) \right \rangle=0 \cdot 0=0.$
This would imply that $C(j)=n_c=0$ and ${\mathcal N}=1$.
In this case, if $p=0.2$, there are about $m=400$ trajectories required to obtain an $\epsilon_r=10\%$ error.  
For ${\mathcal N}>1$, one would need $m/{\mathcal N}=m_u=400$ to get the same error. Here, $m_u$ is called the number of
effectively uncorrelated samples.

In general, $C(j) \neq 0$ except for the limit $j\rightarrow \infty$
as correlation decays.
If a
MC move is rejected at step $j$, then the previous sample is kept and recounted such that sample $j$ is identical to sample $j-1$. Hence, if there are $j$ consecutive rejections, sample $j$ is identical to sample 0 causing  correlation over multiple steps. Even if 
 the $j$-th step is
accepted, it tends to have some similarity with the previous sample. Therefore,
there is an high probability for $h_j=h_{j-1}$, even if the samples are not identical.
The correlations 
lead to a 
sampling output 
$(h_0, h_1, h_2, \ldots)$
with
long rows of consecutive zeros 
and consecutive ones. 

To illustrate this effect with a
mathematical example: suppose that
the MC move has a probability $\pi_R$ to remain unchanged such that $h_j=h_{j-1}$ and a probability $\pi_M=1-\pi_R$ to actually make a move that potentially (but not necessarily) changes the output: the new sample yields $h_j=1$ with a probability $p$ and  $h_j=0$ with a probability $(1-p)$.
As shown in the appendix, for this mathematical model the statistical inefficiency equals:
\begin{align}
    {\mathcal N}=\frac{2-\pi_M}{\pi_M}
    \label{eq:ineffpim}
\end{align}
This shows that for a typical MC acceptance probability of 50\%, the 
effect of rejections alone causes the statistical inefficiency to be equal to 3. 
The situation is usually worse 
in complex systems and also more difficult to identify than merely by the presence of rows of consecutive ones or zeros. For instance inter- and intramolecular changes 
of reactants could temporarily boost or reduce the probability of a transition. 
The same 
kind of fluctuations in the temporary transition probability
can be caused by the local
solvent structure and the position and orientation of catalytic molecules. These describe degrees of freedom 
that are orthogonal to the reaction coordinate.

We can examine this by a slightly more complex model where we assume that 
there are two phases $\alpha$ and $\beta$,
described by the orthogonal degrees of freedom,
which occur with probabilities $P_\alpha$ and $P_\beta=1-P_\alpha$.
Let $p_\alpha$ and $p_\beta$ be the corresponding local crossing probabilities 
along the reaction coordinate
for these phases such that:
$p=P_\alpha p_\alpha+P_\beta p_\beta$.
Analogous to the above, let 
$\pi_\rho$ be 
the chance to not update the phase, and 
 $\pi_\mu=1-\pi_\rho$ be the chance 
 to freshly choose between phase 
 $\alpha$ or $\beta$ with respective 
 probabilities $P_\alpha$ and $P_\beta$.
As shown in the appendix, in this case the statistical inefficiency equals:
\begin{align}
    {\mathcal N}=\frac{2 K_s-\pi_\mu (2 K_s-1)}{\pi_\mu}
    \label{eq:ineffpimx}
\end{align}
where $K_s$ is a system parameter that does not depend on the type of MC move:
\begin{align}
    K_s=\frac{P_\alpha P_\beta (p_\alpha-p_\beta)^2 }{p(1-p)} 
    =\frac{(p-p_\alpha)(p_\beta-p)}{p(1-p)}
    \label{eq:Ks}
\end{align}
Note that $K_s=0$ whenever $p_\alpha=p_\beta$, which 
gives ${\mathcal N}=1$. 
This would be the case if all TIS interfaces are placed 
at isocommittor surfaces,
which partly supports the hypothesis of 
Ref.~\onlinecite{isocom} that stated that path sampling simulations are most efficient
if the reaction coordinate $\lambda$ equals the committor.
However, although this surely minimizes the statistical inefficiencies, the mean path lengths 
in the path ensembles
also depend on 
the choice of the reaction coordinate $\lambda$. 
If this is included in the analysis, the hypothesis 
is at least not generally true.~\cite{TitusRev}

\begin{figure}[ht!]
    \centering
    \includegraphics[width=0.4\textwidth]{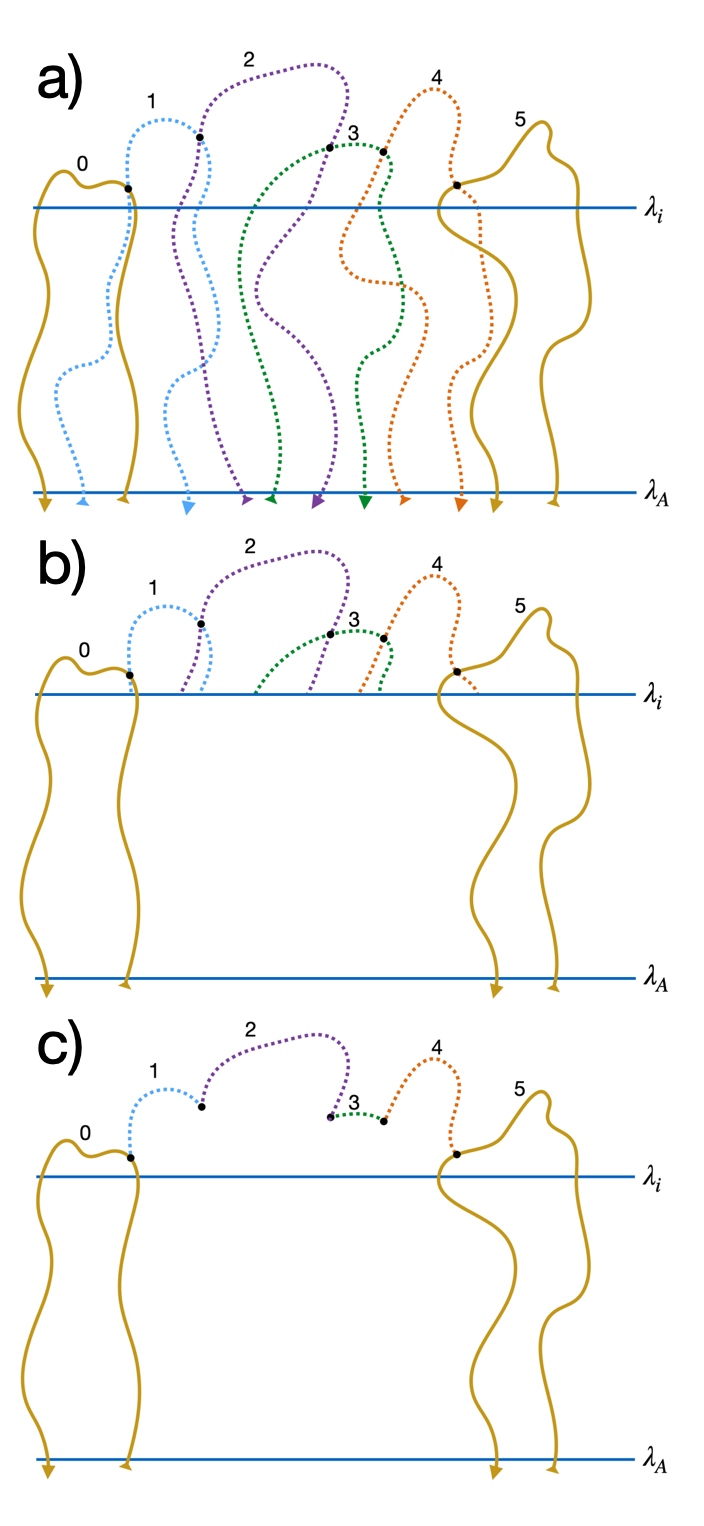}
    \caption{Illustration of wasted MD steps in shooting and WF. Panel a) shows six consecutive paths being generated by the shooting move where only the solid golden paths, with index 0 and 5, are being saved. Panel b) gives an equivalent scenario in the WF algorithm showing that considerable fewer 
    MD steps are needed to obtain the same paths 0 and 5 via $N_s=5$ subtrajectories. Still, WF is not as thrifty as SS and WT since only parts of the subtrajectories, shown in panel c), actually contribute to the sampling progress to get from path 0 to 5. The additional steps in panel b) are seemingly ``wasted'' but still needed for the superdetailed balance relation. SS and WT do not generated  wasted MD steps, but rely on a one-step crossing condition as discussed in the main text.
    }
    \label{fig:MDvsWF}
\end{figure}

Now assume that not all generated paths
are saved and analyzed, but instead only every $N_s$-th path is kept. While this will cause a reduction in the number of samples from $m$ to $m/N_s$, it does not necessarily reduce the number of uncorrelated samples $m_u$ as the statistical inefficiency between
saved samples is also reduced. In particular, the
``remain'' probability between saved samples changes
from $\pi_\rho$ to $\pi_\rho^{N_s}$ and,
therefore, the ``move'' probability
changes from $\pi_\mu$ to $1-\pi_\rho^{N_s}=
1-(1-\pi_\mu)^{N_s}$. The statistical inefficiency
between saved samples is henceforth:
\begin{align}
    {\mathcal N}(N_s)=\frac{2 K_s- (1-(1-\pi_\mu)^{N_s})(2 K_s-1)}
    {1-(1-\pi_\mu)^{N_s}}
    \label{eq:ineffNs}
\end{align}
Eq.~\ref{eq:ineffNs} shows that
the statistical efficiency indeed goes down with 
increasing $N_s$ 
up to an asymptote equal to 1.
Taking the power series up to first order in $\pi_\mu$, we see that the
initial downfall is inversely linear:
\begin{align}
    {\mathcal N}(N_s)
    \approx \frac{2 K_s- N_s \pi_\mu (2 K_s-1)}
    {N_s \pi_\mu} \approx \frac{{\mathcal N}(1)}{N_s}
    \label{eq:ineffNs2}
\end{align}
where we assumed $N_s \pi_\mu \ll 1$.
As a result, saving every $N_s$-th path instead of all paths will not affect much the  post-simulation analysis in terms of accuracy. The reduction in the number of data points from $m$ to $m/N_s$ is compensated by a lower statistical inefficiency such that the number of uncorrelated samples $m_u$ remains nearly unchanged. While this allows for  obvious 
data storage savings, reducing
both the memory and time for writing to disk,
it also paves the way to reduce 
 MD steps as is shown in Fig.~\ref{fig:MDvsWF}.
The figure illustrates a hypothetical MC sequence in path sampling of six consecutive paths, labeled 0 to 5, where the shooting point has an order parameter larger than $\lambda_i$. 
If only every fifth path is saved, only path 0 and 5 are considered as in Fig.~\ref{fig:MDvsWF}-a). 
Although the intermediate paths 
contribute for their
decorrelation, it is clear that many MD steps can be omitted, as exploited by the subtrajectory moves.
Fig.~\ref{fig:MDvsWF}-b) shows a scenario where the same final path is being generated with a set of hypothetical WF
 subtrajectories
 resembling 
the top scenario.
Instead of five full trajectories, only four short subtrajectories and one full trajectory are needed to establish a new full path (path 5) from the old one (path 0).
Based on this principle alone, the relative efficiency gain $\eta$ of 
 subtrajectory moves compared to standard shooting
is expected to be
\begin{align}
    \eta(N_s)=\frac{N_s L_p}{L_p+(N_s-1) L_s}
    \label{eq:eta}
\end{align}
where $L_p$ and $L_s$ are, respectively, the average length of a full path and a subpath.
Still, if we purely focus on the MD steps 
that are required to allow for the progression from path 0
to path 5, even fewer MD steps are needed as shown in 
Fig.~\ref{fig:MDvsWF}-c). Yet,
the ``extra'' (wasted) MD steps in panel b) are required for
the superdetailed balance as discussed in Sec.~\ref{sec:supdetbal}. Wasted MD steps  are avoided in SS and WT where the shooting always
happens at an interface (see Figs.~\ref{fig:3moves}-a)
and b)).
The price to be paid for this is the additional complication with regard to the one-step crossing condition (see Sec.~\ref{sec:one-step}). But even with a slightly higher 
MD waste, the WF move  requires 
considerably
less MD steps 
than standard shooting.

Eq.~\ref{eq:eta} levels off to a 
constant $L_p/L_s$ for increasing $N_s$. %
Likewise,
Eqs.~\ref{eq:ineffNs2} and \ref{eq:ineffNs} show
that the trend ${\mathcal N}(N_s)={\mathcal N}(1)/N_s$ is 
not sustained for increasing $N_s$ as ${\mathcal N}$ ultimately  
levels off to 1. It is henceforth assumed
that while efficiency 
initially increases quite rapidly 
as function of $N_s$, it can not surpass $L_p/L_s$ and ultimately 
even 
decreases 
when  ${\mathcal N}(N_s)$ levels off.
Clearly, for the  $[0^-]$ and  $[0^+]$ ensemble where 
$L_p=L_s$ no gain is expected and one could set $N_s=1$ if data storage latency would not be an issue.
Therefore, as a rule off thumb,
$N_s$ can be set approximately equal to  $L_p/L_s$ such that for $L_p> L_s$ 
the cost of the MC move is less than doubled, while 
Eq.~\ref{eq:eta} reaches more than 
50\% of its anyways unattainable maximum of $L_p/L_s$.

Although the essence of the above analysis is 
correct, 
there  is however a caveat:
rejections leave a much heavier mark on the subtrajectory move than on standard shooting. 
If, for instance, the extension of the fifth and last subpath
in Fig.~\ref{fig:MDvsWF}-b) is rejected,
it would imply a complete reset to the latest accepted full path (path 0) since subpath 4 is not a valid trajectory and extending subpath 4 after the rejection
 would violate detailed balance.
As a result, all MD steps of subpaths 1 to 5 are trashed
as the next move starts from path 0 again.
Instead, 
the MC chain will only fall back to path 4 (assuming path 4 was 
accepted) 
in standard shooting.
It is therefore clear that rejections
in the subtrajectory move approach
should be avoided even more than in the shooting method.
This can be achieved with the high-acceptance procedure that is discussed in the next section.

\section{High-acceptance procedure}
\label{sec:highacc}
As discussed in the previous section, 
a rejection in the subtrajectory moves
implies a large amount of wasted MD steps.
An early rejection 
scheme,
as the one used in TIS and RETIS
with standard shooting (see Sec.~\ref{sec:supdetbal}),
is also not so helpful as a rejection
cannot be made until the 
generation of the
last subtrajectory has been initiated.
It is, therefore, preferable to combine the 
subtrajectory moves with the 
\emph{high-acceptance} scheme.~\cite{riccardi_fast_2017}
The approach uses the following two tricks. 
First, if the final subtrajectory has a backward extension ending in state $B$, the MC move is not directly rejected. Instead, the
extension forward in time is completed and, if it ends in state $A$, the path is time-reversed  providing an
$A \rightarrow B$ path. The consequence is that the time-direction selection probability $P_{\rm sel}({\rm td})$ 
in Eq.~\ref{eq:P1propto}
is no longer 0.5 for all paths as an  $A \rightarrow B$
path can be generated in two ways: either by choosing the
correct time direction immediately, or in reverse.
This implies an extra factor two in the generation 
probabilities $P_{\rm gen}$, in Eqs.~\ref{eq:supdet}
and \ref{eq:ratio}, of
the $A \rightarrow B$ paths compared to 
$A \rightarrow A$ paths.
We henceforth write:
\begin{align}
\frac{
P_{\rm gen}({\rm path}^{(n)} \rightarrow 
{\rm path}^{(o)} \textrm{ via } \overline{\chi})
}{
 P_{\rm gen}({\rm path}^{(o)} \rightarrow 
{\rm path}^{(n)} \textrm{ via } {\chi})
} &=
\frac{
P({\rm path}^{(o)}) q({\rm path}^{(o)})
/M^{(n)}
}{
P({\rm path}^{(n)})
q({\rm path}^{(n)})
/M^{(o)}
}\label{eq:withq}
\end{align}
where
\begin{align}
q({\rm path})=
\begin{cases}
  1  &  \text{ if } {\rm path} \in \{
  A \rightarrow A
  \}\\
  2  &  \text{ if } {\rm path} \in \{
  A \rightarrow B
  \}
\end{cases}
\end{align}

The second trick is to slightly change the sampling distribution. Instead of sampling the correct physical
path distribution, $P(\rm path)$, restrained to the 
path
ensemble's
requirements, 
an alternative path distribution 
$\tilde{P}(\rm path)$
is sampled. From Eqs.~\ref{eq:supdet} and
\ref{eq:withq}, the acceptance probability thus becomes
\begin{align}
P_{\rm acc}=
{\rm min}\left[ 1,
\frac{\tilde{P}({\rm path}^{(n)})
P({\rm path}^{(o)}) q({\rm path}^{(o)}) M^{(o)}
}{
\tilde{P}({\rm path}^{(o)})
P({\rm path}^{(n)}) q({\rm path}^{(n)}) M^{(n)}
}
\right]
\label{eq:supdet_tilde}
\end{align}
and
to maximize the acceptance, we choose 
the sampling distribution in ensemble $[i^+]$ as
\begin{align}
\tilde{P}({\rm path})&=
w_i({\rm path}) {\bf 1}_{[i^+]}({\rm path})
\textrm{ with } 
\nonumber \\
w_i({\rm path}) &=  q({\rm path}) M_{\lambda_i}(
{\rm path}
)
\label{eq:Ptilde}
\end{align}
where ${\bf 1}_{C}(x)$ is the indicator 
function that equals 1 if $x$ is part of set $C$ and 0 otherwise. A subscript $\lambda_i$ was added to the 
last term  $M$, as the number of 
equal probable
possibilities for a first 
shooting, generally depends on the interface $\lambda_i$.
Substituting Eq.~\ref{eq:Ptilde} in 
Eq.~\ref{eq:supdet_tilde} implies that 
with high-acceptance
\begin{align}
    P_{\rm acc}={\bf 1}_{[i^+]}({\rm path}^{(n)}) 
\end{align}
In other words, the new path will always be accepted 
unless the MC move led to a path not obeying the
ensemble's definition: starting at $\lambda_A$, ending at 
$\lambda_A$ or $\lambda_B$, and having at least one crossing with $\lambda_i$. By construction, the crossing
of $\lambda_i$ is always achieved in the subtrajectory moves
if the starting condition at $\lambda_A$ is met. Hence,
the only necessary rejection is when the extension of
the final successful subtrajectory ends at 
$\lambda_B$ in both time-directions.

If no successful subtrajectories were generated after 
$N_s$ attempts,
$s^0$ could be extended. However, 
since this would 
regenerate the old trajectory in deterministic dynamics and 
otherwise a trajectory that is highly correlated with the old
one, it is preferable 
to reject
the move.
Other potential reasons for a rejections
could be due to non-convergence of the atomistic forces
in AIMD level calculations.
Another potential issue is jumpy 
order parameters,~\cite{jumpy} 
such that $M_{\lambda_i}$ can be zero even if the 
path is actually valid. 
This issue is further discussed in Sec.~\ref{sec:WF_move}.

Exact natural averages can still be obtained 
by weighting each sample $j$ with the inverse of 
of 
$w_{i}(j)$. 
For instance,
the estimated local crossing probability, previously
defined by Eq.~\ref{eq:pm}, can now be expressed as
\begin{align}
        p(m) = \frac{
        \sum_{j=0}^{m-1}
        w_{i}(j)^{-1}
        h_j
        }{
        \sum_{j=0}^{m-1}
        w_{i}(j)^{-1}
        }  \approx
    P_A(\lambda_{i+1}|\lambda_{i})
    \label{eq:unw}
\end{align}
The effect of the weighting
implies that different samples have different contribution.
If a sample $j'$ has
a much lower than average 
$w_{i}^{-1}$ 
factor,
the sample could essentially be removed from
Eq.~\ref{eq:unw} without 
significantly affecting the estimate $p(m)$.
Yet, thanks to this sample not being rejected, sample
$j'-1$ is more different than $j'+1$ than it would be
in case that $j'$ was rejected.
This shows the power of the high-acceptance approach.

The improved acceptance in the subtrajectory move
will slightly reduce the acceptance in the replica exchange move. For instance, if a path $j$ from ensemble $[i^+]$
will be exchanged with a path $k$
from ensemble $[(i+1)^+]$, the acceptance becomes~\cite{riccardi_fast_2017}:
\begin{align}
    P_{\rm acc}= 
    {\bf 1}_{[(i+1)^+]}(j) \times
    {\rm min}\left[ 1,
\frac{
w_{i}(k) w_{(i+1)}(j)
}{
w_{i}(j) w_{(i+1)}(k)
}
\right]
\label{eq:REha}
\end{align}
Without high-acceptance, the factor 
in Eq.~\ref{eq:REha}
after the multiplication
sign 
equals 1. This means that whenever $j$, the path originating from 
$[i^+]$ is valid for $[(i+1)^+]$, the swap will be accepted. Note that any path in $[(i+1)^+]$ is also valid 
in $[i^+]$. This lower acceptance
is not dramatic since replica exchange moves
do not require any MD steps.  Therefore,
replica exchange moves have negligible CPU cost. 
The only exception is the 
$[0^-] \leftrightarrow [0^+]$ swap
in which two new paths are generated.
Without high-acceptance, this move is always accepted.
For SS and WT the acceptance remains 100\%, but this is not the case for WF. 
We can solve this problem for WF in RETIS by sampling the 
$[0^-]$ and 
 $[0^+]$ ensembles with the standard shooting method without 
 high acceptance.
 Due to this $w_{0^+}$ 
 and $w_{0^-}$
 equal 1 irrespective to the the paths and 
 swapping between these 
 two ensembles will always be accepted.
 The absence of high-acceptance 
 is  partly
compensated by early rejection 
(see Sec.~\ref{sec:supdetbal}).
Moreover, in these ensembles there is no difference 
between the average path length of a subpath 
and a full 
path, making the subtrajectory moves 
anyways
not so effective for these
ensembles.

The high-acceptance protocol 
eliminates the more serious drawbacks of rejections 
in the subtrajectory moves compared to shooting.
In the next section we discuss how the one-step crossing condition can be met.

\section{One-step crossing condition}
\label{sec:one-step}
As discussed above, 
SS and WT are very thrifty algorithms with
respect to the number of generated MD steps. Yet, the one-step
crossing condition puts a challenge to the
implementation. 
One can eliminate the one-step 
crossing condition via the
new but less  thrifty 
WF algorithm that is 
further discussed in Sec.~\ref{sec:WF_move}.
In this section we discuss a few algorithmic 
solutions
to overcome the one-step
crossing condition in SS and WT.
These two 
approaches assume that 
 one time step in (RE)TIS is effectively also
one MD step. 

The one-step crossing 
can be achieved in different ways. 
The most straightforward way is to generate
velocities from a Maxwell-Boltzmann distribution, execute an  MD step, 
calculate the new order parameter, and if the crossing is established,
then
the two frames comprising the crossing are extended at the side above $\lambda_i$ to create a new subpath.
The problem with this approach is  that, after each velocity
generation, an MD step, and therefore a force calculation is required. Especially if $\lambda_i$
is at a steep slope of the potential energy surface, the two trajectory frames forming the crossing of a given interface might be rather far apart in 
$\lambda$-space. 
In such cases, if one of the two frames is located in the very proximity of the interface, it might be extremely unlikely to re-generate a new one step crossing from the configuration furthest to the $\lambda_i$ interface given a random approach to generate velocities.  

There are essentially two strategies to reduce 
the cost for fulfilling the one-step criterion: i)
generate atom velocities from a Maxwell-Boltzmann distribution and predict the 
next step's order parameter without performing an actual MD step, and ii) generate velocities in a way such 
that the crossing is likely achieved after very
few attempts.
Strategy i) assumes that generating new velocities is rather computationally inexpensive and the expense of the one-step crossing condition
is mostly provided by the force calculation. This is the case for AIMD level simulations as these
typically consist of just a few (hundreds of) atoms, while requiring a high CPU demand for the force calculation.
In large classical MD systems with a significant number of atoms, the velocity generation might actually be 
equally expensive as  a force calculation. In that case, strategy ii) might be preferable.

\subsection{Prediction strategy}
The velocity-Verlet~\cite{FrenkelBook}
MD integrator propagates a phase point $x(t)=(r(t), v(t))$ deterministically to a next phase point 
$x(t+\Delta t)=(r(t+\Delta t), v(t+\Delta t))$.
The integrator is most conveniently expressed via ``intermediate velocities'' at $t+\Delta t/2$:
\begin{align}
    v(t+\Delta t/2)&=v(t)+f(t)\Delta t/(2 m)
    \nonumber \\
    r(t+\Delta t)&=r(t) +v(t+\Delta t/2) \Delta t
    \nonumber \\
    v(t+\Delta t)&=v(t+\Delta t/2)+f(t+\Delta t)\Delta t/(2 m)
\label{eq:vv}
\end{align}
where $m$ is the mass and $f$ are the forces.
We used a simplified notation here, but one should realize that for an $N$ particle system both $r, v$ and $f$ are 3N-dimensional vectors and $m$ is actual a $3N \times 3N$
diagonal mass matrix.
Further, the forces are determined from the positions:
$f(t)=f(r(t))$. 

Eq.~\ref{eq:vv} suggests that one MD step requires two force 
evaluations, 
but this is not the case 
when the steps of Eq.~\ref{eq:vv} are
called repeatedly in a loop.
After the force calculation at the third step, 
required to determine
$v(t+\Delta t)$, the forces are stored such that
these can be used at the first step of the next cycle.
With the same reasoning, if the forces 
are known already  at time $t$ from its previous step, a new force evaluation 
is only needed to determine $v(t+\Delta t)$, but not
$r(t+\Delta t)$. This means that if the order parameter
only depends on geometry, $\lambda=\lambda(r)$,
its value at $t+\Delta t$ can also be determined without the need of doing an actual force calculation.

When testing the one-step crossing for
the selected configuration with randomized velocities, 
a new (single step) MD trajectory is started with no  information available
from previous MD step.
However,
the selected configuration is also part  of the previous subpath, so the corresponding forces could have been known, in principle.
When not available, 
the forces 
can be reobtained 
from the trajectory data
without further electronic structure calculation in AIMD or from the gradient of force field potential in classical MD.
In particular, let $x_1=(r_1, v_1)$ and
$x_2=(r_2, v_2)$ be two consecutive phase points 
of the latest subpath
that define a
crossing. 
This means that $x_2$ follows from $x_1$ through a single MD step and both points are at opposite sides of the interface.
Therefore, both points are viable points for shooting off 
the next subpath. 
By inverting Eq.~\ref{eq:vv} we can derive
\begin{align}
    f_1 =
\frac{2\,m\,\left( r_2-r_1-\Delta t\,{v}_1\right)}{
\Delta t^2
 }, \quad
    f_2 =
\frac{2\,m\,\left( r_1-r_2+\Delta t\,{v}_2\right)}{
\Delta t^2
 } \label{eq:f1f2}
\end{align}
So Eq.~\ref{eq:f1f2} directly provides the forces on the two potential shooting points by reading the trajectory data from the subpath. Given that one of these two points 
is selected as a shooting point and new randomized velocities 
are generated, the coordinates after one MD step
can be determined without any additional force calculation but
using just the first two steps of Eq.~\ref{eq:vv}.
Hence, the value of the order parameter after one step can
be asserted.

If the prediction suggests that a crossing might be achieved,
the MD step is completed
and then the next subtrajectory is generated. 
If the velocities 
do not lead to a crossing, a new velocity randomization
is attempted until the crossing condition is met. As in SS
the shooting point selection has to be maintained,
the computation of Eq~\ref{eq:f1f2} only needs to be done once
for the generation of each subpath.
Naturally, if the MD step integrator is more complex than 
velocity-Verlet
(due to thermostats, 
barostats, constraints, stochasticity), then the prediction becomes more difficult. The method also works best if a MD step is computationally expensive while regenerating velocities is relatively cheap. 
This method is therefore more suitable for simulations with AIMD level. 
The approach has been implemented in the PyRETIS software, and it can be directly used with the CP2K~\cite{CP2K} external MD engine.
Note that 
the use of the
plain velocity-Verlet  MD integrator is rather common in path sampling
since the generation of paths is already thermostated via the shooting move that allows a change of energy, while the individual paths have NVE dynamics.

\subsection{Alternative velocity generation}
The mathematically simple form of 
Eq.~\ref{eq:ratio} is due to the many terms conveniently canceling out.
For instance,
the terms in Eqs.~\ref{eq:P12propto}, $\rho_r(r^{0,3})$, 
$\rho_r(r^{4,3})$, and $\rho_r(r^{4,6})$
in $ P_{\rm gen}({\rm path}^{(o)} \rightarrow 
{\rm path}^{(n)} \textrm{ via } {\chi})$
cancel out with, respectively, 
$\rho_r(r^{3,0})$,   $\rho_r(r^{4,3})$, and    $\rho_r(r^{6,4})$ in 
$P_{\rm gen}({\rm path}^{(n)} \rightarrow 
{\rm path}^{(o)} \textrm{ via } \overline{\chi})$ 
due to the fact that $r^{\alpha,\beta}=r^{\beta, \alpha}$.
However, whereas
consecutive (accepted) subtrajectories share 
a common configuration point, they do not necessarily
share of common phase point as $v^{\alpha,\beta}
\neq v^{\beta, \alpha}$.
Here, $v^{\alpha,\beta}$ refers to the 
velocities
of $s^\beta$ at the configuration point 
$r^{\alpha,\beta}$, and $v^{\beta, \alpha}$
refers to the velocities of $s^\alpha$ at an identical configuration point. These velocities 
have
typically not the same orientation nor amplitude. 
Luckily, the $\rho_v(v^{\alpha,\beta})$ terms still cancel out 
via 
$P_{\rm gen}(v^{\alpha,\beta})=\rho_v(v^{\alpha,\beta})$
and  $\rho(x^{\alpha,\beta})=\rho_r(r^{\alpha,\beta})\rho_v(v^{\alpha,\beta})$ in
Eq.~\ref{eq:PgenPmd}.

Now, suppose that in a $N$ particle system not all 
$3N$ velocity components are
regenerated 
from a Maxwell-Boltzmann distribution,
but some velocities components are kept and
some others are inverted (multiplied with -1).
These two velocity groups do not cancel out
in  Eq.~\ref{eq:PgenPmd} 
as they are not part of $P_{\rm gen}$ which implies that the final results changes from 
$P(s^6)/\rho_r(r^{4,6})$ to
$P(s^6)/[\rho_r(r^{4,6}) \rho_v(u^{4,6})]$
where $u^{4,6}$ are
the velocity components that are either
unchanged or inverted. Since the
equilibrium 
velocity distribution is symmetric $\rho_v(v)=
\rho_v(-v)$, and $u^{4,6}$ is identical  
to $u^{6,4}$ except for some components having different sign, all the $\rho_v(u^{\alpha,\beta})$ terms cancel
in the ratio, Eq.~\ref{eq:ratio}, just like
the $\rho_r(r^{\alpha,\beta})$ terms.

This allows for different strategies. 
For instance, if the dynamics is stochastic, all velocities can simply be inverted. This option was used 
for WT in Ref.~\onlinecite{riccardi_fast_2017}.
Inverting 
the velocities of specific atoms or molecules
whose coordinates determine 
the order parameter could also be effective. The other velocities could be either kept unchanged, randomnized or a combination.
For instance, in protein folding simulations  inverting the velocities of all protein atoms while leaving the velocities of the solvent molecules (partly) unchanged would make 
the sampling  less 
diffusive. Reinspection of Eq.~\ref{eq:vv}
shows that the coordinates of the atoms with the inverted velocities  
are mapped exactly back after 1 MD step 
to the previous coordinates regardless of the velocities of the other atoms.
As a result, the one-step crossing condition is automatically fulfilled. 

This approach requires, however, a single MD step 
resolution at the interface crossing. 
In large molecular systems, it is not desirable to 
save trajectory coordinates every MD step as it 
could  overwhelm  hard disk capacity and will result in a 
loss of effective CPU efficiency due to
time that is spend for  writing to disk.
An adaptive scheme could be adopted when 
the 
frequency of order parameter determination and 
the data retention is intensified whenever
the system approaches an interface. 
Since trajectories can later be swapped in a replica exchange move,
this adaptive approach would have to be 
carried out for all interfaces or, at least, in the proximity of neighboring interfaces.
The latter choice might still lead to path ensembles 
receiving a trajectory missing the right resolution 
at the relevant interface. That part of the trajectory would
then
have to be reintegrated by MD.
While all these issues
can be solved 
in theory, it puts quite some 
challenges to the implementation. 
Moreover, if the integrator is not deterministic, but
involves a thermostat or barostat, the one-step crossing might still not be guaranteed.
Several velocity generation steps might still
 needed.
These challenges lead us to derive the WF move 
that straightforwardly can be implemented in present 
path sampling codes like OpenPathSampling~\cite{OPS1, OPS2} and PyRETIS~\cite{PyRETIS1, PyRETIS2} with, potentially, any MD engine.

\section{Wire Fencing}
\label{sec:WF_move}

\begin{figure}[ht!]
   \includegraphics[width=0.4\textwidth]{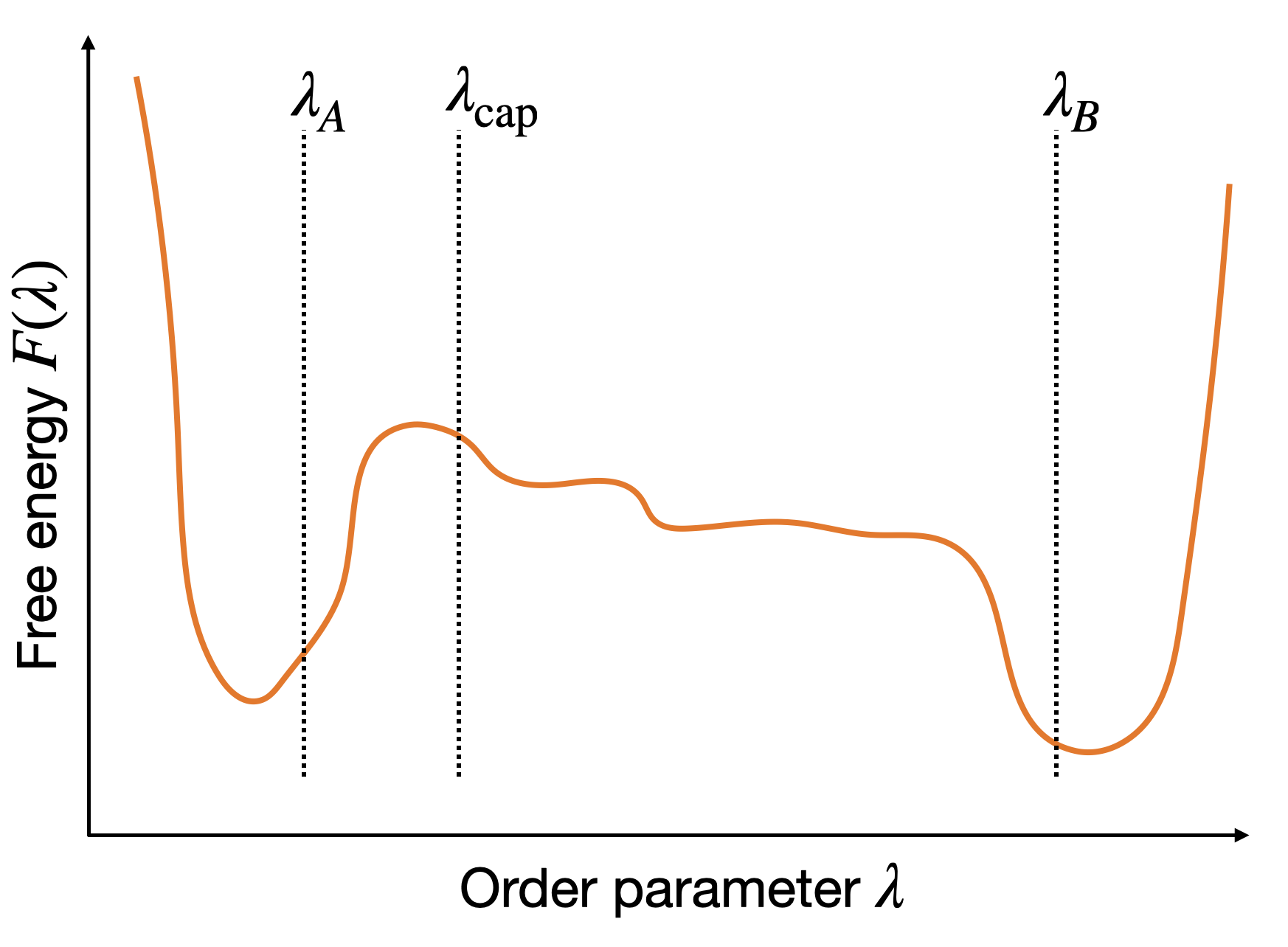}
   \caption{Illustration of an asymmetric barrier where the placement of a cap-interface, $\lambda_{\rm cap}$, in WF can avoid the generation of long subtrajectories and
   too many shooting points being in the 
   basin of attraction of state $B$.
   }
   \label{fig:asymm}
   \end{figure}

Compared to the SS and WT moves, the shooting point selection of the WF move is constructed to avoid the one-step crossing issue altogether. Instead of restricting the shooting point 
to sets of crossing points at an interface,
 WF allows any phase point between 
 the path ensemble's specific ensemble interface, $\lambda_i$, and interface $\lambda_\mathrm{B}$ to be picked.
To increase the 
efficiency
of the WF move in systems with asymmetric free energy barriers (See Fig.~\ref{fig:asymm}), 
the selection range and 
the boundaries of the subtrajectories
can be changed by 
replacing $\lambda_\mathrm{B}$ with an user-defined \emph{cap}-interface, $\lambda_\mathrm{cap}$ with
$\lambda_i<\lambda_\mathrm{cap}\leq\lambda_\mathrm{B}$ value.
   
The presence of a relative flat downhill region after the barrier's maximum and before
a stable product state 
implies that transition paths can become very long.
If accepted, the paths will have
a large fraction of points at the right side of the free energy barrier
from which shooting has a very high chance to generate a failed 
$\lambda_B \rightarrow \lambda_B$ trajectory.
This problem was also addressed 
by the spring-shooting method.\cite{springshoot}

In an AIMD level simulation of 
aqueous silicate condensation,~\cite{Mahmoud_silic} 
this issue was solved by 
defining $\lambda_B$ in the 
RETIS algorithm at the position of 
$\lambda_{\rm cap}$ in the figure.
After the simulation was completed,
all paths reaching $\lambda_B$ were extended in a straight forward MD simulation.
The introduction of the $\lambda_{\rm cap}$
interface make these post-simulation MD extensions 
redundant.

We will first outline
the WF algorithm without a cap-interface 
(or  $\lambda_\mathrm{cap}=\lambda_B$) 
using the high-acceptance protocol.
The introduction of the 
$\lambda_\mathrm{cap}$ only requires
a few modifications that we discuss afterwards.

\begin{enumerate}

\item 
From the old path,
count the number of frames 
$M_{\lambda_i}^{(o)}$ between 
$\lambda_i$ and $\lambda_B$. 
If $M_{\lambda_i}^{(o)}=0$ we immediately reject
the full MC move. Otherwise continue with the next step.

\item 
\label{step:subdiv}
Subdivide the $M_{\lambda_i}^{(o)}$ 
points into groups where each group are the points lying on a segment connecting $\lambda_i$
with $\lambda_B$ or a segment connecting $\lambda_i$ with itself. 

\item  
Select one segment as $s^0$ based on a weighted random selection 
such that each segment has a chance to be selected proportional
to the number of points it has. 

\item Set two counter $n_s$ and $n_a$ equal to zero: $n_s=n_a=0$.
Then start the following loop: step~\ref{step:pickpoint} til
\ref{step:final?}.

\item 
\label{step:pickpoint}
Select at random one of the configuration points of the last subpath, $s^{n_s}$, as the new shooting point.

\item 
Generate random velocities from a Maxwell-Boltzmann distribution.

\item Starting from the configuration point with 
the new velocities, apply the MD integrator to go backward and forward in time
until $\lambda_i$ or $\lambda_\mathrm{B}$ is crossed. 

\item Increase the $n_s$ counter by one: $n_s=n_s+1$.
    
\item If both time-directions crosses $\lambda_\mathrm{B}$, the trial subpath is rejected. In that case, the previous successful 
subpath is kept, $s^{n_s}=s^{n_s-1}$. Go
to step ~\ref{step:final?}. Otherwise, continue with next step.

\item Increase the $n_a$ counter by one: $n_a=n_a+1$.

\item Accept the trial subpath such that it becomes $s^{n_s}$.

\item 
\label{step:final?}
If $n_s<N_s$, 
go to step \ref{step:pickpoint}.
Otherwise, continue with next step.

 \item If no accepted subpaths have been generated, $n_a=0$, stop and reject the move. Otherwise,
 continue with the next step.
    
 \item  Extend the  last subpath $s^{N_s}$
 in both time-directions with MD until $\lambda_\mathrm{A}$ or 
 $\lambda_\mathrm{B}$ is hit. 
 If the path ends at $\lambda_\mathrm{B}$ at both 
 time-directions, the whole MC move is rejected. Otherwise, continue to the next step

\item 
If the path is $\lambda_B \rightarrow \lambda_A$, reverse the time-direction of the path.

\item 
Now a new full path has successfully been established. 
Let $q^{(n)}$ be 2 if it is a $\lambda_A \rightarrow 
\lambda_B$ path. Otherwise, it is 1.
Let $M_{\lambda_i}^{(n)}$ be the number of frames between $\lambda_i$
and $\lambda_B$. The weight-factor of the path is $w^{(n)}=q^{(n)} M_{\lambda_i}^{(n)}$ that is needed for computing proper path ensemble averages, Eq.~\ref{eq:unw}, and for a possible swap move via Eq.~\ref{eq:REha}.
\end{enumerate}

 \begin{figure}[ht!]
   \begin{center}
   \includegraphics[width=0.45\textwidth]{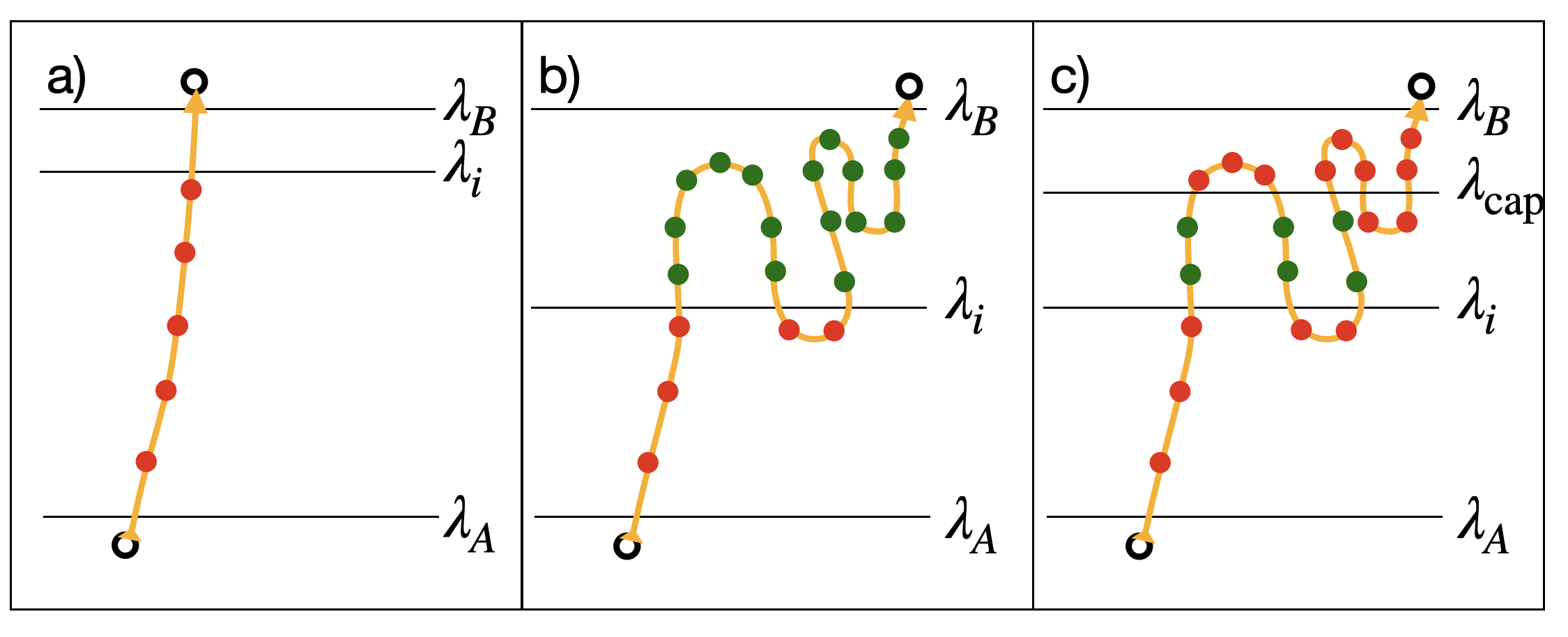}
   \caption{Illustration 
   of the $s^0$ selection from the old path. Selectable shooting points are shown in green, end-points by open black circles, and all other points in red. a) shows the 
   ``jumpy orderparameter'' case 
    that leads to an immediate rejection as no selectable points are present.
    b) and c) show the selectable points
    without and with cap-interface, respectively.
   }
   \label{fig:3scenarios}
    \end{center}
   \end{figure}
The scenario of the potential rejection 
at step 1, is shown if Fig.~\ref{fig:3scenarios}-a) which can occur due
to a jumpy character of the order parameter.~\cite{jumpy}
A typical example is nucleation 
where
the time steps
in path sampling 
is usually chosen to consist of
 many MD steps~\cite{MoroniPRL} for the reason that computing  order parameters for nucleation
is rather costly. As a result, occasionally the order parameter, defined by the size of the largest 
 cluster, can make sudden jumps such that more than one interface is crossed in a single RETIS time step. 

 The path shown in Fig.~\ref{fig:3scenarios}-a)
 is a valid path in $[i^+]$ such that ${\bf 1}_{[i^+]}=1$, but $w_i=0$ since $M_{\lambda_i}=0$. 
 If only WF moves are applied, 
 such a path has zero probability 
 to be generated. Yet, 
 its contribution
 in Eq.~\ref{eq:unw}
 to the average, if hypothetically 
 sampled, would be $w_i^{-1}=\infty$
 and, therefore, the sampling average becomes ill-defined.
 This can be solved by not allowing $w=0$ weights:
\begin{align}
w_i({\rm path}) &=
{\rm min}\left[ 1,
q({\rm path}) M_{\lambda_i}(
{\rm path}
) \right]
\label{eq:Ptilde2}
\end{align}
Introducing this small modification of Eq.~\ref{eq:Ptilde} 
solves the ``division by zero'' problem and
has further no impact of the 
 implementation nor on 
the robustness of the algorithm.
The existence of jumpy trajectories implies that
the WF move  is no longer ergodic. A path like the one in Fig.~\ref{fig:3scenarios}-a) can never be
made from a WF move and, vice versa, it can not be destroyed by the WF move if it is fed as initial path to the algorithm. However, the full sampling remains ergodic due to the replica exchange moves.

Step~\ref{step:subdiv} is further illustrated in Fig.~\ref{fig:3scenarios}-b). We can identify 
two groups of  selectable shooting points (in green), one group of seven point lying on a 
$\lambda_i \rightarrow \lambda_i$ segment and
one group of nine points on a 
$\lambda_i \rightarrow  \lambda_B$ segment.
So these segments 
are selected as $s^0$ with
 a 7/16 and 9/16 probability, respectively.
 In the next step, the points of the selected segment have an equal probability to be selected
 for the first shooting. 
 
 Despite that all the green points
 have the same 1/16 probability to be selected 
 for shooting off the first subpath, the two-step selection process is needed to fix $s^0$.
 With a single step selection,
 it could be possible to first obtain a failed trial
 path $t^1$ that starts from a point 
 at the first group, followed by a successful 
 subtrajectory  that is launched from 
 a point of the second group. This will 
 break the superdetailed balance as it would not be possible to generate $t^1$ from $s^0$ in the reverse path (see the example construction paths in Eqs.~\ref{eq:chi} and \ref{eq:invchi}).

The introduction of the cap-interface changes 
the initial
$s^0$ selection as is shown in
Fig.~\ref{fig:3scenarios}-c)
where, for the same path as panel b),  
there are now three groups of two points 
that can be chosen. Note that not all 
the points between $\lambda_i$ and 
$\lambda_{\rm cap}$ are selectable as 
the points on a $\lambda_{\rm cap} \rightarrow \lambda_{\rm cap}$ segment should be excluded.
The algorithm is further identical as described above with  $\lambda_{\rm cap}$ instead of 
$\lambda_B$ in the main loop 
(step~\ref{step:pickpoint} til
\ref{step:final?}).
Outside the main loop (step 13-15), 
$\lambda_B$ is not replaced by $\lambda_{\rm cap}$
since the final extension always 
shall reach the $A$ or $B$ states.
In the final step (16), $M_{\lambda_i}^{(n)}$ is replaced the number of
frames between $\lambda_i$ and  $\lambda_{\rm cap}$ excluding those on
$\lambda_{\rm cap} \rightarrow \lambda_{\rm cap}$ segments.

\section{Numerical results}
We tested the WF algorithm on three model systems: a simple one-dimensional system for which we can perform  full RETIS simulations with high convergence, and two challenging complex systems based on classical MD and AIMD, where our analysis is more qualitative based on a single path ensemble simulation.
The one-dimensional system describes a single particle in a double-well potential 
that is moving 
following  the underdamped Langevin equation as previously described in Ref.~\onlinecite{TitusRev}.
The purpose of these simulations is to show numerically that the WF method leads indeed to exact results. 
In addition, due to the high degree of convergence
that can be reached, we also draw some conclusions on the efficiency compared to standard shooting. However, it should be taken into account that a larger boost factor is expected for more complex high-dimensional systems.

The other two systems are part of ongoing projects on which we plan to report extensively at later publications. The classical MD system describes 
the thin film breakage in oil-water mixtures based on the studies Ref.~\onlinecite{aaroen_thin_2022, aaroen2021exploring, riccardi2019calcium, riccardi2014structure}.
The system size of this simulation is over 100,000 atoms making the one-step crossing impracticable 
as it requires a stop/restart at every MD step. Instead, 
in our single path ensemble simulation the coordinates
were recorded every 50 MD steps.
The AIMD system describes 
the electron transfer 
between
ruthenium ions in
a redox reaction taking place in liquid water. 
To determine the relative position of the moving electron,
the 
Kohn-Sham orbitals are projected
on maximally localized Wannier Functions~\cite{Wannier} whose centers
can viewed as ``electron positions''. 
This implies that in order to compute 
the order parameter from a configuration point, 
a full electronic structure calculation is required.
A cheap prediction scheme as described in Sec.~\ref{sec:one-step} is therefore not suitable.
For both systems, we show the usefulness of the cap-interface in practical simulations.

\subsection{Double well 1D barrier}
\label{sec:double_well}
Despite the model's simplicity, several popular rare event simulation methods, like forward flux sampling (FFS)~\cite{FFSrev,FFSrev2} and other splitting based 
methods,\cite{split1,split2,RESTART} have shown that they can easily fall into a kind of sampling trap 
when applied to this system
yielding a too low rate 
and non time-symmetric transition paths.\cite{TitusRev}

The double well barrier system consists
of a one-dimensional particle moving in the following potential~\cite{TitusRev}
\begin{equation}
V(z)=z^{4}-2z^{2}
\end{equation}
with underdamped Langevin dynamics. In reduced units,
the Boltzmann constant and mass are set to unity,
$k_B=m=1$, while the temperature and friction coefficient 
are set equal to $T=0.07$ and $\gamma=0.3$.
The equations of motion are propagated using an MD time step equal to $\mathrm{d}t = 0.025$.
In a straightforward MD run, the 
particle will 
mostly oscillate within one of the 
 potential minima at $z = -1$ and $z = 1$, but also (very) infrequently cross the transition state at z = 0.
 During the oscillatory movement, the total energy of the particle will fluctuate by the random force of the Langevin dynamics.
 As a result, the system is effectively two-dimensional in phase space where the velocity can be considered as an orthogonal degree of freedom.
 The reason that FFS and other splitting type methods
 underestimate the crossing rate is due to an insufficient
 sampling of the tail in the velocity distribution.\cite{TitusRev}
 Path sampling methods like RETIS
  that are based on both forward and backward in time
 propagation do not have this issue.

We defined eight RETIS 
interfaces:  $\lambda_A=\lambda_0=-0.99$, 
$\lambda_1=-0.8$,  $\lambda_2=-0.7$,  $\lambda_3=-0.6$,  $\lambda_4=-0.5$,  $\lambda_5=-0.4$,  $\lambda_6=-0.3$, and  $\lambda_B=\lambda_7=1.0$, 
and ran four RETIS simulations using the PyRETIS code~\cite{PyRETIS1, PyRETIS2} consisting of 200,000 cycles.
In all simulations (Shooting, WF*, WF, WF-cap), each
path ensemble either employs only
shooting or only WF as main MC move in addition to replica exchange moves.
In simulation ``Shooting'' all path ensembles employ the shooting move. In the other simulations the
WF move is used for most path ensembles. However, simulation WF* uses normal shooting in the $[0^-]$ ensemble, while simulations WF and WF-cap use the shooting move in both the $[0^-]$ and $[0^+]$ ensemble as was
suggested in Sec.~\ref{sec:supdetbal}.
The
WF-cap simulation uses a cap-interface at $\lambda_{\rm cap}$ = 0.1 
At each cycle, 
all path ensembles are updated with an ensemble move 
(shooting or WF)
or with 
replica exchange moves
with a 50\%-50\% probability. 
In case that a replica exchange move is selected, another 
50\%-50\% probability determines whether the 
$[0^-] \leftrightarrow [0^+]$, $[1^+] \leftrightarrow  [2^+]$,
 $\ldots$, $[5^+] \leftrightarrow  [6^+]$ swaps will be attempted
 or the 
 $[0^+] \leftrightarrow [1^+]$, $[2^+] \leftrightarrow  [3^+]$,
 $\ldots$, $[4^+] \leftrightarrow  [5^+]$ swaps. 
 In the latter case, the $[0^-]$ and $[6^+]$ ensembles simply
 duplicate the previous path (null move).
 In the WF simulations, the number of subpaths was arbitrarily set equal to $N_s=6$ for all path ensembles.

\begin{figure}
    \begin{tikzpicture}
\begin{axis}[
xlabel={RETIS cycles/$10^{3}$},
ylabel={Rate/$10^{-7}$},
legend cell align={left},
legend entries={Kramers, WF*, WF, WF-cap, Shooting},
legend style={
draw=none,
legend pos=south east,
font=\small},
ymax=3.2]
\addplot [solid, ultra thick] coordinates {(0,2.58)(200,2.58)};
\addplot [blue, ultra thick] table {Figures/5-rate_internal/rrun_wf_star.txt};
\addplot [purple, ultra thick] table {Figures/5-rate_internal/rrun_shshwf.txt};
\addplot [teal, dashed, ultra thick] table {Figures/5-rate_internal/rrun_shshwf_cap.txt};
\addplot [magenta, ultra thick] table {Figures/5-rate_internal/rrun_sh.txt};
\end{axis}
\end{tikzpicture}
    \caption{Total running average of computed rate as function of RETIS cycles. }
    \label{fig:internal_rate}
\end{figure}
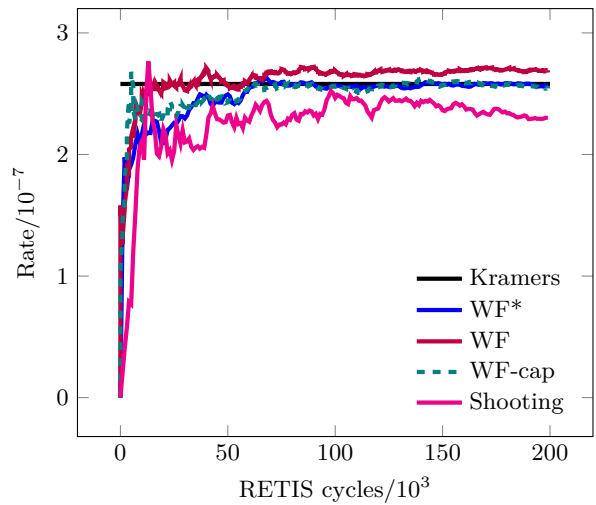

The results are shown in Fig.~\ref{fig:internal_rate}
and in table~\ref{tab:2d}
where they are compared with 
Kramers' theory~\cite{50Kramer} which, for this system, can be considered as a nearly exact reference. 
Fig.~\ref{fig:internal_rate} shows that the 
WF based simulations 
rapidly converge close to the Kramers' value of the rate
confirming the exactness of the superdetailed balance relations
and the correct implementation in the PyRETIS code.
The results based on shooting are further off, 
but have a significant lower computational cost per RETIS cycle (see table~\ref{tab:2d}).

Based on the relative errors from the block averaging analysis and the cost
per cycle, we can compute the CPU efficiency time for each method, shown in the last column.
Based on these numbers, we can see that the WF*, WF and WF-cap simulations are 2.4, 2.5 and 2.7 times more efficient than the simulation in which all path ensembles use 
the standard 
shooting move as their main MC move.
Note that an improvement of more than a factor 2 is rather remarkable given the low dimensionality 
of the system. 
\begin{table}[ht!]
\caption{Simulation data for the double well 1D barrier system.
The cost column describes the total number of calculated MD steps.
The errors are based on block averaging using single standard deviations. The final column shows the 
CPU efficiency times~\cite{TISeff} corresponding to the 
number of required MD steps for obtaining a relative error equal to 1. Simulation ``Shooting'' uses the standard shooting move as main MC move in all path ensembles. The other simulations use the WF move in all 
ensembles except for $[0^-]$ (WF*, WF, WF-cap) and   $[0^+]$ (WF, WF-cap).
WF-cap uses a cap-interface at $\lambda_{\rm cap}=0.1.$
}
\centering
\begin{tabular}{@{}lllll@{}}
\toprule
Simulation  & Rate$/10^{-7}$ & $\epsilon_r$ (\%) & Cost/$10^{7}$
&Cost$\cdot \epsilon_r^2$/$10^{11}$\\ \midrule
Shooting    &2.30 & 6.46  & 5.32  & 222.0 \\
WF*   & 2.58 & 2.19  & 19.56  & 93.9  \\
WF    & 2.69 & 2.28  & 16.98  & 88.3  \\
WF-cap & 2.54 &   2.29       &   15.72   & 82.4    \\ 
Kramers    &  2.58 &  &  \\
\bottomrule
\end{tabular}
\label{tab:2d}
\end{table}

In table~\ref{tab:table1}, we further
examine the acceptance probabilities of the different moves.
It is apparent that in all simulations
the main MC move 
has a nearly 100\% acceptance in the path ensembles where the WF move is employed
thanks to the high-acceptance protocol.
The acceptance is marginally lower at the last path ensembles $[5^+]$
and $[6^+]$ from which there is a higher probability to generate
$\lambda_B \rightarrow \lambda_B$ paths.
The shooting move has a lower acceptance, but has the advantage 
that all swapping moves with the $[0^-]$ ensemble are accepted
if shooting is the main move in both $[0^-]$ and $[0^+]$. 
Since $[0^-]$ can only swap with  $[0^+]$, these are the computational expensive 
$[0^-] \leftrightarrow [0^+]$ swaps.
The other swapping moves are inexpensive as they do not require any MD steps.

Therefore,
an anticipated lower acceptance for these swapping moves in the WF simulations would 
not be dramatic. 
However, even this is not always the case.
At first sight this appears counter-intuitive. Given a pair
of paths in two neighboring ensembles, the standard swap should always have an acceptance probability that is equal or higher
than the acceptance based on Eq.~\ref{eq:REha}. 
However, this effect can be canceled by 
the path distributions not being the same. 
Since the altered path distribution in the high-acceptance
scheme, Eq.~\ref{eq:Ptilde}, 
overrepresents paths with many points between $\lambda_i$ and
$\lambda_B$ or $\lambda_{\rm cap}$, 
the $[i^+]$ path ensemble is likely to contain a higher fraction of
paths crossing $\lambda_{i+1}$. 
From the data of table~\ref{tab:table1}, 
this seems indeed the case
in the majority of path ensembles.
\begin{table}[h!]
\footnotesize
\begin{center}
\caption{Acceptance ratio (\%).
Simulation ``Shooting'' uses the standard shooting move as main MC move in all path ensembles. The other simulations use the WF move in all 
ensembles except for $[0^-]$ (WF*, WF, WF-cap) and   $[0^+]$ (WF, WF-cap).
}
\label{tab:table1}
\begin{tabular}{
l 
l 
l 
l 
l 
l 
l 
l 
l 
}
\cline{1-1}
\cline{2-2}
\cline{3-3}
\cline{4-4}
\cline{5-5}
\cline{6-6}
\cline{7-7}
\cline{8-8}
\cline{9-9}
\multicolumn{1}{|c|}{\multirow{2}{*}{ens.}}&
\multicolumn{2}{c|}{\multirow{1}{*}{Shooting}}&
\multicolumn{2}{c|}{\multirow{1}{*}{WF*}}&
\multicolumn{2}{c|}{\multirow{1}{*}{WF}}&
\multicolumn{2}{c|}{\multirow{1}{*}{WF-cap}}\\
\multicolumn{1}{|c|}{\multirow{2}{*}{}}&
\multicolumn{1}{c}{\multirow{1}{*}{main}}&
\multicolumn{1}{c|}{\multirow{1}{*}{swap}}&
\multicolumn{1}{c}{\multirow{1}{*}{main}}&
\multicolumn{1}{c|}{\multirow{1}{*}{swap}}&
\multicolumn{1}{c}{\multirow{1}{*}{main}}&
\multicolumn{1}{c|}{\multirow{1}{*}{swap}}&
\multicolumn{1}{c}{\multirow{1}{*}{main}}&
\multicolumn{1}{c|}{\multirow{1}{*}{swap}}\\
\cline{1-1}
\cline{2-2}
\cline{3-3}
\cline{4-4}
\cline{5-5}
\cline{6-6}
\cline{7-7}
\cline{8-8}
\cline{9-9}
\multicolumn{1}{|c|}{\multirow{1}{*}{$[0^-]$}}&
\multicolumn{1}{r}{\multirow{1}{*}{84.6}}&
\multicolumn{1}{r|}{\multirow{1}{*}{100.0}}&
\multicolumn{1}{r}{\multirow{1}{*}{84.5}}&
\multicolumn{1}{r|}{\multirow{1}{*}{83.9}}&
\multicolumn{1}{r}{\multirow{1}{*}{84.3}}&
\multicolumn{1}{r|}{\multirow{1}{*}{100.0}}&
\multicolumn{1}{r}{\multirow{1}{*}{84.3}}&
\multicolumn{1}{r|}{\multirow{1}{*}{100.0}}\\
\multicolumn{1}{|c|}{\multirow{1}{*}{$[0^+]$}}&
\multicolumn{1}{r}{\multirow{1}{*}{84.2}}&
\multicolumn{1}{r|}{\multirow{1}{*}{57.8}}&
\multicolumn{1}{r}{\multirow{1}{*}{100.0}}&
\multicolumn{1}{r|}{\multirow{1}{*}{49.0}}&
\multicolumn{1}{r}{\multirow{1}{*}{84.0}}&
\multicolumn{1}{r|}{\multirow{1}{*}{55.6}}&
\multicolumn{1}{r}{\multirow{1}{*}{84.0}}&
\multicolumn{1}{r|}{\multirow{1}{*}{55.8}}\\
\multicolumn{1}{|c|}{\multirow{1}{*}{$[1^+]$}}&
\multicolumn{1}{r}{\multirow{1}{*}{48.8}}&
\multicolumn{1}{r|}{\multirow{1}{*}{15.5}}&
\multicolumn{1}{r}{\multirow{1}{*}{100.0}}&
\multicolumn{1}{r|}{\multirow{1}{*}{17.9}}&
\multicolumn{1}{r}{\multirow{1}{*}{100.0}}&
\multicolumn{1}{r|}{\multirow{1}{*}{16.3}}&
\multicolumn{1}{r}{\multirow{1}{*}{100.0}}&
\multicolumn{1}{r|}{\multirow{1}{*}{16.8}}\\
\multicolumn{1}{|c|}{\multirow{1}{*}{$[2^+]$}}&
\multicolumn{1}{r}{\multirow{1}{*}{37.8}}&
\multicolumn{1}{r|}{\multirow{1}{*}{13.4}}&
\multicolumn{1}{r}{\multirow{1}{*}{100.0}}&
\multicolumn{1}{r|}{\multirow{1}{*}{19.7}}&
\multicolumn{1}{r}{\multirow{1}{*}{100.0}}&
\multicolumn{1}{r|}{\multirow{1}{*}{19.9}}&
\multicolumn{1}{r}{\multirow{1}{*}{100.0}}&
\multicolumn{1}{r|}{\multirow{1}{*}{20.2}}\\
\multicolumn{1}{|c|}{\multirow{1}{*}{$[3^+]$}}&
\multicolumn{1}{r}{\multirow{1}{*}{32.2}}&
\multicolumn{1}{r|}{\multirow{1}{*}{11.5}}&
\multicolumn{1}{r}{\multirow{1}{*}{100.0}}&
\multicolumn{1}{r|}{\multirow{1}{*}{18.0}}&
\multicolumn{1}{r}{\multirow{1}{*}{100.0}}&
\multicolumn{1}{r|}{\multirow{1}{*}{18.5}}&
\multicolumn{1}{r}{\multirow{1}{*}{100.0}}&
\multicolumn{1}{r|}{\multirow{1}{*}{18.4}}\\
\multicolumn{1}{|c|}{\multirow{1}{*}{$[4^+]$}}&
\multicolumn{1}{r}{\multirow{1}{*}{30.1}}&
\multicolumn{1}{r|}{\multirow{1}{*}{12.2}}&
\multicolumn{1}{r}{\multirow{1}{*}{100.0}}&
\multicolumn{1}{r|}{\multirow{1}{*}{20.3}}&
\multicolumn{1}{r}{\multirow{1}{*}{100.0}}&
\multicolumn{1}{r|}{\multirow{1}{*}{20.6}}&
\multicolumn{1}{r}{\multirow{1}{*}{100.0}}&
\multicolumn{1}{r|}{\multirow{1}{*}{20.4}}\\
\multicolumn{1}{|c|}{\multirow{1}{*}{$[5^+]$}}&
\multicolumn{1}{c}{\multirow{1}{*}{30.0}}&
\multicolumn{1}{c|}{\multirow{1}{*}{14.7}}&
\multicolumn{1}{c}{\multirow{1}{*}{99.9}}&
\multicolumn{1}{c|}{\multirow{1}{*}{28.3}}&
\multicolumn{1}{c}{\multirow{1}{*}{99.8}}&
\multicolumn{1}{c|}{\multirow{1}{*}{28.2}}&
\multicolumn{1}{c}{\multirow{1}{*}{100.0}}&
\multicolumn{1}{c|}{\multirow{1}{*}{26.6}}\\
\multicolumn{1}{|c|}{\multirow{1}{*}{$[6^+]$}}&
\multicolumn{1}{c}{\multirow{1}{*}{29.1}}&
\multicolumn{1}{c|}{\multirow{1}{*}{16.7}}&
\multicolumn{1}{c}{\multirow{1}{*}{99.2}}&
\multicolumn{1}{c|}{\multirow{1}{*}{34.2}}&
\multicolumn{1}{c}{\multirow{1}{*}{99.2}}&
\multicolumn{1}{c|}{\multirow{1}{*}{33.9}}&
\multicolumn{1}{c}{\multirow{1}{*}{100.0}}&
\multicolumn{1}{c|}{\multirow{1}{*}{30.8}}\\
\cline{1-1}
\cline{2-2}
\cline{3-3}
\cline{4-4}
\cline{5-5}
\cline{6-6}
\cline{7-7}
\cline{8-8}
\cline{9-9}
\end{tabular}
\end{center}
\end{table}


\subsection{Thin Film Breakage}
\label{sec:hole}
A system of 1100 dodecane molecules layered on a slab of 23936 water molecules is studied in the NPT ensemble via full atom 
TIS simulations
using the GROMACS 2020.1 simulation package \cite{abraham2015gromacs} as external engine. The dodecane molecules are simulated according to the 
OPLS-AA force field \cite{jorgensen_development_1996} and the water molecules with the TIP4p/2005 model \cite{abascal_general_2005}. The preparation of the initial equilibrated system is explained in detail by 
Ref.~\onlinecite{aaroen_thin_2022}. The temperature is set to 300 K and is controlled with a velocity rescaling method \cite{bussi_canonical_2007} employing a coupling time of 0.1 ps. Pressure is controlled by the Berendsen barostat and its normal component is maintained constant at 1 bar, with a time
constant of 1.0 ps and compressibility coefficient of $4.7 \cdot 10^{-5} \mathrm{bar}^{-1}$. The velocity-Verlet algorithm is used to solve the Newton equations of motion with a timestep of 0.002 ps. Periodic boundary conditions are applied in all directions, with the \textit{z} direction being perpendicular to the 2D film. The box size is set to equal box size of $15 \times 15 \times 5.1983$ nm.

The order parameter  of the system is calculated through discretizing the system into $85 \times 85$ tiles in the $x$ and $y$ direction such that the 
order parameter value becomes the number of empty dodecane tiles that also have empty neighbors in the \textit{x} and \textit{y} direction. Such a definition provides a way to measure the presence of low-density regions, in addition to any breakage or ``hole'' formation that occurs within a trajectory. The sensitivity of the order parameter is determined by the specified discretizing size. In our case,
the order parameter values fluctuated between 0 and 5 during an equilibrium run at $T = 300$ K. 
Based on this, we set $\lambda_A=5$. We further defined $\lambda_B=100$ as preliminary analysis showed 
that
from this point on the hole tends to grow further
with negligible chance to close again. 

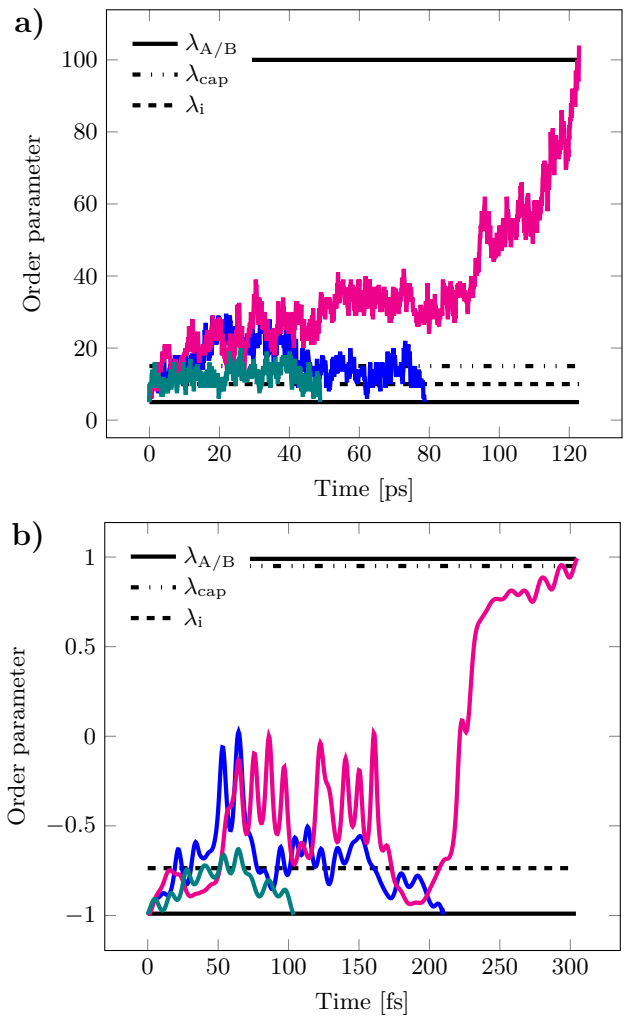
\begin{figure}[ht!]
    \begin{tikzpicture}
\begin{axis}[
xlabel={Time [ps]},
ylabel={Order parameter},
legend cell align={left},
legend entries={$\lambda_\mathrm{A/B}$,$\lambda_\mathrm{cap}$,$\lambda_\mathrm{i}$},
legend style={
draw=none,
legend pos=north west,
font=\small}]
\addplot [solid, ultra thick] coordinates {(0,5)(122.80,5)};
\addplot [loosely dashdotdotted, ultra thick] coordinates {(0,15)(122.80,15)};
\addplot [dashed, ultra thick] coordinates {(0,10)(122.80,10)};
\addplot [solid, ultra thick] coordinates {(0,100)(122.80,100)};
\addplot [blue, ultra thick] table {Figures/1-hole/1-OP_vs_dt/low-247.txt};
\addplot [magenta, ultra thick] table {Figures/1-hole/1-OP_vs_dt/low-687.txt};
\addplot [teal, ultra thick] table {Figures/1-hole/1-OP_vs_dt/low-753.txt};
\end{axis}
\node at (rel axis cs:-0.065,1.05)  {\large\textbf{a)}};
\end{tikzpicture}
    \begin{tikzpicture}
\begin{axis}[
xlabel={Time [fs]},
ylabel={Order parameter},
legend cell align={left},
legend entries={$\lambda_\mathrm{A/B}$,$\lambda_\mathrm{cap}$,$\lambda_\mathrm{i}$},
legend style={
draw=none,
legend pos=north west,
font=\small}]
\addplot [solid, ultra thick] coordinates {(0,-0.99)(304.0,-0.99)};
\addplot [loosely dashdotdotted, ultra thick] coordinates {(0,0.95)(304.0,0.95)};
\addplot [dashed, ultra thick] coordinates {(0,-0.736)(304.0,-0.736)};
\addplot [solid, ultra thick] coordinates {(0,0.99)(304.0,0.99)};
\addplot [blue, ultra thick] table {Figures/6-ruru_opvsdt/44-path.txt};
\addplot [magenta, ultra thick] table {Figures/6-ruru_opvsdt/22-path.txt};
\addplot [teal, ultra thick] table {Figures/6-ruru_opvsdt/49-path.txt};
\end{axis}
\node at (rel axis cs:-0.065,1.05)  {\large\textbf{b)}};
\end{tikzpicture}
    \centering
    \caption{
    Exemplary trajectories from the WF algorithm in the $[i^+]$ path ensemble 
    showing the progress of the order parameter versus
    time. The stable state interfaces $\lambda_A$, $\lambda_B$, the cap-interface $\lambda_{\rm cap}$, and the ensemble interface $\lambda_i$
    are shown as well. The 
    two different panels represent the a)
    classical MD level simulation of the 
    thin film breakage and b) the AIMD level simulations of 
    the ruthenium self-exchange reaction.
    }
    \label{fig:hole_1}
\end{figure}

To obtain an initial reactive trajectory, we ran an equilibrium run at $T = 375$ K until the thin-film broke down.
For our single path ensemble analysis we further defined $\lambda_\mathrm{i} = 10.0$ as the interface that has to be crossed.
In addition, we set the cap-interface 
$\lambda_\mathrm{cap} = 15.0$. 
We then created  1000 trajectories using standard shooting and WF with 
$N_s=10$.
Three exemplary trajectories from the WF simulation are shown in 
Fig.~\ref{fig:hole_1}-a).

From the sample size of 1000 MC moves, the acceptance in WF was equal to 73.4\% and 35.0\% for standard shooting. 
The limited sample size prohibit accurate CPU efficiency analysis, but a qualitative assertion of the sampling effectivity can be obtained 
by viewing the simulated path lengths as function of the MC step.
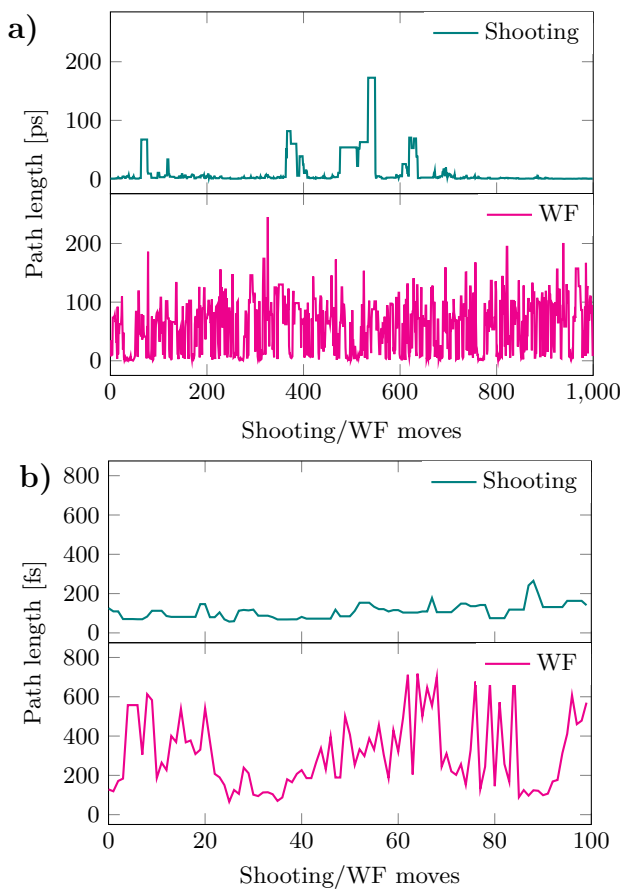
\begin{figure}[ht!]
\begin{tikzpicture}
\begin{groupplot}[
group style={
group name=my plots,
group size=1 by 2,
horizontal sep=0.0cm,
vertical sep=0.0cm,
x descriptions at=edge bottom,
y descriptions at=edge left,
},
legend style={
draw=none,
nodes={scale=1, transform shape},
anchor=north east,
at={(1,1)},
},
width=8cm,
height=4cm,
xlabel=Shooting/WF moves,
xmin=0,ymin=-25,xmax=1000,ymax=285,
]
\nextgroupplot
\addlegendentry {{$\mathrm{Shooting}$}}
\addplot [teal, thick] table {Figures/2-path_length/sh_len_hole.txt};
\nextgroupplot[ylabel={Path length [ps]}, every axis y label/.append style={at=(ticklabel cs:1.0)}]
\addlegendentry {{WF}}
\addplot [magenta, thick] table {Figures/2-path_length/cs_len_hole.txt};
\end{groupplot}
\node at (rel axis cs:-0.18,0.90)  {\large\textbf{a)}};
\end{tikzpicture}
\begin{tikzpicture}
\begin{groupplot}[
group style={
group name=my plots,
group size=1 by 2,
horizontal sep=0.0cm,
vertical sep=0.0cm,
x descriptions at=edge bottom,
y descriptions at=edge left,
},
legend style={
draw=none,
nodes={scale=1, transform shape},
anchor=north east,
at={(1,1)},
},
width=8cm,
height=4cm,
xlabel=Shooting/WF moves,
xmin=0,ymin=-50,xmax=100,ymax=875,
]
\nextgroupplot
\addlegendentry {{$\mathrm{Shooting}$}}
\addplot [teal, thick] table {Figures/2-path_length/sh_len_ruru.txt};
\nextgroupplot[ylabel={Path length [fs]}, every axis y label/.append style={at=(ticklabel cs:1.0)}]
\addlegendentry {{WF}}
\addplot [magenta, thick] table {Figures/2-path_length/cs_len_ruru.txt};
\end{groupplot}
\node at (rel axis cs:-0.15,0.90)  {\large\textbf{b)}};
\end{tikzpicture}
    \centering
    \caption{Path length vs MC move for WF and standard shooting for
    a) classical MD system of thin film breakage and b) AIMD system of 
    ruthenium self-exchange reaction.
    }
    \label{fig:path_length}
\end{figure}

Fig.~\ref{fig:path_length}-a) shows that the WF sampling
has much more frequent transitions between long and short paths whereas shooting is mostly stuck in  the short path domain. Once the shooting move manages to produce a long path, the path remains in the MC chain due to
a long series of rejections (e.g. around step 500 where the same path length remains for a number of steps due to rejections). This indicates that the shooting move is struggling to properly sample  path space. Even if the acceptance is not extremely low for the short paths, it fails to 
make regular switches to the longer paths. Moreover, if a long path is generated, 
the subsequent moves are likely rejected such 
 that other longer paths are not likely found.

\subsection{Ruthenium-Ruthenium Self-Exchange Reaction}
\label{sec:ruru}

We studied self-exchange reaction between two ruthenium ions 
in aqueous solution described by the following chemical reaction
\begin{equation}
    \mathrm{Ru^{2+}} + \mathrm{Ru^{3+}} \rightarrow \mathrm{Ru^{3+}} + \mathrm{Ru^{2+}}
\end{equation}
The simulation system consisted of two ruthenium ions, 
63 $\mathrm{H_2O}$ molecules and one OH$^{-}$ ion. 
The dynamics were propagated using NVE velocity-Verlet and the 
CP2K~\cite{CP2K} simulation package.
The effect of temperature was introduced 
via the randomization of velocities 
from a 
Maxwell-Boltzmann distribution at a temperature of 300 K.
We used a time step of 0.5 fs and periodic boundary conditions were 
applied to a cubic box with an edge length of 12.4138 \AA. Further simulation details on functional and basis sets are explained in Ref.~\onlinecite{tiwari_reactive_2016}. 

To monitor the reaction progress, the electron transfer has been ``followed'' by transforming the occupied Kohn-Sham orbitals~\cite{KS} into maximally localized Wannier functions (MLWF)~\cite{Wannier} and computing the distance between the
center of these localized functions (X) describing the moving electron
to each of the ruthenium ions. The order parameter of the system 
is then defined as
\begin{equation}
\lambda
=\frac{\left(d_{\mathrm{Ru}-X}-d_{\mathrm{Ru}^{\prime}-X}\right)}{d_{\mathrm{Ru}-\mathrm{Ru}^{\prime}}}
\label{eq:lambdaRU}
\end{equation}
where $d_{\mathrm{Ru}-X}$ is the distance between X and the initial ruthenium electron donor, $d_{\mathrm{Ru}^{\prime}-X}$ is the distance between X and the initial ruthenium electron acceptor and $d_{\mathrm{Ru}-\mathrm{Ru}^{\prime}}$ is the distance between the two ruthenium ions in the system. In this formulation, $\lambda_{\mathrm{}}=-1$ and 
$\lambda_{\mathrm{}}=+1$ define the 
reactant state and product state, respectively. 
$\mathrm{Ru^{2+}}$/$\mathrm{Ru^{3+}}$ have 5/6 d-electrons and $\mathrm{H_{2}O}$/$\mathrm{OH^{-}}$ have 8 valence electrons.
This means there are a total of 523 MLWFs in the system.
The order parameter, Eq.~\ref{eq:lambdaRU},
requires the location X of the transferring electron, which is one of the centers of these 523 MLWFs.
To identify which is X, each Wannier center is being linked to either a ruthenium or oxygen atom that is closest. 
Then, if one ruthenium ion has 6 associated MLWFs, X is set to be the one that is the farthest away from this ruthenium ion.
In the case that both ruthenium ions have 5 associated MLWFs, one of the 
oxygens has an excess MLWF (9 instead of 8), and 
the center that is farthest away from this oxygen is set as X.

To qualitatively compare standard shooting and WF for this system we run 
two single path ensemble simulations representing $[i^+]$ with $\lambda_i=-0.736$, $\lambda_A=-0.99$, and $\lambda_B=+0.99$. 
The value for $\lambda_i=-0.736$ was chosen from preliminary runs where we aimed for a 20\% probability that a path ends up at state $B$.
In the WF simulation, an additional $\lambda_{\rm cap}=+0.95$ was set to
avoid $\lambda_B \rightarrow \lambda_B$ rejections due to the selection of shooting points lying within the basin of attraction of state $B$.
Here, we only applied a rather modest number of subtrajectories $N_s=2$.
Higher performances might be obtained with a larger number of subpaths.
Exemplary trajectories of the WF simulation are show in Fig~\ref{fig:hole_1}-b).

Due to the relatively low value of $N_s$, the subpath contribution to the total WF computational cost is only 15\%. The acceptance probability increased from the shooting move's 48\% to WF's 96\%.
Similarly to the classical MD system, the WF simulation seems to show a
better sample exploration when we look at the path length as function of the MC step (Fig.~\ref{fig:path_length}-b)).  
The standard shooting algorithm seems not to be 
able
to produce any paths 
larger than 300 fs.
The WF algorithm, however, started with a short initial path but was able to quickly
move up to to the 600 fs range and making regular transitions between the shorter and 
longer paths. So also here, the sampling quality of the WF algorithm appears 
substantially superior to the one of standard shooting.

\section{Concluding Remarks}
We reviewed the recently developed
subtrajectory moves 
stone skipping (SS)
and web throwing (WT) and added a new member to this group: wire fencing (WF).
These moves are more efficient than the standard shooting move which has been the main MC move for path sampling simulations during the last two decades.
The subtrajectory moves proceed from a complete old path to a complete new path
via a series of intermediate short paths (subpaths/subtrajectories).
While this increases the average cost of a MC step, the correlations between paths are substantially reduced leading to a lower statistical inefficiency.
The use of shorter paths resembles approximate path sampling methods like 
PPTIS or milestoning. 
However, the subtrajectory moves are still exact like standard shooting
as they are based on mathematically rigorous superdetailed balance 
relations.
The approach is preferably combined with a high-acceptance protocol in which 
the sampling distribution of the paths is adjusted in order 
to maximize the acceptance of newly generated trajectories. The effect of the biased distribution is undone
in the post-simulation analysis using appropriate reweighting.
The SS and WT move, however, require a one-step crossing condition which complicates their implementation and we discussed several solutions for this 
issue. The new WF does not rely on 
the one-step crossing condition and is therefore the most practical solution to
the aforementioned problem even if it is slightly more wasteful than SS and WT.
The WF move is in particular useful when the path sampling code 
uses an external MD engine and/or when the computation of the order parameter is costly.
We showed the exactness 
and the efficiency gain
of the WF approach in a RETIS simulation where  the transition rate 
of an underdamped Langevin particle in a double well potential has been computed and compared with the  analytical Kramers' expression.
Thereafter, we showed qualitatively how the WF move performs in a classical MD 
system describing the thin film breaking process and in an AIMD level system
describing an electron transfer process between ruthenium ions in aqueous solution. In both cases, the WF move seems to allow a faster sampling through path space than standard shooting, which was concluded from the rapid switches that WF made between the shorter and longer paths.

\begin{acknowledgments}
We acknowledge funding from the Research Council of Norway (toppforsk project Theolight, grant no. 275506)  and
computational resources from NOTUR (project nn9254k).
\end{acknowledgments}

\section*{Author Declarations}
The authors have no conflicts to disclose.

\section*{Data Availability Statement}
The data that support the findings of this study are available from the corresponding author upon reasonable request.

The 
algorithmic developments
have been implemented
in the current pyretis2.dev version (the current release is 
PyRETIS-2~\cite{PyRETIS2}) 
and will be included in the forthcoming main release (PyRETIS-3).
The code is already available at https://gitlab.com/pyretis following the FAIR principle for scientific data and software and data.~\cite{2019envisioning, lamprecht2020towards}

\appendix

\section{Analytical expressions for the statistical inefficiency in model systems}
\label{app:notional}
Sec.~\ref{sec:statineff} introduces a model where at each MC move $j$ there is a chance of $\pi_R$ that the state 
of the system remains essentially unchanged and a chance of  $\pi_M=1-\pi_R$ to ``throw a dice''. The latter implies that at step $j$ the output value of $h_j$ equals 1 with a probability $p$ and 0 with a probability 
$1-p$. Let us consider the conditional probability that $h_j=0$ given that $h_0=0$: $P(h_j=0|h_0=0)$. We can distinguish two scenarios. Scenario 1 relates to the case that all $j$ moves implied a ``remain'' and therefore $h_j=h_0=0$.
Scenario 2 is related to the situation that at least once the dice was thrown. In this scenario we have that 
$h_j$ is either 1 or 0 with respective probabilities $p$ and $1-p$. The probability of having scenario 1 equals
$\pi_R^j$ and that of scenario 2 equals $1-\pi_R^j$. Therefore,
\begin{align}
    P(h_j=0|h_0=0)&=\pi_R^j+(1-\pi_R^j) (1-p) \nonumber \\
    &= (1-p) + p \pi_R^j
    \label{eq:condp0}
\end{align}
Likewise, we can derive all the other conditional probabilities:
\begin{align}
    P(h_j=1|h_0=0)&=(1-\pi_R^j) p=p - p \pi_R^j \nonumber \\
    P(h_j=0|h_0=1)&=(1-\pi_R^j)  (1-p)=(1-p) + (1-p) \pi_R^j \nonumber \\
    P(h_j=1|h_0=1)&=\pi_R^j+(1-\pi_R^j) p = p + (1-p) \pi_R^j 
\label{eq:condp}
\end{align}
Let us call 
$p_{kl}=P(h_j=k \wedge h_0=l )=P(h_j=k|h_0=l) P(h_0=l)$.
From Eqs. \ref{eq:condp0} and \ref{eq:condp} we can derive:
\begin{align}
    p_{00}&= (1-p)^2 + p (1-p)\pi_R^j  \nonumber \\
    p_{10}&=p(1-p) - p (1-p) \pi_R^j=p_{01}\nonumber \\
    p_{11}&= p^2 + p(1-p)  \pi_R^j 
\label{eq:pkl}
\end{align}
and from this we can compute
\begin{align}
 \left \langle (h_0-p) (h_j-p) \right \rangle &=
 p_{00} p^2 - p_{10} (1-p) p \nonumber \\
 &-p_{01} (1-p) p+p_{11} (1-p)^2 
\label{eq:csub}
\end{align}
In the above expression, all the $\pi_R$-independent terms cancel.  This is expected since we know the result equals 0 if $\pi_R=0$. The remaining $\pi_R$-dependent terms sum up to
\begin{align}
 &p (1-p) \pi_R^j \left [
 p^2+2 p(1-p)+(1-p)^2
 \right] \nonumber \\
 &=p (1-p) \pi_R^j \left [
 p+ (1-p)
 \right]^2= p (1-p) \pi_R^j
\label{eq:csub2}
\end{align}
From Eqs.~\ref{eq:nc}, \ref{eq:s2}, and \ref{eq:csub2} we derive that
\begin{align}
C(j)=\pi_R^j \Rightarrow n_c= \frac{\pi_R}{1-\pi_R}
\label{eq:Cnc}
\end{align}
and via Eq.~\ref{eq:spm2}:
\begin{align}
{\mathcal N}=1+2 \frac{\pi_R}{1-\pi_R} = \frac{1+\pi_R}{1-\pi_R}
\label{eq:ineffpir}
\end{align}
As $\pi_M=1-\pi_R$, \ref{eq:ineffpir} is equivalent to Eq.~\ref{eq:ineffpim} of Sec.~\ref{sec:statineff}.

In the second model we assume $\pi_R=0$, but there are two phases $x=\alpha, \beta$ that have have,
respectively,
probabilities 
$P_\alpha$ and $P_\beta$ and local crossing probabilities $p_\alpha$ and $p_\beta$. Let 
$\pi_{\rho}=1-\pi_{\mu}$ be the probability that the MC maintains the previous phase. 
The inverse probability  $\pi_{\mu}$ implies throwing the dice to determine the phase $x$
such that the selection 
probability
for $x$ corresponds to $P_\alpha$ and $P_\beta=1-P_\alpha$. After the phase $x$ is set, $h_j$ will be set to 1 or 0 with respective probabilities $p_x$ and $(1-p_x)$. Given that the phase of the first sample
equals $x_0=x$, the chance that the $j$-th sample has the same or opposite phase equals, respectively, $\pi_{\rho}^j+(1-\pi_{\rho}^j) P_x$ and $(1-\pi_{\rho}^j) (1-P_x)$.
This leads to following conditional probabilities akin Eqs.~\ref{eq:condp0} and \ref{eq:condp}:
\begin{align}
P(h_j=0|x_0=x)&=\pi_\rho^j (1-p_x)+(1-\pi_\rho^j) (1-p)
\nonumber \\
&=\pi_\rho^j (p-p_x)+(1-p) \nonumber \\
&= \pi_\rho^j P_y (p_y-p_x)+(1-p) \nonumber \\
P(h_j=1|x_0=x)&=\pi_\rho^j p_x + (1-\pi_\rho^j) p
\nonumber \\
&=\pi_\rho^j(p_x-p)+p \nonumber \\
&=\pi_\rho^j P_y (p_x-p_y)+p
\label{eq:Pcond2}
\end{align}
where $y \in (\alpha, \beta)$ and $y\neq x$.
Hence, analogous to Eqs.~\ref{eq:pkl}
\begin{align}
    p_{k0}&= \sum_{x=\alpha,\beta} P_x (1-p_x)
    P(h_j=k|x_0=x)\nonumber \\
    p_{k1}&= \sum_{x=\alpha,\beta} P_x p_x
    P(h_j=k|x_0=x)\nonumber \\
\label{eq:pkl2}
\end{align}
which leads to
\begin{align}
    p_{00}&= (1-p)^2 + \pi_\rho^j \sum_x P_x P_y (1-p_x) (p_y-p_x) \nonumber \\
    &= (1-p)^2 + \pi_\rho^j P_\alpha P_\beta
    (p_\alpha-p_\beta)^2
    \nonumber \\
     p_{10}&= p(1-p) + \pi_\rho^j \sum_x P_x P_y (1-p_x) (p_x-p_y)  \nonumber \\
     &= p(1-p) - \pi_\rho^j P_\alpha P_\beta 
     (p_\alpha-p_\beta)^2 =p_{01}\nonumber \\
      p_{11}&=p^2 + \pi_\rho^j \sum_x P_x
      P_y
      p_x (p_x-p_y) \nonumber \\
      &=p^2 + \pi_\rho^j P_\alpha P_\beta
    (p_\alpha-p_\beta)^2
\label{eq:pkl3}
\end{align}
Analogous to Eqs.~\ref{eq:csub} and \ref{eq:csub2}
we find that
\begin{align}
 \left \langle (h_0-p) (h_j-p) \right \rangle &=
\pi_\rho^j P_\alpha P_\beta (p_\alpha-p_\beta)^2 
\end{align}
and like Eq.~\ref{eq:Cnc}:
\begin{align}
C(j)  &=\frac{P_\alpha P_\beta (p_\alpha-p_\beta)^2 }{p(1-p)} \pi_\rho^j 
=K_{s} \pi_\rho^j 
\nonumber \\
\Rightarrow n_c&=
K_{s} 
\frac{\pi_\rho}{1-\pi_\rho}=  K_{s} 
\frac{1-\pi_\mu}{\pi_\mu} \label{eq:nc2phase}
\end{align}
where we used $\pi_\mu=1-\pi_\rho$ and Eq.~\ref{eq:Ks}. 
Substition of Eq.~\ref{eq:nc2phase} in Eq.~\ref{eq:spm2}
leads to Eq.~\ref{eq:ineffpimx}.

\bibliographystyle{apsrev4-1}
\bibliography{literature.bib}

\begin{thebibliography}{64}%
\makeatletter
\providecommand \@ifxundefined [1]{%
 \@ifx{#1\undefined}
}%
\providecommand \@ifnum [1]{%
 \ifnum #1\expandafter \@firstoftwo
 \else \expandafter \@secondoftwo
 \fi
}%
\providecommand \@ifx [1]{%
 \ifx #1\expandafter \@firstoftwo
 \else \expandafter \@secondoftwo
 \fi
}%
\providecommand \natexlab [1]{#1}%
\providecommand \enquote  [1]{``#1''}%
\providecommand \bibnamefont  [1]{#1}%
\providecommand \bibfnamefont [1]{#1}%
\providecommand \citenamefont [1]{#1}%
\providecommand \href@noop [0]{\@secondoftwo}%
\providecommand \href [0]{\begingroup \@sanitize@url \@href}%
\providecommand \@href[1]{\@@startlink{#1}\@@href}%
\providecommand \@@href[1]{\endgroup#1\@@endlink}%
\providecommand \@sanitize@url [0]{\catcode `\\12\catcode `\$12\catcode
  `\&12\catcode `\#12\catcode `\^12\catcode `\_12\catcode `\%12\relax}%
\providecommand \@@startlink[1]{}%
\providecommand \@@endlink[0]{}%
\providecommand \url  [0]{\begingroup\@sanitize@url \@url }%
\providecommand \@url [1]{\endgroup\@href {#1}{\urlprefix }}%
\providecommand \urlprefix  [0]{URL }%
\providecommand \Eprint [0]{\href }%
\providecommand \doibase [0]{http://dx.doi.org/}%
\providecommand \selectlanguage [0]{\@gobble}%
\providecommand \bibinfo  [0]{\@secondoftwo}%
\providecommand \bibfield  [0]{\@secondoftwo}%
\providecommand \translation [1]{[#1]}%
\providecommand \BibitemOpen [0]{}%
\providecommand \bibitemStop [0]{}%
\providecommand \bibitemNoStop [0]{.\EOS\space}%
\providecommand \EOS [0]{\spacefactor3000\relax}%
\providecommand \BibitemShut  [1]{\csname bibitem#1\endcsname}%
\let\auto@bib@innerbib\@empty
\bibitem [{\citenamefont {Lindorff-Larsen}\ \emph {et~al.}(2011)\citenamefont
  {Lindorff-Larsen}, \citenamefont {Piana}, \citenamefont {Dror},\ and\
  \citenamefont {Shaw}}]{howfast}%
  \BibitemOpen
  \bibfield  {author} {\bibinfo {author} {\bibfnamefont {K.}~\bibnamefont
  {Lindorff-Larsen}}, \bibinfo {author} {\bibfnamefont {S.}~\bibnamefont
  {Piana}}, \bibinfo {author} {\bibfnamefont {R.~O.}\ \bibnamefont {Dror}}, \
  and\ \bibinfo {author} {\bibfnamefont {D.~E.}\ \bibnamefont {Shaw}},\ }\href
  {\doibase 10.1126/science.1208351} {\bibfield  {journal} {\bibinfo  {journal}
  {Science}\ }\textbf {\bibinfo {volume} {334}},\ \bibinfo {pages} {517}
  (\bibinfo {year} {2011})}\BibitemShut {NoStop}%
\bibitem [{\citenamefont {Shaw}\ \emph {et~al.}(2021)\citenamefont {Shaw},
  \citenamefont {Adams}, \citenamefont {Azaria}, \citenamefont {Bank},
  \citenamefont {Batson}, \citenamefont {Bell}, \citenamefont {Bergdorf},
  \citenamefont {Bhatt}, \citenamefont {Butts}, \citenamefont {Correia},
  \citenamefont {Dirks}, \citenamefont {Dror}, \citenamefont {Eastwood},
  \citenamefont {Edwards}, \citenamefont {Even}, \citenamefont {Feldmann},
  \citenamefont {Fenn}, \citenamefont {Fenton}, \citenamefont {Forte},
  \citenamefont {Gagliardo}, \citenamefont {Gill}, \citenamefont {Gorlatova},
  \citenamefont {Greskamp}, \citenamefont {Grossman}, \citenamefont
  {Gullingsrud}, \citenamefont {Harper}, \citenamefont {Hasenplaugh},
  \citenamefont {Heily}, \citenamefont {Heshmat}, \citenamefont {Hunt},
  \citenamefont {Ierardi}, \citenamefont {Iserovich}, \citenamefont {Jackson},
  \citenamefont {Johnson}, \citenamefont {Kirk}, \citenamefont {Klepeis},
  \citenamefont {Kuskin}, \citenamefont {Mackenzie}, \citenamefont {Mader},
  \citenamefont {McGowen}, \citenamefont {McLaughlin}, \citenamefont {Moraes},
  \citenamefont {Nasr}, \citenamefont {Nociolo}, \citenamefont {O'Donnell},
  \citenamefont {Parker}, \citenamefont {Peticolas}, \citenamefont {Pocina},
  \citenamefont {Predescu}, \citenamefont {Quan}, \citenamefont {Salmon},
  \citenamefont {Schwink}, \citenamefont {Shim}, \citenamefont {Siddique},
  \citenamefont {Spengler}, \citenamefont {Szalay}, \citenamefont {Tabladillo},
  \citenamefont {Tartler}, \citenamefont {Taube}, \citenamefont {Theobald},
  \citenamefont {Towles}, \citenamefont {Vick}, \citenamefont {Wang},
  \citenamefont {Wazlowski}, \citenamefont {Weingarten}, \citenamefont
  {Williams},\ and\ \citenamefont {Yuh}}]{Anton3}%
  \BibitemOpen
  \bibfield  {author} {\bibinfo {author} {\bibfnamefont {D.~E.}\ \bibnamefont
  {Shaw}}, \bibinfo {author} {\bibfnamefont {P.~J.}\ \bibnamefont {Adams}},
  \bibinfo {author} {\bibfnamefont {A.}~\bibnamefont {Azaria}}, \bibinfo
  {author} {\bibfnamefont {J.~A.}\ \bibnamefont {Bank}}, \bibinfo {author}
  {\bibfnamefont {B.}~\bibnamefont {Batson}}, \bibinfo {author} {\bibfnamefont
  {A.}~\bibnamefont {Bell}}, \bibinfo {author} {\bibfnamefont {M.}~\bibnamefont
  {Bergdorf}}, \bibinfo {author} {\bibfnamefont {J.}~\bibnamefont {Bhatt}},
  \bibinfo {author} {\bibfnamefont {J.~A.}\ \bibnamefont {Butts}}, \bibinfo
  {author} {\bibfnamefont {T.}~\bibnamefont {Correia}}, \bibinfo {author}
  {\bibfnamefont {R.~M.}\ \bibnamefont {Dirks}}, \bibinfo {author}
  {\bibfnamefont {R.~O.}\ \bibnamefont {Dror}}, \bibinfo {author}
  {\bibfnamefont {M.~P.}\ \bibnamefont {Eastwood}}, \bibinfo {author}
  {\bibfnamefont {B.}~\bibnamefont {Edwards}}, \bibinfo {author} {\bibfnamefont
  {A.}~\bibnamefont {Even}}, \bibinfo {author} {\bibfnamefont {P.}~\bibnamefont
  {Feldmann}}, \bibinfo {author} {\bibfnamefont {M.}~\bibnamefont {Fenn}},
  \bibinfo {author} {\bibfnamefont {C.~H.}\ \bibnamefont {Fenton}}, \bibinfo
  {author} {\bibfnamefont {A.}~\bibnamefont {Forte}}, \bibinfo {author}
  {\bibfnamefont {J.}~\bibnamefont {Gagliardo}}, \bibinfo {author}
  {\bibfnamefont {G.}~\bibnamefont {Gill}}, \bibinfo {author} {\bibfnamefont
  {M.}~\bibnamefont {Gorlatova}}, \bibinfo {author} {\bibfnamefont
  {B.}~\bibnamefont {Greskamp}}, \bibinfo {author} {\bibfnamefont
  {J.}~\bibnamefont {Grossman}}, \bibinfo {author} {\bibfnamefont
  {J.}~\bibnamefont {Gullingsrud}}, \bibinfo {author} {\bibfnamefont
  {A.}~\bibnamefont {Harper}}, \bibinfo {author} {\bibfnamefont
  {W.}~\bibnamefont {Hasenplaugh}}, \bibinfo {author} {\bibfnamefont
  {M.}~\bibnamefont {Heily}}, \bibinfo {author} {\bibfnamefont {B.~C.}\
  \bibnamefont {Heshmat}}, \bibinfo {author} {\bibfnamefont {J.}~\bibnamefont
  {Hunt}}, \bibinfo {author} {\bibfnamefont {D.~J.}\ \bibnamefont {Ierardi}},
  \bibinfo {author} {\bibfnamefont {L.}~\bibnamefont {Iserovich}}, \bibinfo
  {author} {\bibfnamefont {B.~L.}\ \bibnamefont {Jackson}}, \bibinfo {author}
  {\bibfnamefont {N.~P.}\ \bibnamefont {Johnson}}, \bibinfo {author}
  {\bibfnamefont {M.~M.}\ \bibnamefont {Kirk}}, \bibinfo {author}
  {\bibfnamefont {J.~L.}\ \bibnamefont {Klepeis}}, \bibinfo {author}
  {\bibfnamefont {J.~S.}\ \bibnamefont {Kuskin}}, \bibinfo {author}
  {\bibfnamefont {K.~M.}\ \bibnamefont {Mackenzie}}, \bibinfo {author}
  {\bibfnamefont {R.~J.}\ \bibnamefont {Mader}}, \bibinfo {author}
  {\bibfnamefont {R.}~\bibnamefont {McGowen}}, \bibinfo {author} {\bibfnamefont
  {A.}~\bibnamefont {McLaughlin}}, \bibinfo {author} {\bibfnamefont {M.~A.}\
  \bibnamefont {Moraes}}, \bibinfo {author} {\bibfnamefont {M.~H.}\
  \bibnamefont {Nasr}}, \bibinfo {author} {\bibfnamefont {L.~J.}\ \bibnamefont
  {Nociolo}}, \bibinfo {author} {\bibfnamefont {L.}~\bibnamefont {O'Donnell}},
  \bibinfo {author} {\bibfnamefont {A.}~\bibnamefont {Parker}}, \bibinfo
  {author} {\bibfnamefont {J.~L.}\ \bibnamefont {Peticolas}}, \bibinfo {author}
  {\bibfnamefont {G.}~\bibnamefont {Pocina}}, \bibinfo {author} {\bibfnamefont
  {C.}~\bibnamefont {Predescu}}, \bibinfo {author} {\bibfnamefont
  {T.}~\bibnamefont {Quan}}, \bibinfo {author} {\bibfnamefont {J.~K.}\
  \bibnamefont {Salmon}}, \bibinfo {author} {\bibfnamefont {C.}~\bibnamefont
  {Schwink}}, \bibinfo {author} {\bibfnamefont {K.~S.}\ \bibnamefont {Shim}},
  \bibinfo {author} {\bibfnamefont {N.}~\bibnamefont {Siddique}}, \bibinfo
  {author} {\bibfnamefont {J.}~\bibnamefont {Spengler}}, \bibinfo {author}
  {\bibfnamefont {T.}~\bibnamefont {Szalay}}, \bibinfo {author} {\bibfnamefont
  {R.}~\bibnamefont {Tabladillo}}, \bibinfo {author} {\bibfnamefont
  {R.}~\bibnamefont {Tartler}}, \bibinfo {author} {\bibfnamefont {A.~G.}\
  \bibnamefont {Taube}}, \bibinfo {author} {\bibfnamefont {M.}~\bibnamefont
  {Theobald}}, \bibinfo {author} {\bibfnamefont {B.}~\bibnamefont {Towles}},
  \bibinfo {author} {\bibfnamefont {W.}~\bibnamefont {Vick}}, \bibinfo {author}
  {\bibfnamefont {S.~C.}\ \bibnamefont {Wang}}, \bibinfo {author}
  {\bibfnamefont {M.}~\bibnamefont {Wazlowski}}, \bibinfo {author}
  {\bibfnamefont {M.~J.}\ \bibnamefont {Weingarten}}, \bibinfo {author}
  {\bibfnamefont {J.~M.}\ \bibnamefont {Williams}}, \ and\ \bibinfo {author}
  {\bibfnamefont {K.~A.}\ \bibnamefont {Yuh}},\ }in\ \href {\doibase
  10.1145/3458817.3487397} {\emph {\bibinfo {booktitle} {Proceedings of the
  International Conference for High Performance Computing, Networking, Storage
  and Analysis}}},\ \bibinfo {series and number} {SC '21}\ (\bibinfo
  {publisher} {Association for Computing Machinery},\ \bibinfo {address} {New
  York, NY, USA},\ \bibinfo {year} {2021})\BibitemShut {NoStop}%
\bibitem [{\citenamefont {Goldberg}\ \emph {et~al.}(1990)\citenamefont
  {Goldberg}, \citenamefont {Semisotnov}, \citenamefont {Friguet},
  \citenamefont {Kuwajima}, \citenamefont {Ptitsyn},\ and\ \citenamefont
  {Sugai}}]{slowfold}%
  \BibitemOpen
  \bibfield  {author} {\bibinfo {author} {\bibfnamefont {M.~E.}\ \bibnamefont
  {Goldberg}}, \bibinfo {author} {\bibfnamefont {G.~V.}\ \bibnamefont
  {Semisotnov}}, \bibinfo {author} {\bibfnamefont {B.}~\bibnamefont {Friguet}},
  \bibinfo {author} {\bibfnamefont {K.}~\bibnamefont {Kuwajima}}, \bibinfo
  {author} {\bibfnamefont {O.~B.}\ \bibnamefont {Ptitsyn}}, \ and\ \bibinfo
  {author} {\bibfnamefont {S.}~\bibnamefont {Sugai}},\ }\href {\doibase
  https://doi.org/10.1016/0014-5793(90)80703-L} {\bibfield  {journal} {\bibinfo
   {journal} {FEBS Lett.}\ }\textbf {\bibinfo {volume} {263}},\ \bibinfo
  {pages} {51} (\bibinfo {year} {1990})}\BibitemShut {NoStop}%
\bibitem [{\citenamefont {Peters}(2017)}]{BaronBook}%
  \BibitemOpen
  \bibfield  {author} {\bibinfo {author} {\bibfnamefont {B.}~\bibnamefont
  {Peters}},\ }\href@noop {} {\emph {\bibinfo {title} {Reaction rate theory and
  rare events}}}\ (\bibinfo  {publisher} {Elsevier},\ \bibinfo {address}
  {Amsterdam, Netherlands},\ \bibinfo {year} {2017})\BibitemShut {NoStop}%
\bibitem [{\citenamefont {van Erp}\ \emph {et~al.}(2003)\citenamefont {van
  Erp}, \citenamefont {Moroni},\ and\ \citenamefont {Bolhuis}}]{TIS}%
  \BibitemOpen
  \bibfield  {author} {\bibinfo {author} {\bibfnamefont {T.~S.}\ \bibnamefont
  {van Erp}}, \bibinfo {author} {\bibfnamefont {D.}~\bibnamefont {Moroni}}, \
  and\ \bibinfo {author} {\bibfnamefont {P.~G.}\ \bibnamefont {Bolhuis}},\
  }\href@noop {} {\bibfield  {journal} {\bibinfo  {journal} {J. Chem. Phys.}\
  }\textbf {\bibinfo {volume} {118}},\ \bibinfo {pages} {7762} (\bibinfo {year}
  {2003})}\BibitemShut {NoStop}%
\bibitem [{\citenamefont {{van Erp}}(2007)}]{RETIS}%
  \BibitemOpen
  \bibfield  {author} {\bibinfo {author} {\bibfnamefont {T.}~\bibnamefont {{van
  Erp}}},\ }\href@noop {} {\bibfield  {journal} {\bibinfo  {journal} {Phys.
  Rev. Lett.}\ }\textbf {\bibinfo {volume} {98}},\ \bibinfo {pages} {268301}
  (\bibinfo {year} {2007})}\BibitemShut {NoStop}%
\bibitem [{\citenamefont {Dellago}\ \emph
  {et~al.}(1998{\natexlab{a}})\citenamefont {Dellago}, \citenamefont {Bolhuis},
  \citenamefont {Csajka},\ and\ \citenamefont {Chandler}}]{TPS98}%
  \BibitemOpen
  \bibfield  {author} {\bibinfo {author} {\bibfnamefont {C.}~\bibnamefont
  {Dellago}}, \bibinfo {author} {\bibfnamefont {P.~G.}\ \bibnamefont
  {Bolhuis}}, \bibinfo {author} {\bibfnamefont {F.~S.}\ \bibnamefont {Csajka}},
  \ and\ \bibinfo {author} {\bibfnamefont {D.}~\bibnamefont {Chandler}},\
  }\href@noop {} {\bibfield  {journal} {\bibinfo  {journal} {J. Chem. Phys.}\
  }\textbf {\bibinfo {volume} {108}},\ \bibinfo {pages} {1964} (\bibinfo {year}
  {1998}{\natexlab{a}})}\BibitemShut {NoStop}%
\bibitem [{\citenamefont {Cabriolu}\ \emph {et~al.}(2017)\citenamefont
  {Cabriolu}, \citenamefont {Refsnes}, \citenamefont {Bolhuis},\ and\
  \citenamefont {van Erp}}]{Raffa}%
  \BibitemOpen
  \bibfield  {author} {\bibinfo {author} {\bibfnamefont {R.}~\bibnamefont
  {Cabriolu}}, \bibinfo {author} {\bibfnamefont {K.~M.~S.}\ \bibnamefont
  {Refsnes}}, \bibinfo {author} {\bibfnamefont {P.~G.}\ \bibnamefont
  {Bolhuis}}, \ and\ \bibinfo {author} {\bibfnamefont {T.~S.}\ \bibnamefont
  {van Erp}},\ }\href {\doibase {10.1063/1.4989844}} {\bibfield  {journal}
  {\bibinfo  {journal} {J. Chem. Phys.}\ }\textbf {\bibinfo {volume} {{147}}},\
  \bibinfo {pages} {152722} (\bibinfo {year} {{2017}})}\BibitemShut {NoStop}%
\bibitem [{\citenamefont {Arjun}\ and\ \citenamefont
  {Bolhuis}(2020)}]{methanehydrate}%
  \BibitemOpen
  \bibfield  {author} {\bibinfo {author} {\bibfnamefont {A.}~\bibnamefont
  {Arjun}}\ and\ \bibinfo {author} {\bibfnamefont {P.~G.}\ \bibnamefont
  {Bolhuis}},\ }\href@noop {} {\bibfield  {journal} {\bibinfo  {journal} {J.
  Phys. Chem. B}\ }\textbf {\bibinfo {volume} {124}},\ \bibinfo {pages} {8099}
  (\bibinfo {year} {2020})}\BibitemShut {NoStop}%
\bibitem [{\citenamefont {Moqadam}\ \emph {et~al.}(2018)\citenamefont
  {Moqadam}, \citenamefont {Lervik}, \citenamefont {Riccardi}, \citenamefont
  {Venkatraman}, \citenamefont {Alsberg},\ and\ \citenamefont {van
  Erp}}]{PNASwater}%
  \BibitemOpen
  \bibfield  {author} {\bibinfo {author} {\bibfnamefont {M.}~\bibnamefont
  {Moqadam}}, \bibinfo {author} {\bibfnamefont {A.}~\bibnamefont {Lervik}},
  \bibinfo {author} {\bibfnamefont {E.}~\bibnamefont {Riccardi}}, \bibinfo
  {author} {\bibfnamefont {V.}~\bibnamefont {Venkatraman}}, \bibinfo {author}
  {\bibfnamefont {B.~K.}\ \bibnamefont {Alsberg}}, \ and\ \bibinfo {author}
  {\bibfnamefont {T.~S.}\ \bibnamefont {van Erp}},\ }\href@noop {} {\bibfield
  {journal} {\bibinfo  {journal} {Proc. Natl. Acad. Sci. USA}\ }\textbf
  {\bibinfo {volume} {{115}}},\ \bibinfo {pages} {E4569} (\bibinfo {year}
  {2018})}\BibitemShut {NoStop}%
\bibitem [{\citenamefont {Eigen}\ and\ \citenamefont
  {de~Maeyer}(1958)}]{eigen1958}%
  \BibitemOpen
  \bibfield  {author} {\bibinfo {author} {\bibfnamefont {M.}~\bibnamefont
  {Eigen}}\ and\ \bibinfo {author} {\bibfnamefont {L.}~\bibnamefont
  {de~Maeyer}},\ }\href {\doibase 10.1098/rspa.1958.0208} {\bibfield  {journal}
  {\bibinfo  {journal} {Proc. R. Soc. Lond. A Math. Phys. Sci.}\ }\textbf
  {\bibinfo {volume} {247}},\ \bibinfo {pages} {505} (\bibinfo {year}
  {1958})}\BibitemShut {NoStop}%
\bibitem [{\citenamefont {Natzle}\ and\ \citenamefont
  {Moore}(1985)}]{natzle1985}%
  \BibitemOpen
  \bibfield  {author} {\bibinfo {author} {\bibfnamefont {W.~C.}\ \bibnamefont
  {Natzle}}\ and\ \bibinfo {author} {\bibfnamefont {C.~B.}\ \bibnamefont
  {Moore}},\ }\href {\doibase 10.1021/j100258a035} {\bibfield  {journal}
  {\bibinfo  {journal} {J. Phys. Chem.}\ }\textbf {\bibinfo {volume} {89}},\
  \bibinfo {pages} {2605} (\bibinfo {year} {1985})}\BibitemShut {NoStop}%
\bibitem [{\citenamefont {Moroni}\ \emph {et~al.}(2003)\citenamefont {Moroni},
  \citenamefont {Bolhuis},\ and\ \citenamefont {{van Erp}}}]{PPTIS}%
  \BibitemOpen
  \bibfield  {author} {\bibinfo {author} {\bibfnamefont {D.}~\bibnamefont
  {Moroni}}, \bibinfo {author} {\bibfnamefont {P.}~\bibnamefont {Bolhuis}}, \
  and\ \bibinfo {author} {\bibfnamefont {T.}~\bibnamefont {{van Erp}}},\
  }\href@noop {} {\bibfield  {journal} {\bibinfo  {journal} {J. Chem. Phys.}\
  }\textbf {\bibinfo {volume} {120}},\ \bibinfo {pages} {1044} (\bibinfo {year}
  {2003})}\BibitemShut {NoStop}%
\bibitem [{\citenamefont {Faradjian}\ and\ \citenamefont
  {Elber}(2004)}]{Milestoning}%
  \BibitemOpen
  \bibfield  {author} {\bibinfo {author} {\bibfnamefont {A.~K.}\ \bibnamefont
  {Faradjian}}\ and\ \bibinfo {author} {\bibfnamefont {R.}~\bibnamefont
  {Elber}},\ }\href@noop {} {\bibfield  {journal} {\bibinfo  {journal} {{J.
  Chem. Phys.}}\ }\textbf {\bibinfo {volume} {{120}}},\ \bibinfo {pages}
  {10880} (\bibinfo {year} {{2004}})}\BibitemShut {NoStop}%
\bibitem [{\citenamefont {Roet}\ \emph {et~al.}(2022)\citenamefont {Roet},
  \citenamefont {Zhang},\ and\ \citenamefont {van Erp}}]{infRETIS}%
  \BibitemOpen
  \bibfield  {author} {\bibinfo {author} {\bibfnamefont {S.}~\bibnamefont
  {Roet}}, \bibinfo {author} {\bibfnamefont {D.~T.}\ \bibnamefont {Zhang}}, \
  and\ \bibinfo {author} {\bibfnamefont {T.~S.}\ \bibnamefont {van Erp}},\
  }\href {\doibase 10.48550/ARXIV.2205.12663} {\enquote {\bibinfo {title}
  {Exchanging replicas with unequal cost, infinitely and permanently.
  {P}reprint at https://arxiv.org/abs/2205.12663},}\ } (\bibinfo {year}
  {2022})\BibitemShut {NoStop}%
\bibitem [{\citenamefont {Dellago}\ \emph
  {et~al.}(1998{\natexlab{b}})\citenamefont {Dellago}, \citenamefont
  {Bolhuis},\ and\ \citenamefont {Chandler}}]{shoot}%
  \BibitemOpen
  \bibfield  {author} {\bibinfo {author} {\bibfnamefont {C.}~\bibnamefont
  {Dellago}}, \bibinfo {author} {\bibfnamefont {P.~G.}\ \bibnamefont
  {Bolhuis}}, \ and\ \bibinfo {author} {\bibfnamefont {D.}~\bibnamefont
  {Chandler}},\ }\href {\doibase 10.1063/1.476378} {\bibfield  {journal}
  {\bibinfo  {journal} {The Journal of Chemical Physics}\ }\textbf {\bibinfo
  {volume} {108}},\ \bibinfo {pages} {9236} (\bibinfo {year}
  {1998}{\natexlab{b}})},\ \Eprint
  {http://arxiv.org/abs/https://doi.org/10.1063/1.476378}
  {https://doi.org/10.1063/1.476378} \BibitemShut {NoStop}%
\bibitem [{\citenamefont {Riccardi}\ \emph {et~al.}(2017)\citenamefont
  {Riccardi}, \citenamefont {Dahlen},\ and\ \citenamefont {van
  Erp}}]{riccardi_fast_2017}%
  \BibitemOpen
  \bibfield  {author} {\bibinfo {author} {\bibfnamefont {E.}~\bibnamefont
  {Riccardi}}, \bibinfo {author} {\bibfnamefont {O.}~\bibnamefont {Dahlen}}, \
  and\ \bibinfo {author} {\bibfnamefont {T.~S.}\ \bibnamefont {van Erp}},\
  }\href@noop {} {\bibfield  {journal} {\bibinfo  {journal} {J. Phys. Chem.
  Lett.}\ }\textbf {\bibinfo {volume} {8}},\ \bibinfo {pages} {4456} (\bibinfo
  {year} {2017})}\BibitemShut {NoStop}%
\bibitem [{\citenamefont {Swenson}\ \emph
  {et~al.}(2019{\natexlab{a}})\citenamefont {Swenson}, \citenamefont {Prinz},
  \citenamefont {Noe}, \citenamefont {Chodera},\ and\ \citenamefont
  {Bolhuis}}]{OPS1}%
  \BibitemOpen
  \bibfield  {author} {\bibinfo {author} {\bibfnamefont {D.~W.~H.}\
  \bibnamefont {Swenson}}, \bibinfo {author} {\bibfnamefont {J.-H.}\
  \bibnamefont {Prinz}}, \bibinfo {author} {\bibfnamefont {F.}~\bibnamefont
  {Noe}}, \bibinfo {author} {\bibfnamefont {J.~D.}\ \bibnamefont {Chodera}}, \
  and\ \bibinfo {author} {\bibfnamefont {P.~G.}\ \bibnamefont {Bolhuis}},\
  }\href {\doibase 10.1021/acs.jctc.8b00626} {\bibfield  {journal} {\bibinfo
  {journal} {J. Chem. Theory Comput.}\ }\textbf {\bibinfo {volume} {15}},\
  \bibinfo {pages} {813} (\bibinfo {year} {2019}{\natexlab{a}})}\BibitemShut
  {NoStop}%
\bibitem [{\citenamefont {Swenson}\ \emph
  {et~al.}(2019{\natexlab{b}})\citenamefont {Swenson}, \citenamefont {Prinz},
  \citenamefont {Noe}, \citenamefont {Chodera},\ and\ \citenamefont
  {Bolhuis}}]{OPS2}%
  \BibitemOpen
  \bibfield  {author} {\bibinfo {author} {\bibfnamefont {D.~W.~H.}\
  \bibnamefont {Swenson}}, \bibinfo {author} {\bibfnamefont {J.-H.}\
  \bibnamefont {Prinz}}, \bibinfo {author} {\bibfnamefont {F.}~\bibnamefont
  {Noe}}, \bibinfo {author} {\bibfnamefont {J.~D.}\ \bibnamefont {Chodera}}, \
  and\ \bibinfo {author} {\bibfnamefont {P.~G.}\ \bibnamefont {Bolhuis}},\
  }\href@noop {} {\bibfield  {journal} {\bibinfo  {journal} {J. Chem. Theory
  Comput.}\ }\textbf {\bibinfo {volume} {15}},\ \bibinfo {pages} {837}
  (\bibinfo {year} {2019}{\natexlab{b}})}\BibitemShut {NoStop}%
\bibitem [{\citenamefont {Lervik}\ \emph {et~al.}(2017)\citenamefont {Lervik},
  \citenamefont {Riccardi},\ and\ \citenamefont {van Erp}}]{PyRETIS1}%
  \BibitemOpen
  \bibfield  {author} {\bibinfo {author} {\bibfnamefont {A.}~\bibnamefont
  {Lervik}}, \bibinfo {author} {\bibfnamefont {E.}~\bibnamefont {Riccardi}}, \
  and\ \bibinfo {author} {\bibfnamefont {T.~S.}\ \bibnamefont {van Erp}},\
  }\href@noop {} {\bibfield  {journal} {\bibinfo  {journal} {J. Comput. Chem.}\
  }\textbf {\bibinfo {volume} {38}},\ \bibinfo {pages} {2439} (\bibinfo {year}
  {2017})}\BibitemShut {NoStop}%
\bibitem [{\citenamefont {Riccardi}\ \emph {et~al.}(2020)\citenamefont
  {Riccardi}, \citenamefont {Lervik}, \citenamefont {Roet}, \citenamefont
  {Aaroen},\ and\ \citenamefont {van Erp}}]{PyRETIS2}%
  \BibitemOpen
  \bibfield  {author} {\bibinfo {author} {\bibfnamefont {E.}~\bibnamefont
  {Riccardi}}, \bibinfo {author} {\bibfnamefont {A.}~\bibnamefont {Lervik}},
  \bibinfo {author} {\bibfnamefont {S.}~\bibnamefont {Roet}}, \bibinfo {author}
  {\bibfnamefont {O.}~\bibnamefont {Aaroen}}, \ and\ \bibinfo {author}
  {\bibfnamefont {T.~S.}\ \bibnamefont {van Erp}},\ }\href@noop {} {\bibfield
  {journal} {\bibinfo  {journal} {J. Comput. Chem.}\ }\textbf {\bibinfo
  {volume} {41}},\ \bibinfo {pages} {370} (\bibinfo {year} {2020})}\BibitemShut
  {NoStop}%
\bibitem [{\citenamefont {Abraham}\ \emph
  {et~al.}(2015{\natexlab{a}})\citenamefont {Abraham}, \citenamefont {Murtola},
  \citenamefont {Schulz}, \citenamefont {Páll}, \citenamefont {Smith},
  \citenamefont {Hess},\ and\ \citenamefont {Lindahl}}]{GROMACS}%
  \BibitemOpen
  \bibfield  {author} {\bibinfo {author} {\bibfnamefont {M.~J.}\ \bibnamefont
  {Abraham}}, \bibinfo {author} {\bibfnamefont {T.}~\bibnamefont {Murtola}},
  \bibinfo {author} {\bibfnamefont {R.}~\bibnamefont {Schulz}}, \bibinfo
  {author} {\bibfnamefont {S.}~\bibnamefont {Páll}}, \bibinfo {author}
  {\bibfnamefont {J.~C.}\ \bibnamefont {Smith}}, \bibinfo {author}
  {\bibfnamefont {B.}~\bibnamefont {Hess}}, \ and\ \bibinfo {author}
  {\bibfnamefont {E.}~\bibnamefont {Lindahl}},\ }\href@noop {} {\bibfield
  {journal} {\bibinfo  {journal} {SoftwareX}\ }\textbf {\bibinfo {volume}
  {1-2}},\ \bibinfo {pages} {19} (\bibinfo {year}
  {2015}{\natexlab{a}})}\BibitemShut {NoStop}%
\bibitem [{\citenamefont {Plimpton}(1995)}]{lammps1995}%
  \BibitemOpen
  \bibfield  {author} {\bibinfo {author} {\bibfnamefont {S.}~\bibnamefont
  {Plimpton}},\ }\href {\doibase 10.1006/jcph.1995.1039} {\bibfield  {journal}
  {\bibinfo  {journal} {J. Comput. Phys.}\ }\textbf {\bibinfo {volume} {117}},\
  \bibinfo {pages} {1} (\bibinfo {year} {1995})}\BibitemShut {NoStop}%
\bibitem [{\citenamefont {Eastman}\ \emph {et~al.}(2013)\citenamefont
  {Eastman}, \citenamefont {Friedrichs}, \citenamefont {Chodera}, \citenamefont
  {Radmer}, \citenamefont {Bruns}, \citenamefont {Ku}, \citenamefont
  {Beauchamp}, \citenamefont {Lane}, \citenamefont {Wang}, \citenamefont
  {Shukla}, \citenamefont {Tye}, \citenamefont {Houston}, \citenamefont
  {Stich}, \citenamefont {Klein}, \citenamefont {Shirts},\ and\ \citenamefont
  {Pande}}]{openmm}%
  \BibitemOpen
  \bibfield  {author} {\bibinfo {author} {\bibfnamefont {P.}~\bibnamefont
  {Eastman}}, \bibinfo {author} {\bibfnamefont {M.~S.}\ \bibnamefont
  {Friedrichs}}, \bibinfo {author} {\bibfnamefont {J.~D.}\ \bibnamefont
  {Chodera}}, \bibinfo {author} {\bibfnamefont {R.~J.}\ \bibnamefont {Radmer}},
  \bibinfo {author} {\bibfnamefont {C.~M.}\ \bibnamefont {Bruns}}, \bibinfo
  {author} {\bibfnamefont {J.~P.}\ \bibnamefont {Ku}}, \bibinfo {author}
  {\bibfnamefont {K.~A.}\ \bibnamefont {Beauchamp}}, \bibinfo {author}
  {\bibfnamefont {T.~J.}\ \bibnamefont {Lane}}, \bibinfo {author}
  {\bibfnamefont {L.-P.}\ \bibnamefont {Wang}}, \bibinfo {author}
  {\bibfnamefont {D.}~\bibnamefont {Shukla}}, \bibinfo {author} {\bibfnamefont
  {T.}~\bibnamefont {Tye}}, \bibinfo {author} {\bibfnamefont {M.}~\bibnamefont
  {Houston}}, \bibinfo {author} {\bibfnamefont {T.}~\bibnamefont {Stich}},
  \bibinfo {author} {\bibfnamefont {C.}~\bibnamefont {Klein}}, \bibinfo
  {author} {\bibfnamefont {M.~R.}\ \bibnamefont {Shirts}}, \ and\ \bibinfo
  {author} {\bibfnamefont {V.~S.}\ \bibnamefont {Pande}},\ }\href@noop {}
  {\bibfield  {journal} {\bibinfo  {journal} {J. Chem. Theory Comput.}\
  }\textbf {\bibinfo {volume} {9}},\ \bibinfo {pages} {461} (\bibinfo {year}
  {2013})}\BibitemShut {NoStop}%
\bibitem [{\citenamefont {Hutter}\ \emph {et~al.}(2014)\citenamefont {Hutter},
  \citenamefont {Iannuzzi}, \citenamefont {Schiffmann},\ and\ \citenamefont
  {VandeVondele}}]{CP2K}%
  \BibitemOpen
  \bibfield  {author} {\bibinfo {author} {\bibfnamefont {J.}~\bibnamefont
  {Hutter}}, \bibinfo {author} {\bibfnamefont {M.}~\bibnamefont {Iannuzzi}},
  \bibinfo {author} {\bibfnamefont {F.}~\bibnamefont {Schiffmann}}, \ and\
  \bibinfo {author} {\bibfnamefont {J.}~\bibnamefont {VandeVondele}},\ }\href
  {\doibase 10.1002/wcms.1159} {\bibfield  {journal} {\bibinfo  {journal}
  {WileyWIREs Comput Mol Sci}\ }\textbf {\bibinfo {volume} {4}},\ \bibinfo
  {pages} {15} (\bibinfo {year} {2014})}\BibitemShut {NoStop}%
\bibitem [{\citenamefont {Stukowski}(2012)}]{nucleation1}%
  \BibitemOpen
  \bibfield  {author} {\bibinfo {author} {\bibfnamefont {A.}~\bibnamefont
  {Stukowski}},\ }\href@noop {} {\bibfield  {journal} {\bibinfo  {journal}
  {Model. Simul. Mater. Sci. Eng.}\ }\textbf {\bibinfo {volume} {20}},\
  \bibinfo {pages} {045021} (\bibinfo {year} {2012})}\BibitemShut {NoStop}%
\bibitem [{\citenamefont {Winczewski}\ \emph {et~al.}(2016)\citenamefont
  {Winczewski}, \citenamefont {Dziedzic},\ and\ \citenamefont
  {Rybicki}}]{CPUnucleation2}%
  \BibitemOpen
  \bibfield  {author} {\bibinfo {author} {\bibfnamefont {S.}~\bibnamefont
  {Winczewski}}, \bibinfo {author} {\bibfnamefont {J.}~\bibnamefont
  {Dziedzic}}, \ and\ \bibinfo {author} {\bibfnamefont {J.}~\bibnamefont
  {Rybicki}},\ }\href@noop {} {\bibfield  {journal} {\bibinfo  {journal}
  {Comput. Phys. Commun.}\ }\textbf {\bibinfo {volume} {198}},\ \bibinfo
  {pages} {128} (\bibinfo {year} {2016})}\BibitemShut {NoStop}%
\bibitem [{\citenamefont {Metropolis}\ \emph {et~al.}(1953)\citenamefont
  {Metropolis}, \citenamefont {Rosenbluth}, \citenamefont {Rosenbluth},
  \citenamefont {Teller},\ and\ \citenamefont {Teller}}]{Metroplois}%
  \BibitemOpen
  \bibfield  {author} {\bibinfo {author} {\bibfnamefont {N.}~\bibnamefont
  {Metropolis}}, \bibinfo {author} {\bibfnamefont {A.}~\bibnamefont
  {Rosenbluth}}, \bibinfo {author} {\bibfnamefont {M.}~\bibnamefont
  {Rosenbluth}}, \bibinfo {author} {\bibfnamefont {A.}~\bibnamefont {Teller}},
  \ and\ \bibinfo {author} {\bibfnamefont {E.}~\bibnamefont {Teller}},\
  }\href@noop {} {\bibfield  {journal} {\bibinfo  {journal} {J. Chem. Phys.}\
  }\textbf {\bibinfo {volume} {21}},\ \bibinfo {pages} {1087} (\bibinfo {year}
  {1953})}\BibitemShut {NoStop}%
\bibitem [{\citenamefont {Frenkel}\ and\ \citenamefont
  {Smit}(2002)}]{FrenkelBook}%
  \BibitemOpen
  \bibfield  {author} {\bibinfo {author} {\bibfnamefont {D.}~\bibnamefont
  {Frenkel}}\ and\ \bibinfo {author} {\bibfnamefont {B.}~\bibnamefont {Smit}},\
  }\href@noop {} {\emph {\bibinfo {title} {Understanding molecular simulations
  from algorithms to applications}}}\ (\bibinfo  {publisher} {Academic press},\
  \bibinfo {address} {San Diego, California, U.S.A.},\ \bibinfo {year}
  {2002})\BibitemShut {NoStop}%
\bibitem [{\citenamefont {Siepmann}\ and\ \citenamefont
  {Frenkel}(1992)}]{Siepmann92}%
  \BibitemOpen
  \bibfield  {author} {\bibinfo {author} {\bibfnamefont {J.~I.}\ \bibnamefont
  {Siepmann}}\ and\ \bibinfo {author} {\bibfnamefont {D.}~\bibnamefont
  {Frenkel}},\ }\href@noop {} {\bibfield  {journal} {\bibinfo  {journal} {Mol.
  Phys.}\ }\textbf {\bibinfo {volume} {75}},\ \bibinfo {pages} {59} (\bibinfo
  {year} {1992})}\BibitemShut {NoStop}%
\bibitem [{\citenamefont {Vlugt}\ \emph {et~al.}(1999)\citenamefont {Vlugt},
  \citenamefont {Krishna},\ and\ \citenamefont {Smit}}]{Vlugt99}%
  \BibitemOpen
  \bibfield  {author} {\bibinfo {author} {\bibfnamefont {T.}~\bibnamefont
  {Vlugt}}, \bibinfo {author} {\bibfnamefont {R.}~\bibnamefont {Krishna}}, \
  and\ \bibinfo {author} {\bibfnamefont {B.}~\bibnamefont {Smit}},\ }\href@noop
  {} {\bibfield  {journal} {\bibinfo  {journal} {J. Phys. Chem. B}\ }\textbf
  {\bibinfo {volume} {103}},\ \bibinfo {pages} {1102} (\bibinfo {year}
  {1999})}\BibitemShut {NoStop}%
\bibitem [{\citenamefont {Hastings}(1970)}]{Hastings}%
  \BibitemOpen
  \bibfield  {author} {\bibinfo {author} {\bibfnamefont {W.}~\bibnamefont
  {Hastings}},\ }\href {\doibase 10.2307/2334940} {\bibfield  {journal}
  {\bibinfo  {journal} {Biometrika}\ }\textbf {\bibinfo {volume} {57}},\
  \bibinfo {pages} {97} (\bibinfo {year} {1970})}\BibitemShut {NoStop}%
\bibitem [{\citenamefont {{van Erp}}(2012)}]{TitusRev}%
  \BibitemOpen
  \bibfield  {author} {\bibinfo {author} {\bibfnamefont {T.}~\bibnamefont {{van
  Erp}}},\ }\href@noop {} {\bibfield  {journal} {\bibinfo  {journal} {Adv.
  Chem. Phys.}\ }\textbf {\bibinfo {volume} {151}},\ \bibinfo {pages} {27}
  (\bibinfo {year} {2012})}\BibitemShut {NoStop}%
\bibitem [{\citenamefont {Ghysels}\ \emph {et~al.}(2021)\citenamefont
  {Ghysels}, \citenamefont {Roet}, \citenamefont {Davoudi},\ and\ \citenamefont
  {van Erp}}]{permeability}%
  \BibitemOpen
  \bibfield  {author} {\bibinfo {author} {\bibfnamefont {A.}~\bibnamefont
  {Ghysels}}, \bibinfo {author} {\bibfnamefont {S.}~\bibnamefont {Roet}},
  \bibinfo {author} {\bibfnamefont {S.}~\bibnamefont {Davoudi}}, \ and\
  \bibinfo {author} {\bibfnamefont {T.~S.}\ \bibnamefont {van Erp}},\ }\href
  {\doibase 10.1103/PhysRevResearch.3.033068} {\bibfield  {journal} {\bibinfo
  {journal} {Phys. Rev. Research}\ }\textbf {\bibinfo {volume} {3}},\ \bibinfo
  {pages} {033068} (\bibinfo {year} {2021})}\BibitemShut {NoStop}%
\bibitem [{\citenamefont {Dellago}\ and\ \citenamefont {Bolhuis}(2004)}]{actE}%
  \BibitemOpen
  \bibfield  {author} {\bibinfo {author} {\bibfnamefont {C.}~\bibnamefont
  {Dellago}}\ and\ \bibinfo {author} {\bibfnamefont {P.~G.}\ \bibnamefont
  {Bolhuis}},\ }\href@noop {} {\bibfield  {journal} {\bibinfo  {journal} {Mol.
  Simu.}\ }\textbf {\bibinfo {volume} {30}},\ \bibinfo {pages} {795} (\bibinfo
  {year} {2004})}\BibitemShut {NoStop}%
\bibitem [{\citenamefont {van Erp}\ and\ \citenamefont {Bolhuis}(2005)}]{Elab}%
  \BibitemOpen
  \bibfield  {author} {\bibinfo {author} {\bibfnamefont {T.}~\bibnamefont {van
  Erp}}\ and\ \bibinfo {author} {\bibfnamefont {P.}~\bibnamefont {Bolhuis}},\
  }\href@noop {} {\bibfield  {journal} {\bibinfo  {journal} {J. Comput. Phys.}\
  }\textbf {\bibinfo {volume} {205}},\ \bibinfo {pages} {157} (\bibinfo {year}
  {2005})}\BibitemShut {NoStop}%
\bibitem [{\citenamefont {van Erp}(2006)}]{TISeff}%
  \BibitemOpen
  \bibfield  {author} {\bibinfo {author} {\bibfnamefont {T.~S.}\ \bibnamefont
  {van Erp}},\ }\href@noop {} {\bibfield  {journal} {\bibinfo  {journal} {J.
  Chem. Phys.}\ }\textbf {\bibinfo {volume} {125}},\ \bibinfo {pages} {174106}
  (\bibinfo {year} {2006})}\BibitemShut {NoStop}%
\bibitem [{\citenamefont {Ferrenberg}\ and\ \citenamefont
  {Swendsen}(1989)}]{WHAM}%
  \BibitemOpen
  \bibfield  {author} {\bibinfo {author} {\bibfnamefont {A.}~\bibnamefont
  {Ferrenberg}}\ and\ \bibinfo {author} {\bibfnamefont {R.}~\bibnamefont
  {Swendsen}},\ }\href@noop {} {\bibfield  {journal} {\bibinfo  {journal}
  {{Phys. Rev. Lett.}}\ }\textbf {\bibinfo {volume} {{63}}},\ \bibinfo {pages}
  {1195} (\bibinfo {year} {{1989}})}\BibitemShut {NoStop}%
\bibitem [{\citenamefont {van Erp}\ \emph {et~al.}(2016)\citenamefont {van
  Erp}, \citenamefont {Moqadam}, \citenamefont {Riccardi},\ and\ \citenamefont
  {Lervik}}]{predictive}%
  \BibitemOpen
  \bibfield  {author} {\bibinfo {author} {\bibfnamefont {T.~S.}\ \bibnamefont
  {van Erp}}, \bibinfo {author} {\bibfnamefont {M.}~\bibnamefont {Moqadam}},
  \bibinfo {author} {\bibfnamefont {E.}~\bibnamefont {Riccardi}}, \ and\
  \bibinfo {author} {\bibfnamefont {A.}~\bibnamefont {Lervik}},\ }\href@noop {}
  {\bibfield  {journal} {\bibinfo  {journal} {J. Chem. Theory Comput.}\
  }\textbf {\bibinfo {volume} {{12}}},\ \bibinfo {pages} {5398} (\bibinfo
  {year} {{2016}})}\BibitemShut {NoStop}%
\bibitem [{\citenamefont {Rogal}\ \emph {et~al.}(2010)\citenamefont {Rogal},
  \citenamefont {Lechner}, \citenamefont {Juraszek}, \citenamefont {Ensing},\
  and\ \citenamefont {Bolhuis}}]{Rogal}%
  \BibitemOpen
  \bibfield  {author} {\bibinfo {author} {\bibfnamefont {J.}~\bibnamefont
  {Rogal}}, \bibinfo {author} {\bibfnamefont {W.}~\bibnamefont {Lechner}},
  \bibinfo {author} {\bibfnamefont {J.}~\bibnamefont {Juraszek}}, \bibinfo
  {author} {\bibfnamefont {B.}~\bibnamefont {Ensing}}, \ and\ \bibinfo {author}
  {\bibfnamefont {P.~G.}\ \bibnamefont {Bolhuis}},\ }\href@noop {} {\bibfield
  {journal} {\bibinfo  {journal} {J. Comp. Phys.}\ }\textbf {\bibinfo {volume}
  {{133}}},\ \bibinfo {pages} {174109} (\bibinfo {year} {{2010}})}\BibitemShut
  {NoStop}%
\bibitem [{\citenamefont {Vanden-Eijnden}\ \emph {et~al.}(2008)\citenamefont
  {Vanden-Eijnden}, \citenamefont {Venturoli}, \citenamefont {Ciccotti},\ and\
  \citenamefont {Elber}}]{isocom}%
  \BibitemOpen
  \bibfield  {author} {\bibinfo {author} {\bibfnamefont {E.}~\bibnamefont
  {Vanden-Eijnden}}, \bibinfo {author} {\bibfnamefont {M.}~\bibnamefont
  {Venturoli}}, \bibinfo {author} {\bibfnamefont {G.}~\bibnamefont {Ciccotti}},
  \ and\ \bibinfo {author} {\bibfnamefont {R.}~\bibnamefont {Elber}},\
  }\href@noop {} {\bibfield  {journal} {\bibinfo  {journal} {J. Comp. Phys.}\
  }\textbf {\bibinfo {volume} {129}},\ \bibinfo {pages} {174102} (\bibinfo
  {year} {2008})}\BibitemShut {NoStop}%
\bibitem [{\citenamefont {Haji-Akbari}(2018)}]{jumpy}%
  \BibitemOpen
  \bibfield  {author} {\bibinfo {author} {\bibfnamefont {A.}~\bibnamefont
  {Haji-Akbari}},\ }\href {\doibase 10.1063/1.5018303} {\bibfield  {journal}
  {\bibinfo  {journal} {J. Chem. Phys.}\ }\textbf {\bibinfo {volume} {149}},\
  \bibinfo {pages} {072303} (\bibinfo {year} {2018})}\BibitemShut {NoStop}%
\bibitem [{\citenamefont {Brotzakis}\ and\ \citenamefont
  {Bolhuis}(2016)}]{springshoot}%
  \BibitemOpen
  \bibfield  {author} {\bibinfo {author} {\bibfnamefont {Z.~F.}\ \bibnamefont
  {Brotzakis}}\ and\ \bibinfo {author} {\bibfnamefont {P.~G.}\ \bibnamefont
  {Bolhuis}},\ }\href@noop {} {\bibfield  {journal} {\bibinfo  {journal} {J.
  Chem. Phys.}\ }\textbf {\bibinfo {volume} {145}},\ \bibinfo {pages} {164112}
  (\bibinfo {year} {2016})}\BibitemShut {NoStop}%
\bibitem [{\citenamefont {Moqadam}\ \emph {et~al.}(2017)\citenamefont
  {Moqadam}, \citenamefont {Riccardi}, \citenamefont {Trinh}, \citenamefont
  {Lervik},\ and\ \citenamefont {van Erp}}]{Mahmoud_silic}%
  \BibitemOpen
  \bibfield  {author} {\bibinfo {author} {\bibfnamefont {M.}~\bibnamefont
  {Moqadam}}, \bibinfo {author} {\bibfnamefont {E.}~\bibnamefont {Riccardi}},
  \bibinfo {author} {\bibfnamefont {T.~T.}\ \bibnamefont {Trinh}}, \bibinfo
  {author} {\bibfnamefont {A.}~\bibnamefont {Lervik}}, \ and\ \bibinfo {author}
  {\bibfnamefont {T.~S.}\ \bibnamefont {van Erp}},\ }\href@noop {} {\bibfield
  {journal} {\bibinfo  {journal} {Phys. Chem. Chem. Phys.}\ }\textbf {\bibinfo
  {volume} {19}},\ \bibinfo {pages} {13361} (\bibinfo {year}
  {2017})}\BibitemShut {NoStop}%
\bibitem [{\citenamefont {Moroni}\ \emph {et~al.}(2005)\citenamefont {Moroni},
  \citenamefont {ten Wolde},\ and\ \citenamefont {Bolhuis}}]{MoroniPRL}%
  \BibitemOpen
  \bibfield  {author} {\bibinfo {author} {\bibfnamefont {D.}~\bibnamefont
  {Moroni}}, \bibinfo {author} {\bibfnamefont {P.~R.}\ \bibnamefont {ten
  Wolde}}, \ and\ \bibinfo {author} {\bibfnamefont {P.~G.}\ \bibnamefont
  {Bolhuis}},\ }\href@noop {} {\bibfield  {journal} {\bibinfo  {journal} {Phys.
  Rev. Lett.}\ }\textbf {\bibinfo {volume} {94}},\ \bibinfo {pages} {235703}
  (\bibinfo {year} {2005})}\BibitemShut {NoStop}%
\bibitem [{\citenamefont {Aarøen}\ \emph {et~al.}(2022)\citenamefont
  {Aarøen}, \citenamefont {Riccardi}, \citenamefont {Erp},\ and\ \citenamefont
  {Sletmoen}}]{aaroen_thin_2022}%
  \BibitemOpen
  \bibfield  {author} {\bibinfo {author} {\bibfnamefont {O.}~\bibnamefont
  {Aarøen}}, \bibinfo {author} {\bibfnamefont {E.}~\bibnamefont {Riccardi}},
  \bibinfo {author} {\bibfnamefont {T.~S.~v.}\ \bibnamefont {Erp}}, \ and\
  \bibinfo {author} {\bibfnamefont {M.}~\bibnamefont {Sletmoen}},\ }\href@noop
  {} {\bibfield  {journal} {\bibinfo  {journal} {Colloids and Surfaces A:
  Physicochemical and Engineering Aspects}\ }\textbf {\bibinfo {volume}
  {632}},\ \bibinfo {pages} {127808} (\bibinfo {year} {2022})}\BibitemShut
  {NoStop}%
\bibitem [{\citenamefont {Aar{\o}en}\ \emph {et~al.}(2021)\citenamefont
  {Aar{\o}en}, \citenamefont {Riccardi},\ and\ \citenamefont
  {Sletmoen}}]{aaroen2021exploring}%
  \BibitemOpen
  \bibfield  {author} {\bibinfo {author} {\bibfnamefont {O.}~\bibnamefont
  {Aar{\o}en}}, \bibinfo {author} {\bibfnamefont {E.}~\bibnamefont {Riccardi}},
  \ and\ \bibinfo {author} {\bibfnamefont {M.}~\bibnamefont {Sletmoen}},\
  }\href@noop {} {\bibfield  {journal} {\bibinfo  {journal} {RSC advances}\
  }\textbf {\bibinfo {volume} {11}},\ \bibinfo {pages} {8730} (\bibinfo {year}
  {2021})}\BibitemShut {NoStop}%
\bibitem [{\citenamefont {Riccardi}\ and\ \citenamefont
  {Tichelkamp}(2019)}]{riccardi2019calcium}%
  \BibitemOpen
  \bibfield  {author} {\bibinfo {author} {\bibfnamefont {E.}~\bibnamefont
  {Riccardi}}\ and\ \bibinfo {author} {\bibfnamefont {T.}~\bibnamefont
  {Tichelkamp}},\ }\href@noop {} {\bibfield  {journal} {\bibinfo  {journal}
  {Colloids and Surfaces A: Physicochemical and Engineering Aspects}\ }\textbf
  {\bibinfo {volume} {573}},\ \bibinfo {pages} {246} (\bibinfo {year}
  {2019})}\BibitemShut {NoStop}%
\bibitem [{\citenamefont {Riccardi}\ \emph {et~al.}(2014)\citenamefont
  {Riccardi}, \citenamefont {Kovalchuk}, \citenamefont {Mehandzhiyski},\ and\
  \citenamefont {Grimes}}]{riccardi2014structure}%
  \BibitemOpen
  \bibfield  {author} {\bibinfo {author} {\bibfnamefont {E.}~\bibnamefont
  {Riccardi}}, \bibinfo {author} {\bibfnamefont {K.}~\bibnamefont {Kovalchuk}},
  \bibinfo {author} {\bibfnamefont {A.~Y.}\ \bibnamefont {Mehandzhiyski}}, \
  and\ \bibinfo {author} {\bibfnamefont {B.~A.}\ \bibnamefont {Grimes}},\
  }\href@noop {} {\bibfield  {journal} {\bibinfo  {journal} {Journal of
  dispersion science and technology}\ }\textbf {\bibinfo {volume} {35}},\
  \bibinfo {pages} {1018} (\bibinfo {year} {2014})}\BibitemShut {NoStop}%
\bibitem [{\citenamefont {Marzari}\ \emph {et~al.}(2012)\citenamefont
  {Marzari}, \citenamefont {Mostofi}, \citenamefont {Yates}, \citenamefont
  {Souza},\ and\ \citenamefont {Vanderbilt}}]{Wannier}%
  \BibitemOpen
  \bibfield  {author} {\bibinfo {author} {\bibfnamefont {N.}~\bibnamefont
  {Marzari}}, \bibinfo {author} {\bibfnamefont {A.~A.}\ \bibnamefont
  {Mostofi}}, \bibinfo {author} {\bibfnamefont {J.~R.}\ \bibnamefont {Yates}},
  \bibinfo {author} {\bibfnamefont {I.}~\bibnamefont {Souza}}, \ and\ \bibinfo
  {author} {\bibfnamefont {D.}~\bibnamefont {Vanderbilt}},\ }\href {\doibase
  10.1103/RevModPhys.84.1419} {\bibfield  {journal} {\bibinfo  {journal} {Rev.
  Mod. Phys.}\ }\textbf {\bibinfo {volume} {84}},\ \bibinfo {pages} {1419}
  (\bibinfo {year} {2012})}\BibitemShut {NoStop}%
\bibitem [{\citenamefont {Allen}\ \emph {et~al.}(2009)\citenamefont {Allen},
  \citenamefont {Valeriani},\ and\ \citenamefont {ten Wolde}}]{FFSrev}%
  \BibitemOpen
  \bibfield  {author} {\bibinfo {author} {\bibfnamefont {R.~J.}\ \bibnamefont
  {Allen}}, \bibinfo {author} {\bibfnamefont {C.}~\bibnamefont {Valeriani}}, \
  and\ \bibinfo {author} {\bibfnamefont {P.~R.}\ \bibnamefont {ten Wolde}},\
  }\href@noop {} {\bibfield  {journal} {\bibinfo  {journal} {J. Phys.-Condes.
  Matter}\ }\textbf {\bibinfo {volume} {21}},\ \bibinfo {pages} {463102}
  (\bibinfo {year} {2009})}\BibitemShut {NoStop}%
\bibitem [{\citenamefont {Escobedo}\ \emph {et~al.}(2009)\citenamefont
  {Escobedo}, \citenamefont {Borrero},\ and\ \citenamefont {Araque}}]{FFSrev2}%
  \BibitemOpen
  \bibfield  {author} {\bibinfo {author} {\bibfnamefont {F.~A.}\ \bibnamefont
  {Escobedo}}, \bibinfo {author} {\bibfnamefont {E.~E.}\ \bibnamefont
  {Borrero}}, \ and\ \bibinfo {author} {\bibfnamefont {J.~C.}\ \bibnamefont
  {Araque}},\ }\href@noop {} {\bibfield  {journal} {\bibinfo  {journal} {J.
  Phys.-Condes. Matter}\ }\textbf {\bibinfo {volume} {21}},\ \bibinfo {pages}
  {333101} (\bibinfo {year} {2009})}\BibitemShut {NoStop}%
\bibitem [{\citenamefont {Booth}\ and\ \citenamefont
  {Hendricks}(1984)}]{split1}%
  \BibitemOpen
  \bibfield  {author} {\bibinfo {author} {\bibfnamefont {T.~E.}\ \bibnamefont
  {Booth}}\ and\ \bibinfo {author} {\bibfnamefont {J.~S.}\ \bibnamefont
  {Hendricks}},\ }\href@noop {} {\bibfield  {journal} {\bibinfo  {journal}
  {{Nucl. Techno.-Fus.}}\ }\textbf {\bibinfo {volume} {{5}}},\ \bibinfo {pages}
  {90} (\bibinfo {year} {{1984}})}\BibitemShut {NoStop}%
\bibitem [{\citenamefont {Melnik-Melnikov}\ and\ \citenamefont
  {Dekhtyaruk}(2000)}]{split2}%
  \BibitemOpen
  \bibfield  {author} {\bibinfo {author} {\bibfnamefont {P.}~\bibnamefont
  {Melnik-Melnikov}}\ and\ \bibinfo {author} {\bibfnamefont {E.}~\bibnamefont
  {Dekhtyaruk}},\ }\href@noop {} {\bibfield  {journal} {\bibinfo  {journal}
  {{Probab. Eng. Eng. Mech.}}\ }\textbf {\bibinfo {volume} {{15}}},\ \bibinfo
  {pages} {125} (\bibinfo {year} {{2000}})}\BibitemShut {NoStop}%
\bibitem [{\citenamefont {Villenaltamirano}\ and\ \citenamefont
  {Villenaltamirano}(1991)}]{RESTART}%
  \BibitemOpen
  \bibfield  {author} {\bibinfo {author} {\bibfnamefont {M.}~\bibnamefont
  {Villenaltamirano}}\ and\ \bibinfo {author} {\bibfnamefont {J.}~\bibnamefont
  {Villenaltamirano}},\ }in\ \href@noop {} {\emph {\bibinfo {booktitle}
  {Queueing, Performance and Control in Atm}}},\ \bibinfo {series}
  {North-Holland Studies in Telecommunication}, Vol.~\bibinfo {volume} {15},\
  \bibinfo {editor} {edited by\ \bibinfo {editor} {\bibfnamefont {J.~W.}\
  \bibnamefont {Cohen}}\ and\ \bibinfo {editor} {\bibfnamefont {C.~D.}\
  \bibnamefont {Pack}}}\ (\bibinfo  {publisher} {Elsevier Science Publ B V},\
  \bibinfo {address} {Amsterdam},\ \bibinfo {year} {1991})\ pp.\ \bibinfo
  {pages} {71--76},\ \bibinfo {note} {13th International Teletraffic Congress (
  ITC-13 ), Copenhagen, Denmark, Jun 19-26, 1991}\BibitemShut {NoStop}%
\bibitem [{\citenamefont {Hanggi}\ \emph {et~al.}(1990)\citenamefont {Hanggi},
  \citenamefont {Talkner},\ and\ \citenamefont {Borkovec}}]{50Kramer}%
  \BibitemOpen
  \bibfield  {author} {\bibinfo {author} {\bibfnamefont {P.}~\bibnamefont
  {Hanggi}}, \bibinfo {author} {\bibfnamefont {P.}~\bibnamefont {Talkner}}, \
  and\ \bibinfo {author} {\bibfnamefont {M.}~\bibnamefont {Borkovec}},\
  }\href@noop {} {\bibfield  {journal} {\bibinfo  {journal} {Rev. Mod. Phys.}\
  }\textbf {\bibinfo {volume} {62}},\ \bibinfo {pages} {251} (\bibinfo {year}
  {1990})}\BibitemShut {NoStop}%
\bibitem [{\citenamefont {Abraham}\ \emph
  {et~al.}(2015{\natexlab{b}})\citenamefont {Abraham}, \citenamefont {Murtola},
  \citenamefont {Schulz}, \citenamefont {P{\'a}ll}, \citenamefont {Smith},
  \citenamefont {Hess},\ and\ \citenamefont {Lindahl}}]{abraham2015gromacs}%
  \BibitemOpen
  \bibfield  {author} {\bibinfo {author} {\bibfnamefont {M.~J.}\ \bibnamefont
  {Abraham}}, \bibinfo {author} {\bibfnamefont {T.}~\bibnamefont {Murtola}},
  \bibinfo {author} {\bibfnamefont {R.}~\bibnamefont {Schulz}}, \bibinfo
  {author} {\bibfnamefont {S.}~\bibnamefont {P{\'a}ll}}, \bibinfo {author}
  {\bibfnamefont {J.~C.}\ \bibnamefont {Smith}}, \bibinfo {author}
  {\bibfnamefont {B.}~\bibnamefont {Hess}}, \ and\ \bibinfo {author}
  {\bibfnamefont {E.}~\bibnamefont {Lindahl}},\ }\href@noop {} {\bibfield
  {journal} {\bibinfo  {journal} {SoftwareX}\ }\textbf {\bibinfo {volume}
  {1}},\ \bibinfo {pages} {19} (\bibinfo {year}
  {2015}{\natexlab{b}})}\BibitemShut {NoStop}%
\bibitem [{\citenamefont {Jorgensen}\ \emph {et~al.}(1996)\citenamefont
  {Jorgensen}, \citenamefont {Maxwell},\ and\ \citenamefont
  {Tirado-Rives}}]{jorgensen_development_1996}%
  \BibitemOpen
  \bibfield  {author} {\bibinfo {author} {\bibfnamefont {W.~L.}\ \bibnamefont
  {Jorgensen}}, \bibinfo {author} {\bibfnamefont {D.~S.}\ \bibnamefont
  {Maxwell}}, \ and\ \bibinfo {author} {\bibfnamefont {J.}~\bibnamefont
  {Tirado-Rives}},\ }\href@noop {} {\bibfield  {journal} {\bibinfo  {journal}
  {Journal of the American Chemical Society}\ }\textbf {\bibinfo {volume}
  {118}},\ \bibinfo {pages} {11225} (\bibinfo {year} {1996})}\BibitemShut
  {NoStop}%
\bibitem [{\citenamefont {Abascal}\ and\ \citenamefont
  {Vega}(2005)}]{abascal_general_2005}%
  \BibitemOpen
  \bibfield  {author} {\bibinfo {author} {\bibfnamefont {J.~L.~F.}\
  \bibnamefont {Abascal}}\ and\ \bibinfo {author} {\bibfnamefont
  {C.}~\bibnamefont {Vega}},\ }\href {\doibase 10.1063/1.2121687} {\bibfield
  {journal} {\bibinfo  {journal} {J. Chem. Phys.}\ }\textbf {\bibinfo {volume}
  {123}},\ \bibinfo {pages} {234505} (\bibinfo {year} {2005})}\BibitemShut
  {NoStop}%
\bibitem [{\citenamefont {Bussi}\ \emph {et~al.}(2007)\citenamefont {Bussi},
  \citenamefont {Donadio},\ and\ \citenamefont
  {Parrinello}}]{bussi_canonical_2007}%
  \BibitemOpen
  \bibfield  {author} {\bibinfo {author} {\bibfnamefont {G.}~\bibnamefont
  {Bussi}}, \bibinfo {author} {\bibfnamefont {D.}~\bibnamefont {Donadio}}, \
  and\ \bibinfo {author} {\bibfnamefont {M.}~\bibnamefont {Parrinello}},\
  }\href {\doibase 10.1063/1.2408420} {\bibfield  {journal} {\bibinfo
  {journal} {J. Chem. Phys.}\ }\textbf {\bibinfo {volume} {126}},\ \bibinfo
  {pages} {014101} (\bibinfo {year} {2007})}\BibitemShut {NoStop}%
\bibitem [{\citenamefont {Tiwari}\ and\ \citenamefont
  {Ensing}(2016)}]{tiwari_reactive_2016}%
  \BibitemOpen
  \bibfield  {author} {\bibinfo {author} {\bibfnamefont {A.}~\bibnamefont
  {Tiwari}}\ and\ \bibinfo {author} {\bibfnamefont {B.}~\bibnamefont
  {Ensing}},\ }\href@noop {} {\bibfield  {journal} {\bibinfo  {journal}
  {Faraday Discussions}\ }\textbf {\bibinfo {volume} {195}},\ \bibinfo {pages}
  {291} (\bibinfo {year} {2016})}\BibitemShut {NoStop}%
\bibitem [{\citenamefont {Kohn}\ and\ \citenamefont {Sham}(1965)}]{KS}%
  \BibitemOpen
  \bibfield  {author} {\bibinfo {author} {\bibfnamefont {W.}~\bibnamefont
  {Kohn}}\ and\ \bibinfo {author} {\bibfnamefont {L.~J.}\ \bibnamefont
  {Sham}},\ }\href {\doibase 10.1103/PhysRev.140.A1133} {\bibfield  {journal}
  {\bibinfo  {journal} {Phys. Rev.}\ }\textbf {\bibinfo {volume} {140}},\
  \bibinfo {pages} {A1133} (\bibinfo {year} {1965})}\BibitemShut {NoStop}%
\bibitem [{\citenamefont {Riccardi}\ \emph {et~al.}(2019)\citenamefont
  {Riccardi}, \citenamefont {Pantano},\ and\ \citenamefont
  {Potestio}}]{2019envisioning}%
  \BibitemOpen
  \bibfield  {author} {\bibinfo {author} {\bibfnamefont {E.}~\bibnamefont
  {Riccardi}}, \bibinfo {author} {\bibfnamefont {S.}~\bibnamefont {Pantano}}, \
  and\ \bibinfo {author} {\bibfnamefont {R.}~\bibnamefont {Potestio}},\
  }\href@noop {} {\bibfield  {journal} {\bibinfo  {journal} {Interface Focus}\
  }\textbf {\bibinfo {volume} {9}},\ \bibinfo {pages} {20190005} (\bibinfo
  {year} {2019})}\BibitemShut {NoStop}%
\bibitem [{\citenamefont {Lamprecht}\ \emph {et~al.}(2020)\citenamefont
  {Lamprecht}, \citenamefont {Garcia}, \citenamefont {Kuzak}, \citenamefont
  {Martinez}, \citenamefont {Arcila}, \citenamefont {Martin Del~Pico},
  \citenamefont {Dominguez Del~Angel}, \citenamefont {Van De~Sandt},
  \citenamefont {Ison}, \citenamefont {Martinez} \emph
  {et~al.}}]{lamprecht2020towards}%
  \BibitemOpen
  \bibfield  {author} {\bibinfo {author} {\bibfnamefont {A.-L.}\ \bibnamefont
  {Lamprecht}}, \bibinfo {author} {\bibfnamefont {L.}~\bibnamefont {Garcia}},
  \bibinfo {author} {\bibfnamefont {M.}~\bibnamefont {Kuzak}}, \bibinfo
  {author} {\bibfnamefont {C.}~\bibnamefont {Martinez}}, \bibinfo {author}
  {\bibfnamefont {R.}~\bibnamefont {Arcila}}, \bibinfo {author} {\bibfnamefont
  {E.}~\bibnamefont {Martin Del~Pico}}, \bibinfo {author} {\bibfnamefont
  {V.}~\bibnamefont {Dominguez Del~Angel}}, \bibinfo {author} {\bibfnamefont
  {S.}~\bibnamefont {Van De~Sandt}}, \bibinfo {author} {\bibfnamefont
  {J.}~\bibnamefont {Ison}}, \bibinfo {author} {\bibfnamefont {P.~A.}\
  \bibnamefont {Martinez}},  \emph {et~al.},\ }\href@noop {} {\bibfield
  {journal} {\bibinfo  {journal} {Data Science}\ }\textbf {\bibinfo {volume}
  {3}},\ \bibinfo {pages} {37} (\bibinfo {year} {2020})}\BibitemShut {NoStop}%
\end{thebibliography}%

\clearpage 
    \includegraphics[width=0.4\textwidth]{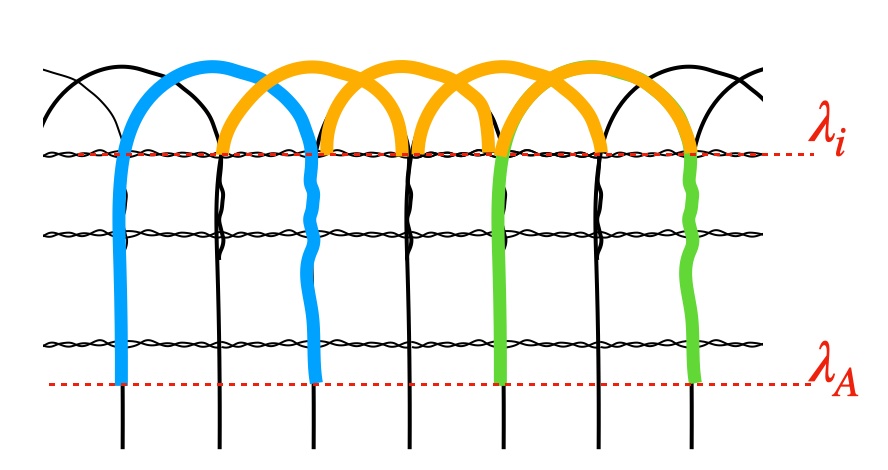}\\
    \centering
    TOC

\end{document}